\documentclass[fleqn,usenatbib]{mnras}

\usepackage{color}
\usepackage{newtxtext,newtxmath,comment}

\usepackage[T1]{fontenc}

\DeclareRobustCommand{\VAN}[3]{#2}
\let\VANthebibliography\thebibliography
\def\thebibliography{\DeclareRobustCommand{\VAN}[3]{##3}\VANthebibliography}

\usepackage{graphicx}	
\usepackage{amsmath}	

\title[HCF4]{HCF4 long}

\newcommand\cm{{\rm\thinspace cm}}

\newcommand\erg{{\rm\thinspace erg}}

\newcommand\K{{\rm\thinspace K}}
\newcommand\keV{{\rm\thinspace keV}}
\newcommand\km{{\rm\thinspace km}}

\newcommand\Mpc{{\rm\thinspace Mpc}}
\newcommand\Msun{\hbox{$\rm\thinspace M_{\odot}$}}

\newcommand\pc{{\rm\thinspace pc}}

\newcommand\s{{\rm\thinspace s}}
\newcommand\yr{{\rm\thinspace yr}}

\newcommand\cmsq{\hbox{$\cm^2\,$}}
\newcommand\cmcu{\hbox{$\cm^3\,$}}

\newcommand\pcmcuK{\hbox{$\cm^{-3}\K\,$}}

\newcommand\ergcmcups{\hbox{$\erg\cm^3\ps\,$}}

\newcommand\ergpcmsqps{\hbox{$\erg\cm^{-2}\s^{-1}\,$}}

\newcommand\ergps{\hbox{$\erg\s^{-1}\,$}}

\newcommand\kmpspmpc{\hbox{$\km\,\s^{-1}\Mpc^{-1}\,$}}

\newcommand\Msunpyr{\hbox{$\Msun\yr^{-1}\,$}}

\newcommand\pcmsq{\hbox{$\cm^{-2}\,$}}
\newcommand\pcmK{\hbox{$\cm^{-3}\K$}}

\newcommand\ps{\hbox{$\s^{-1}\,$}}
\newcommand\psqcm{\hbox{$\cm^{-2}\,$}}

\title[Hidden Cooling Flows IV]{Hidden Cooling Flows IV: More Details on  Centaurus and the Efficiency of AGN Feedback in Clusters}

\author[A. C. Fabian et al.]{
A. C. Fabian,$^{1}$\thanks{E-mail: acf@ast.cam.ac.uk }, G.J. Ferland$^{3}$, J.S. Sanders$^{2}$,  H.R. Russell$^{4}$, B.R. McNamara$^{5}$, C. Pinto$^{6}$, \newauthor J. Hlavacek-Larrondo$^7$, S.A. Walker$^{8}$, L.R. Ivey$^{1}$ and M. McDonald$^{9,10}$
\\
$^{1}$Institute of Astronomy, University of Cambridge, Madingley Road, Cambridge CB3 0HA, UK\\
$^{2} $Department of Physics, University of Kentucky, Lexington KY 40506, USA\\
$^{3} $Max-Planck-Institut fur extraterrestrische Physik, Giessenbachstrasse 1, 85748 Garching, Germany\\
$^{4} $School of Physics \& Astronomy, University of Nottingham, University Park, Nottingham NG7 2RD, UK\\
$^{5} $Department of Physics and Astronomy, University of Waterloo, 200 University Avenue West, Waterloo, ON N2L 3G1, Canada\\
$^{6}$INAF-IASF Palermo, Via U. La Malfa 153, I-90146 Palermo, Italy\\
$^{7}$D\'{e}partement de Physique, Universit\'{e} de Montr\'{e}al, Succ. Centre-Ville, Montr\'{e}al, Qu\'{e}bec, H3C 3J7, Canada\\
$^{8}$Department of Physics and Astronomy, The University of Alabama in Huntsville, Huntsville, AL 35899, USA\\ 
$^{9}$Department of Physics, Massachusetts Institute of Technology, Cambridge, MA 02139, USA\\
$^{10}$Kavli Institute for Astrophysics and Space Research, Massachusetts Institute of Technology, 77 Massachusetts Avenue, Cambridge, MA 02139, USA\\
}
\date{Accepted XXX. Received YYY; in original form ZZZ}
\pubyear{2024}
\begin{document}
\label{firstpage}
\pagerange{\pageref{firstpage}--\pageref{lastpage}}
\maketitle
\begin{abstract}
Cooling flows are common in galaxy clusters which have cool cores. The soft X-ray emission below 1 keV from the flows  is mostly absorbed by cold dusty gas within the central cooling sites. Further evidence for this process is presented here through a more detailed analysis of the nearby Centaurus cluster and some additional clusters. Predictions of JWST near and mid-infrared spectra from cooling gas are presented. [NeVI] emission at $7.65\mu$m should be an important diagnostic of gas cooling between 6 and $1.5\times 10^5\K$. The emerging overall picture of hidden cooling flows is explored. The efficiency of AGN feedback in reducing the total cooling rate in cool cores is shown to be above 50 percent for many clusters but is rarely above 90 per cent. The reduction is mostly in outer gas.  Cooling dominates in elliptical galaxies and galaxy groups which have mass flow rates below about $15\Msunpyr$ and in some massive clusters where  rates can exceed $1000\Msunpyr.$ 
\end{abstract}

\begin{keywords}
galaxies: clusters: intracluster medium
\end{keywords}


\section{Introduction}
We have found that cooling flows are operating in most early-type galaxies \citep[HCFI, II, III,] [] {2022MNRAS.515.3336F, 2023MNRAS.521.1794F, 2023MNRAS.524..716F}. The rapidly cooling, soft X-ray part of these flows is mostly hidden from direct view by photoelectric absorption in surrounding cold gas, much of which may be the product of the cooling process. Such Hidden Cooling Flows (HCF) have mass cooling rates of $1-5\Msunpyr$ in typical elliptical galaxies rising to $10-20\Msunpyr$ in Brightest Group Galaxies (BGG) to $20-100\Msunpyr$ in Brightest Cluster Galaxies (BCGs). Some exceptional distant clusters have higher rates of $\dot M\sim 1000\Msunpyr$ or more. 

The realization that X-ray absorbing gas must be present in cool core clusters began in 1991 \citep{White1991} and its development thereafter is discussed in HCF1. It was initially added  outside  the soft X-ray emission region, but later work developed the intrinsic absorption model where the cold gas is dispersed through it \citep{Allen1997}. Our spectral fits produce contour plots of mass cooling rate (Mdot) against intrinsic column density ($N_{\rm H}$). The allowed regions in most objects avoid $(0,0)$ in those plots, demonstrating that the absorbed emission model is a better fit than one without absorption.

Here we re-analyse the cooling flow in the nearby Centaurus cluster (redshift $z=0.0104$) using a smaller aperture for the XMM Reflection Grating Spectrometers (RGS) data used than for the initial analysis presented in HCF1. We then briefly analyse spectral and imaging data of Centaurus using Chandra data. 

We next compare our results on 29 clusters, groups and elliptical galaxies, including those from 10 more clusters presented in the Appendix,  with mass cooling rates from Chandra imaging and star formation rates tabulated by \citep{McDonald2018}. A picture emerges in which the Hidden Mass Cooling Rates (HCR) are in approximate agreement with rates obtained from Chandra imaging (IMR), assuming an outer  radius of the cooling flow corresponding to where the  cooling time is  3 Gyr, provided neither rate exceeds $15\Msunpyr$. Higher values of IMR tend to have lower HCR by about a factor of 3  compared to the IMR until it exceeds $100\Msunpyr$. We propose in this paper that these trends are due to the effect of AGN Feedback. 

All cooling flows imply significant accumulations of cooled gas if they operate over several Gyr.  The rate of normal star formation typically fails by a factor of about 20. We have addressed this issue in HCF1-III and returned to it in \citep{Fabian24}. Low-mass star formation is the most plausible solution \citep{Fabian1982,Sarazin1983,Fabian24} and is briefly explored further in Section 5.1.

It is widely assumed in the literature that AGN feedback is  efficient in heating intracluster gas so that no cooling flows occur \citep{Mo10}. Here we find that the efficiency in clusters ranges from 50 to 95 per cent. It appears to be much lower, i.e. it is inefficient, in elliptical galaxies and groups of galaxies. 

\section{The Centaurus Cluster}
\subsection{The XMM RGS analysis} 

The RGS provides the spectrum from an effective slit of typically 90, 95 and 99 percent in the cross-dispersion direction, corresponding to 0.8, 1.7 and 3.4 arcmin width. \cite{Sanders2008} show that for Centaurus most of the lines from low temperature gas are captured within the 90 per cent aperture. We therefore use that here. RGS 1 and 2 have dead chips covering 10.4--13.8 \AA\ and 20--24 \AA\ respectively. 
\begin{figure}
    \centering    
\includegraphics[width=0.48\textwidth]{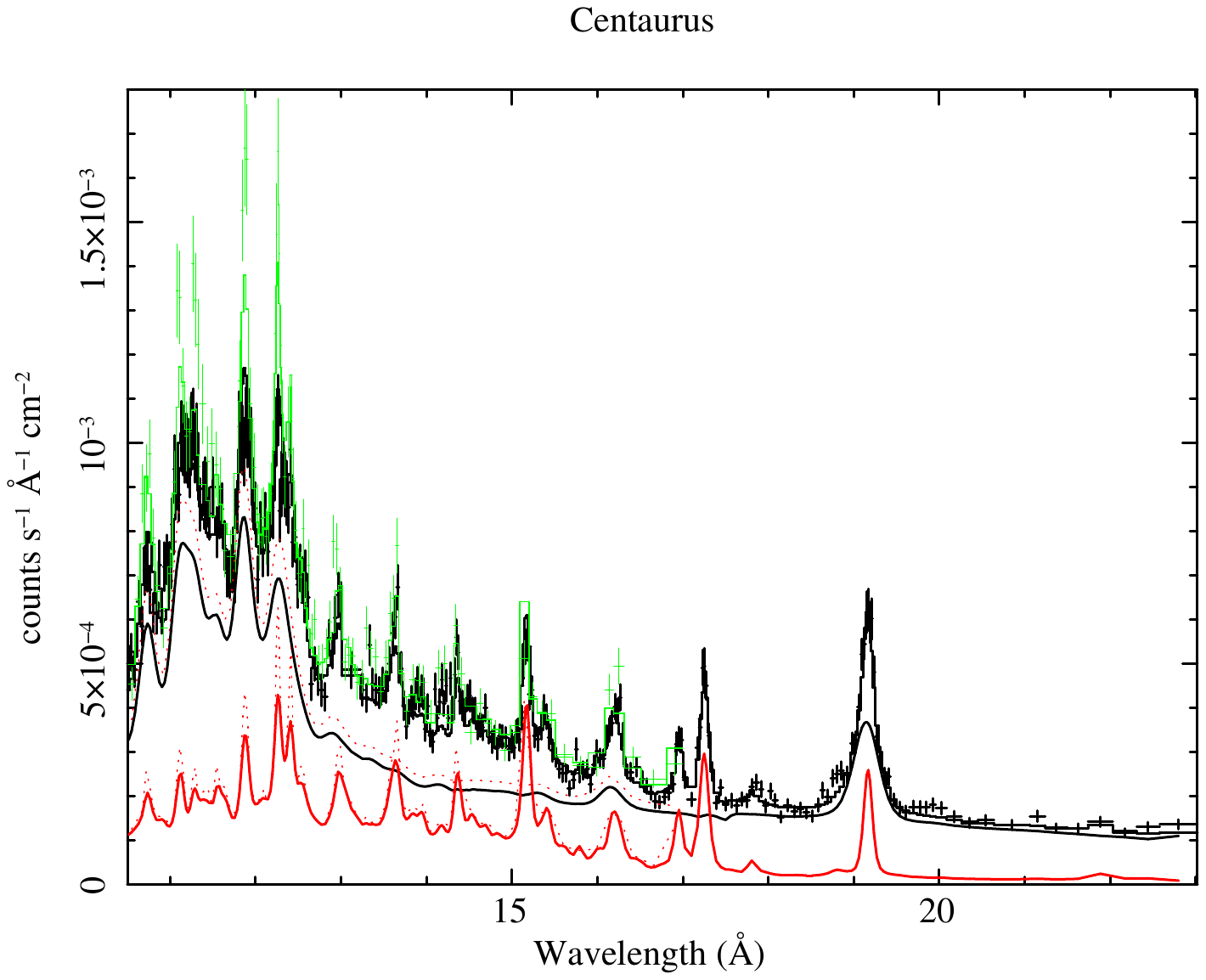} 
\includegraphics[width=0.48\textwidth]{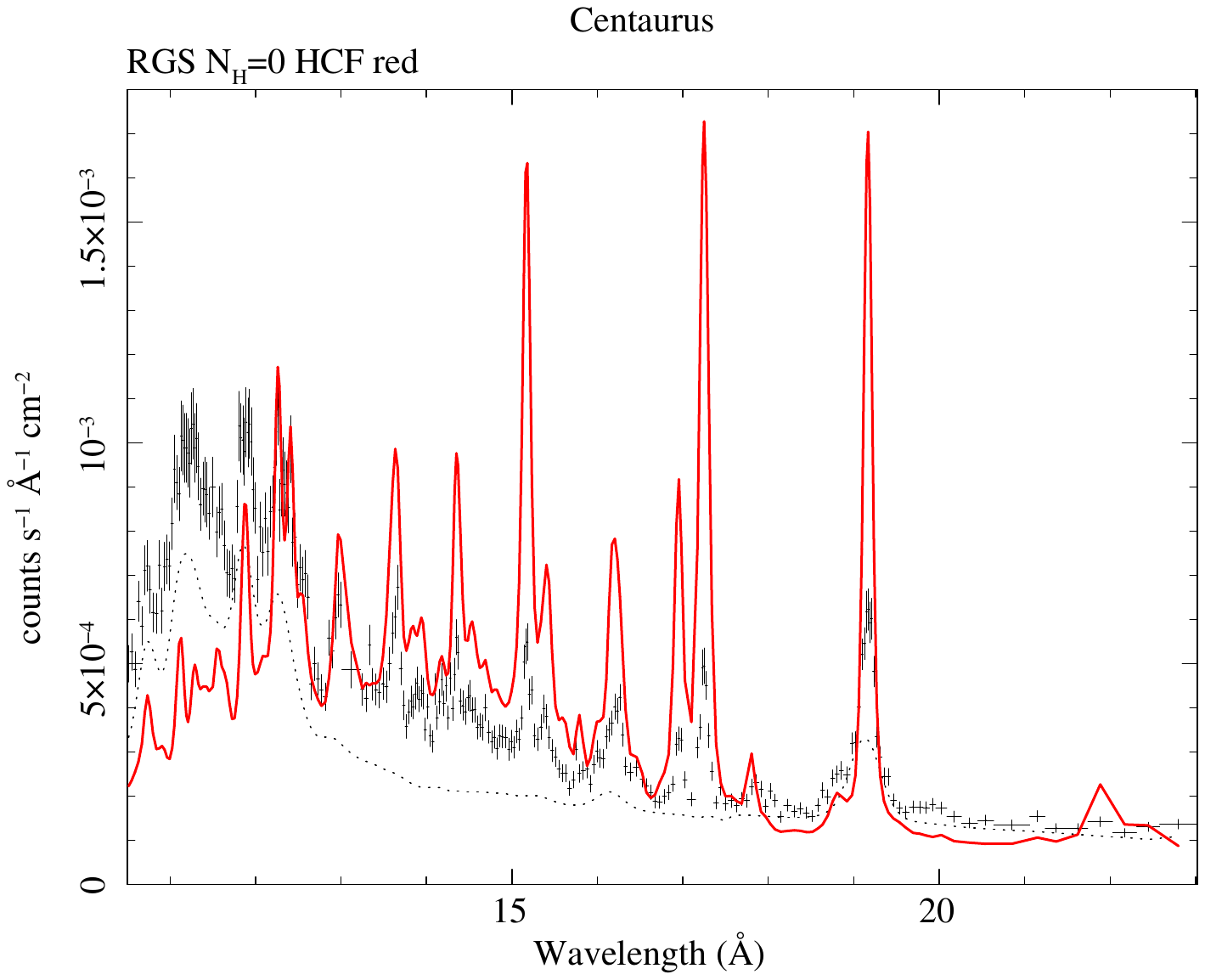}
    \caption{  a) XMM 90\% RGS spectrum of the Centaurus cluster in first and second order (black and green) with the Hidden Cooling Flow component in red. b) The same first order spectrum is in a) but with the intrinsic column density set to zero. This is what the cooling flow component would look like with no absorption.  }
\end{figure}

\begin{figure}
    \centering    
\includegraphics[width=0.45\textwidth]{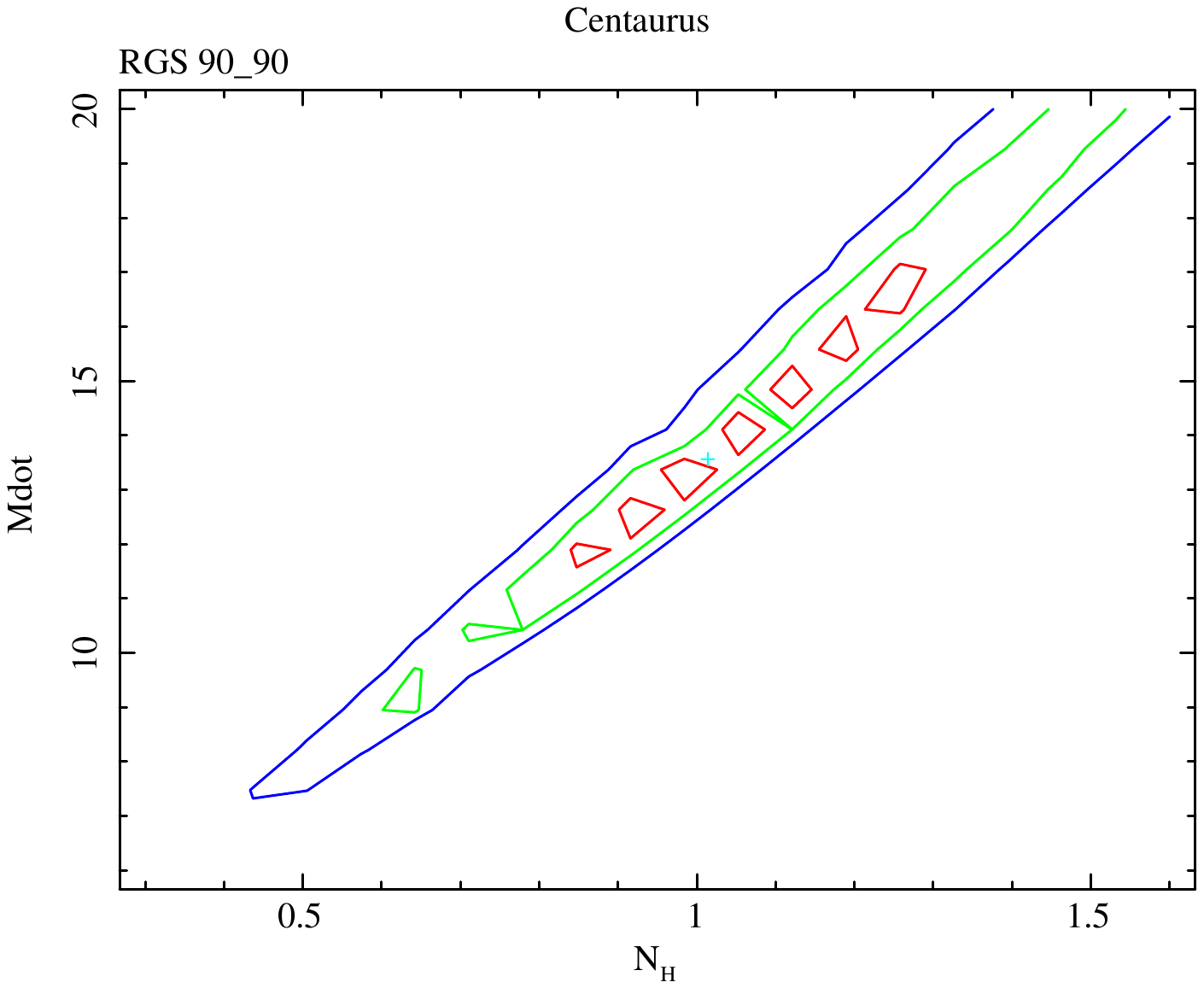} 
\includegraphics[width=0.45\textwidth]{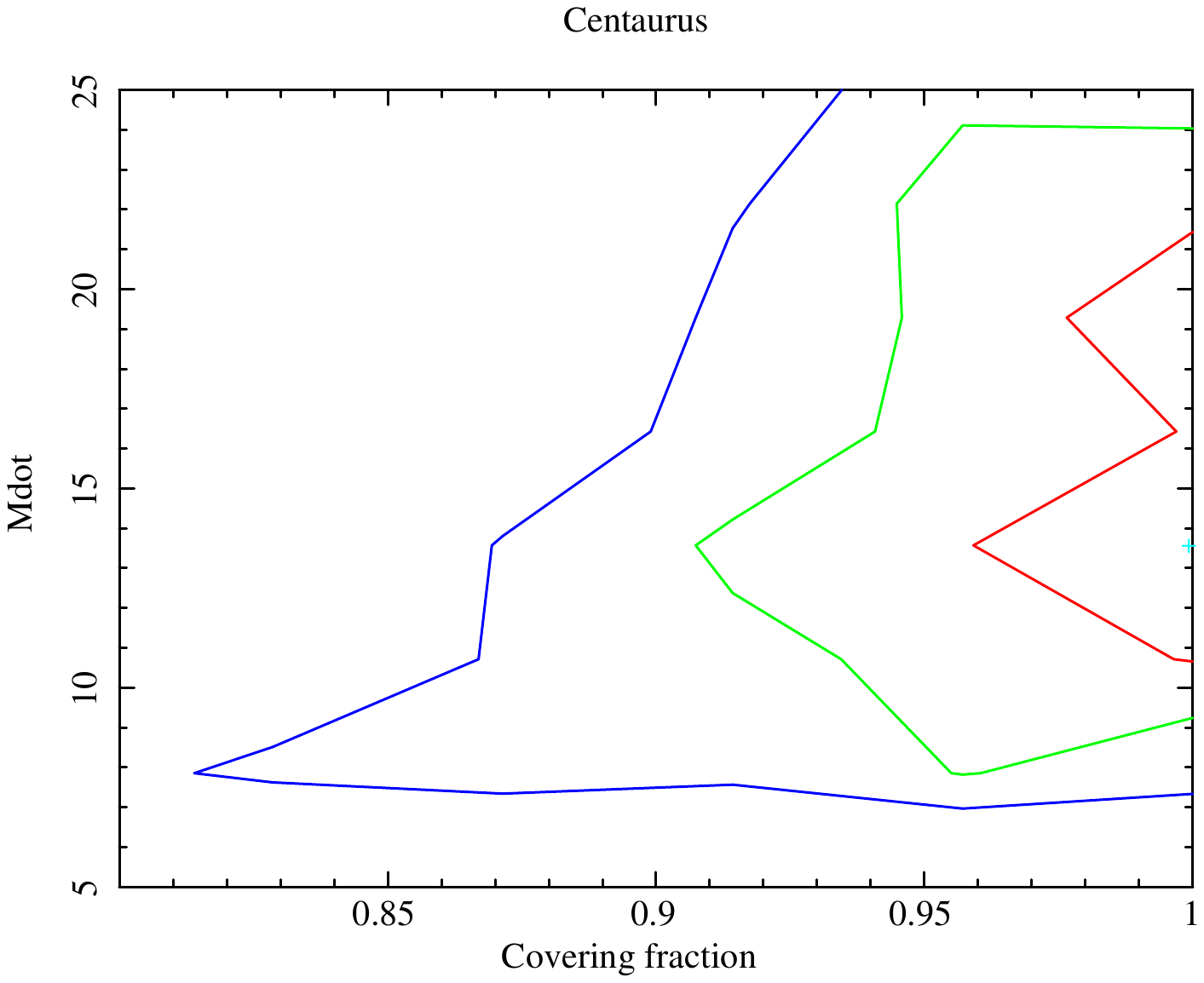} 
\includegraphics[width=0.45\textwidth]{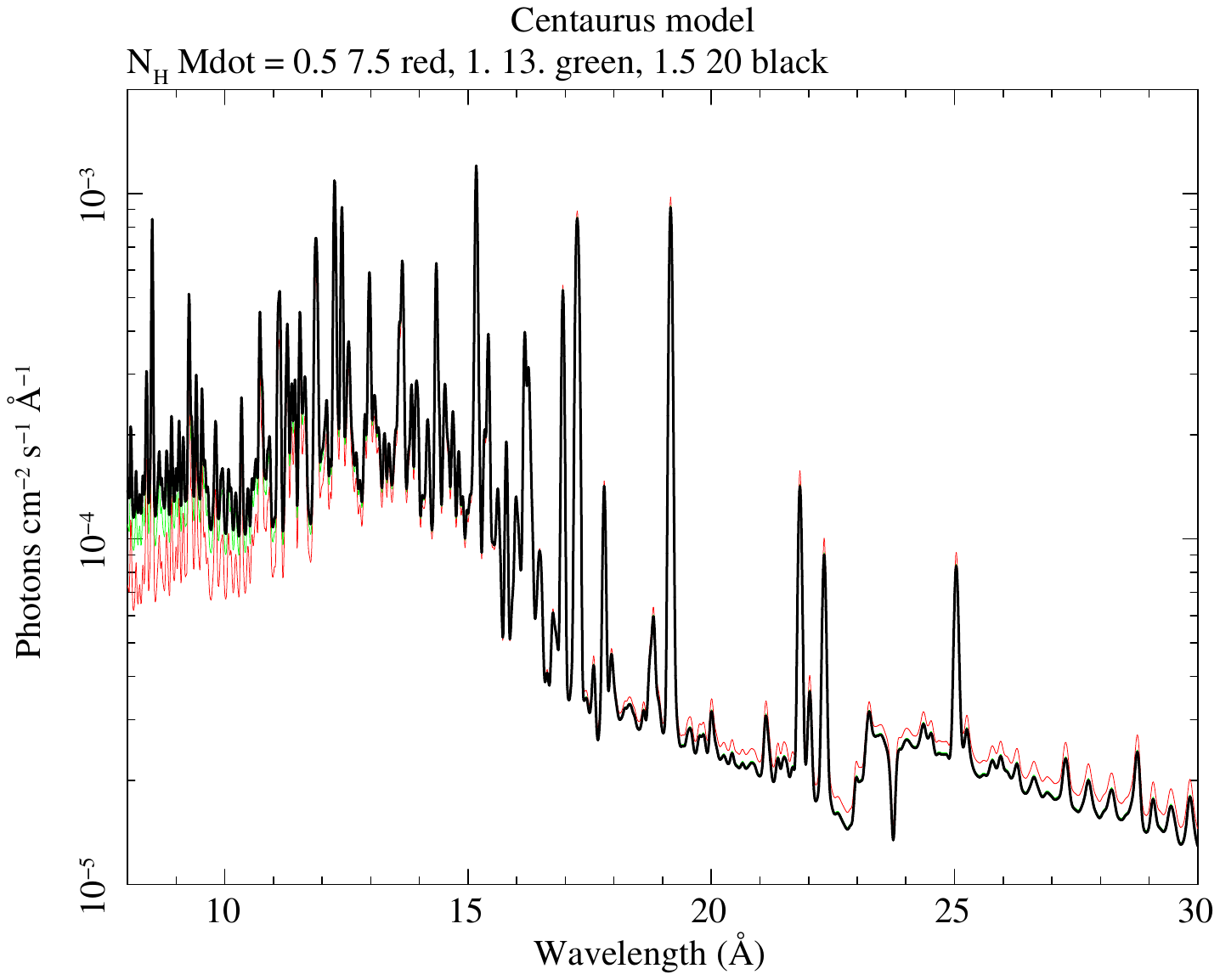}
    \caption{ a) Allowed region in the column density ($N_{\rm H}$ in units of $10^{22}\cmsq$) - mass cooling rate (Mdot in units of $\Msunpyr$) plane for the spectrum in Fig 1a). b) Similar region for the covering fraction and c) Model spectra for 3 fits along the allowed region in 2 a). Note that they are very similar above 12\AA.   }
\end{figure}

\begin{figure}
    \centering    
\includegraphics[width=0.48\textwidth]{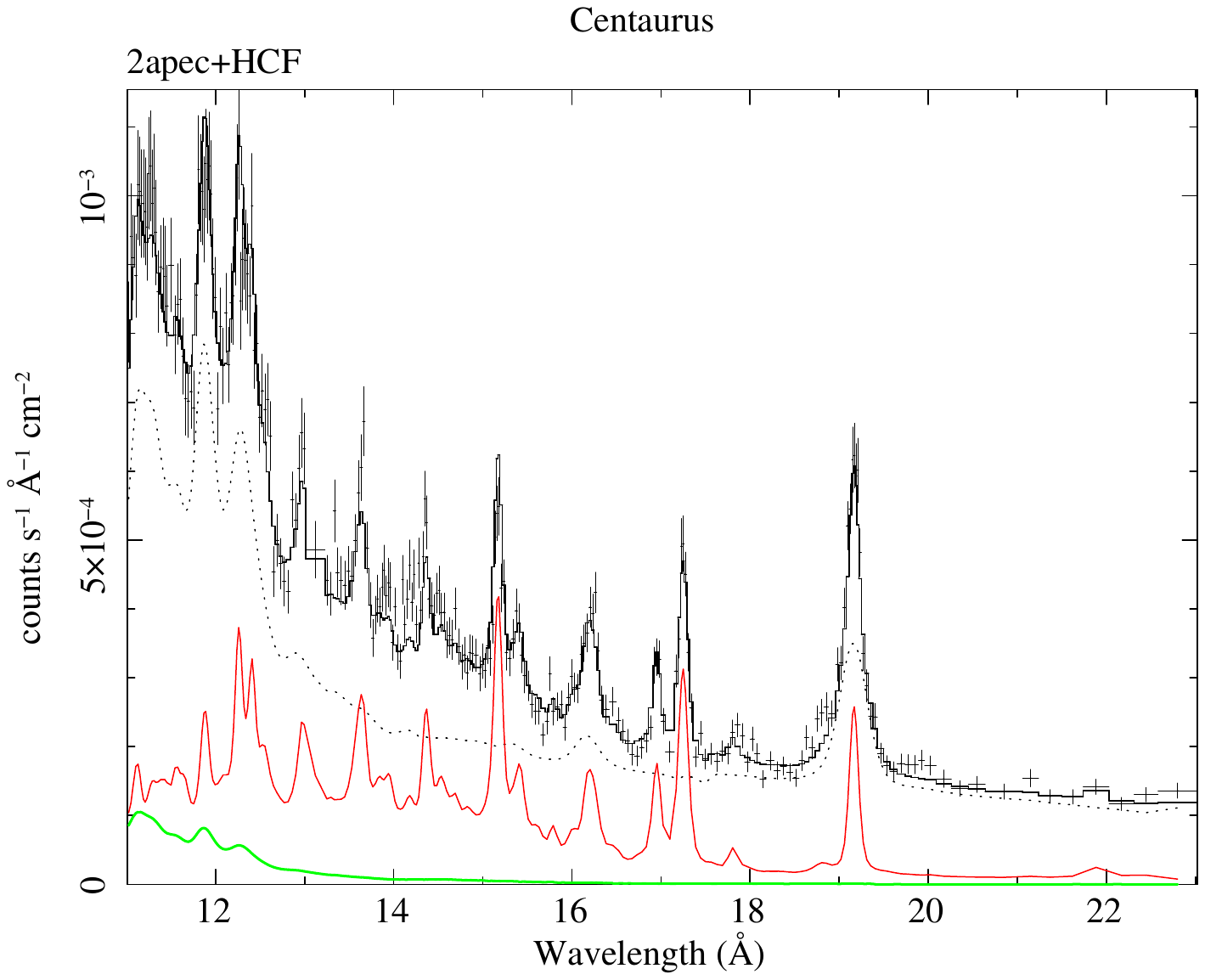} 
\includegraphics[width=0.48\textwidth]{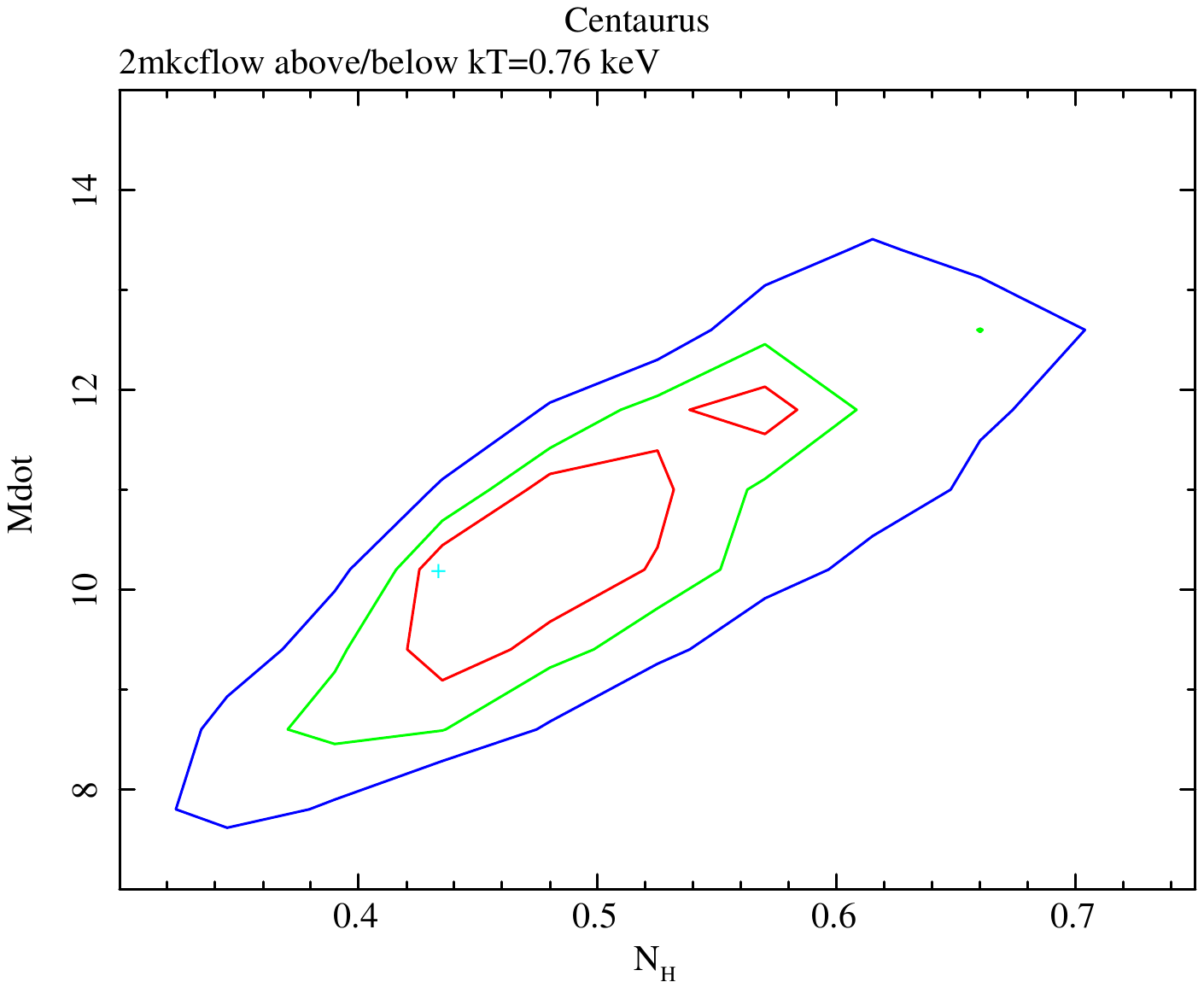}
\includegraphics[width=0.48\textwidth]{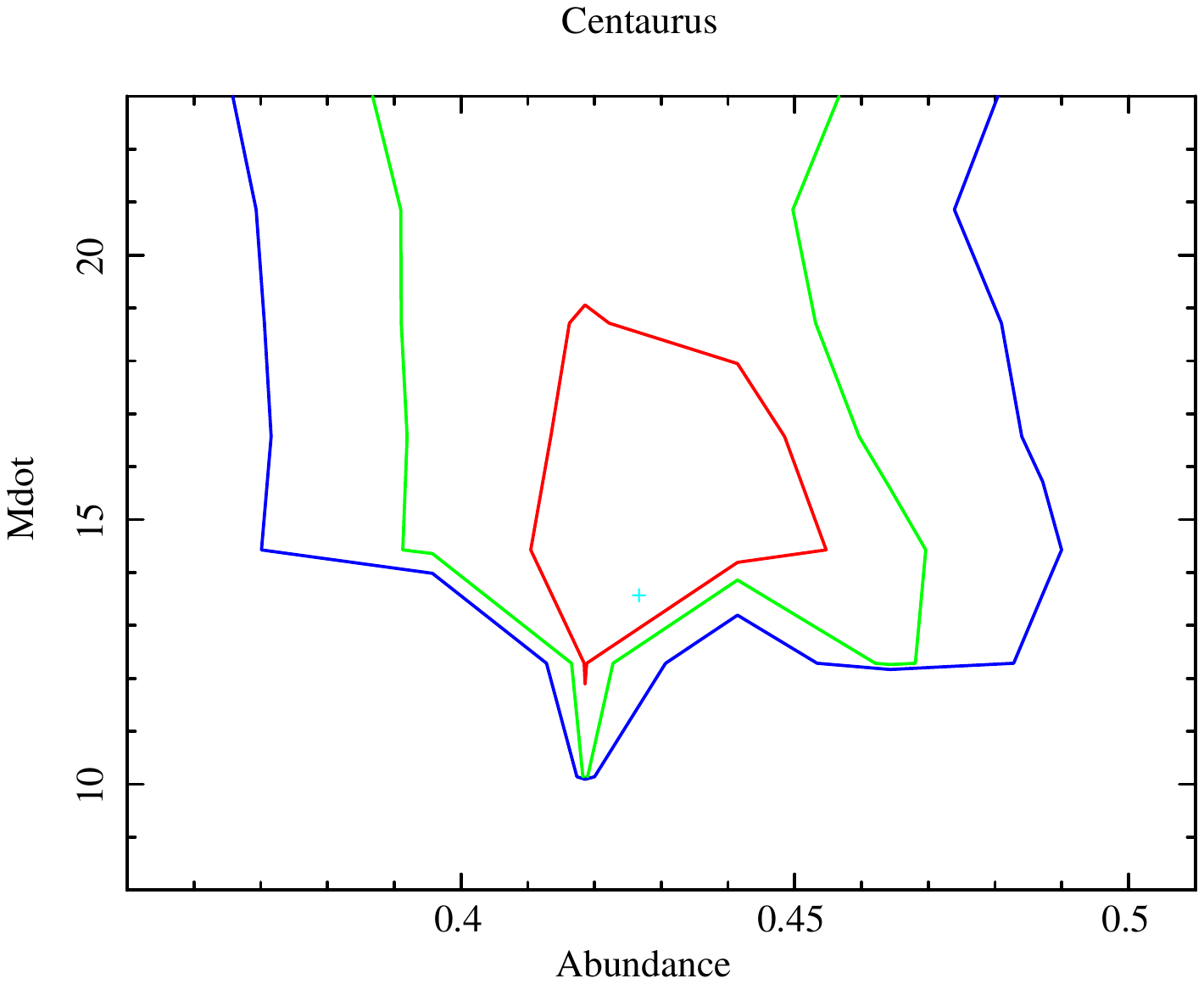}
    \caption{ Top: Centaurus RGS spectrum fitted with emission from the back side (green). Middle: Mdot vs column density $N_{\rm H}$ when the cooling flow is absorbed by a higher column density below 0.76 keV. Bottom: Mass flow rate versus abundance. }
\end{figure}

The corrected spectrum is shown in Fig. 1 with the best fitting Hidden Cooling Flow model (HCF) applied. The total model consists of Galactic absorption applied overall to a single temperature \textsc{APEC} component plus a multilayer absorption model \textsc{ mlayerz} ($(1-\exp(-\sigma N_{\rm H}))/(\sigma N_{\rm H}))$, where $\sigma$ is the photoelectric cross section at the relevant energy and $N_{\rm H}$ is the total column density) applied to a cooling flow model \textsc{mkcflow}. The higher temperature of the cooling flow is fixed to equal that of the single temperature component and the lower temperature is $0.1\keV$. As the spectrum is from an extended object, gaussian smoothing is applied separately to the 2 components. Fig. 1a  shows the spectrum and  best fitting components. Fig. 2b shows the underlying cooling flow component without the multilayered absorption. It is clear that the cooling flow would have been immediately obvious in the first RGS observations had there been no excess absorption.

Fig. 2a shows the mass cooling rate (Mdot) as a function of the total column density $N_{\rm H}$ and Fig. 2b shows Mdot versus Covering Fraction, which is mostly greater than $0.9$. 
Fig. 2c shows the model absorbed spectra for three ($N_{\rm H}$, Mdot) pairs along the narrow best fitting region of Fig. 2a. 
The 3 spectra are very similar above 12\AA. The divergence at shorter wavelengths  is due to the absorption becoming optically thin. The behaviour above 12\AA\ is due to the \textsc{mlayerz} absorption approximating to a powerlaw in energy as $\sigma N_{\rm H}$ increases rather than the usual simple exponential characteristic of single-layer absorption.

If we were dealing with a perfect spectrum of the complete region out to several arcmin then we can determine a single value of Mdot. However we have  a view that is restricted in a complicated manner. Instead we rely on the Far InfraRed (FIR) emission detected by Herschel \citep{Mittal2011} and equate that to the absorbed soft X-ray flux as discussed in HCFI, which gives an Mdot  of about $14\Msunpyr$ and $N_{\rm H}\sim 10^{22}\pcmsq$.

The above model does not include the emission from the hot intracluster gas on the other side of the absorbing core. To accurately include this we would need to know the extent and geometry of the absorbing material. In the absence of that knowledge we assume that half of the \textsc{APEC} component originates from the front and the other half is behind the core, absorbed by a total column density of $N_{\rm H}$. The resulting backside emission is shown in green in Fig. 3. and has a negligible effect on the hidden cooling flow above 13\AA. At shorter wavelengths it can be confused with the uncertain emission from larger radii.

A final issue that we investigated is connected with the simplicity of the \textsc{mlayerz} model used on what is likely to be a very complex region. Is the absorption  the same for all temperatures? In essence, we assume that the same complete cooling flow lies behind each interleaved layer of absorbing matter. To test this we split the cooling flow model in two, the first part cooling from the outer temperature to a lower one, $T_{\rm low}$ the second part then cooling from $T_{\rm low}$ to 0.1 keV. $\dot M$ is the same for both parts and $N_{\rm H}$ allowed to be different. The resulting fit for $\dot M$ vs the  value of $N_{\rm H}$  for the hotter flow is shown in Fig. 3b. The value of $N_{\rm H}$ for the cooler flow is 2.5 times larger. We checked for this effect in a few other clusters and found that it was not needed in 2A 0335+096 but did improve the fit for HCG62, where  $T_{\rm low}=0.35\keV$ was required. The overall value of $\dot M$ is slightly reduced. Lastly, Fig. 3c shows the mass cooling rate as a function of abundance $Z$.

\subsection{Chandra observations of the Centaurus cool core}

\begin{figure}
    \centering    
\includegraphics[width=0.48\textwidth]{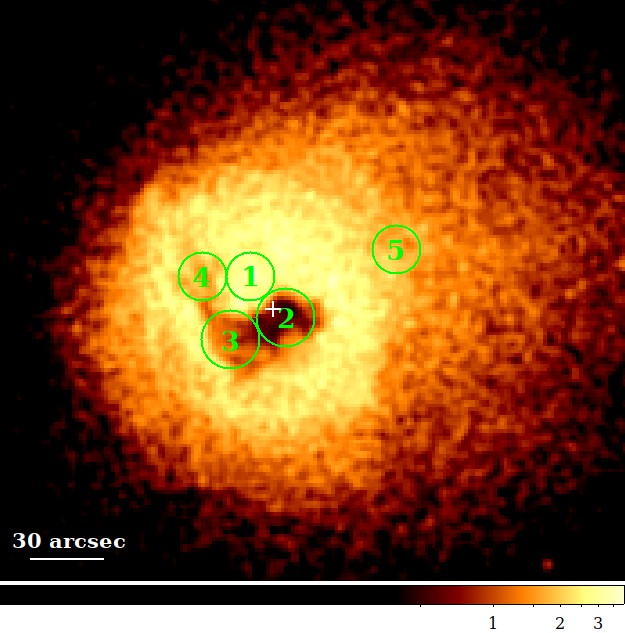} 
\includegraphics[width=0.48\textwidth]{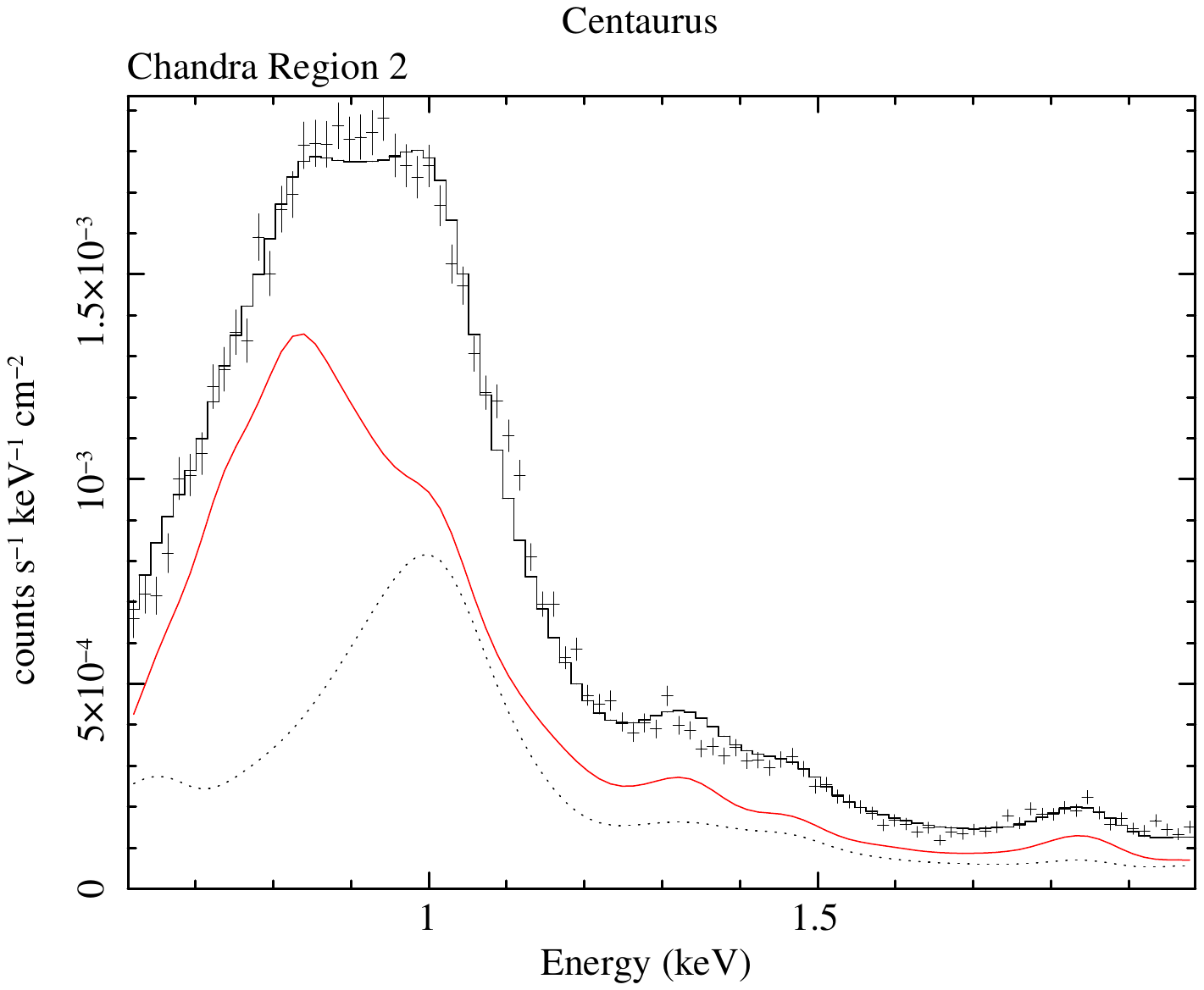} 
\includegraphics[width=0.48\textwidth]{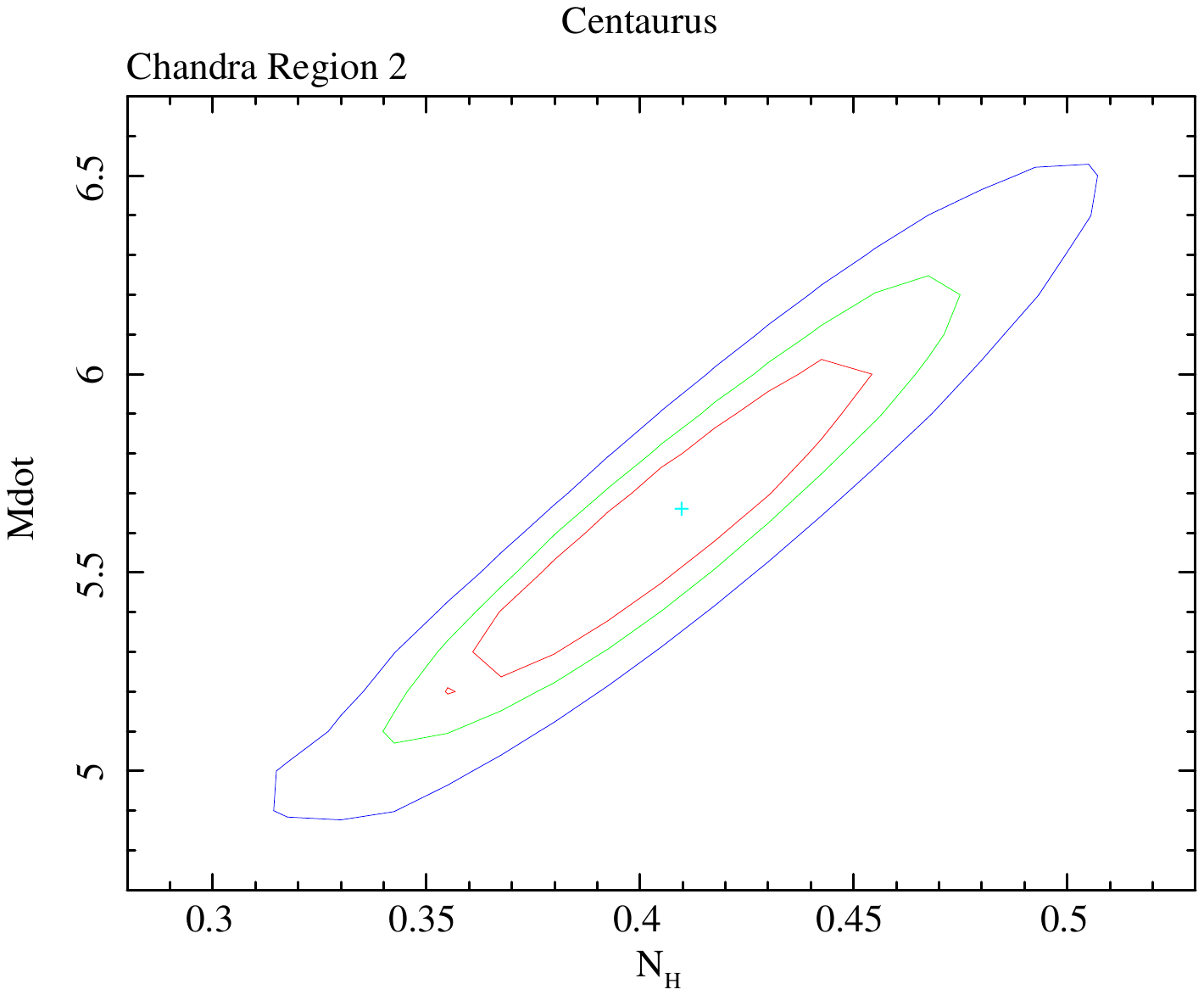} 
    \caption{ a) Chandra image of spectral colour (ratio of 1.05--2 keV divided by 0.65--0.95 keV) in the centre of the Centaurus cluster. b) Chandra Spectrum extracted from 12 arcsec radius region 2 in a) (with nucleus excised). the red line is the HCF. Bottom: HCR vs $N_{\rm H}$ for the spectrum shown above.}
\end{figure}

Chandra has observed the Centaurus cluster several times \citep{Crawford05, Sanders2016}. Using 280ks of early data, taken in 2000-2004 when the absorption due to material built up on the Optical Blocking Filter \citep{Plucinsky22} was minimal,   we made a colour image of the core (Fig. 4a shows the ratio of hard (1.05 - 2 kev) to soft mid-band (0.65 - 0.95 keV) X-rays smoothed with a 2D gaussian with sigma = 5 arcsec) . The nucleus is marked by the pale cross. The Chandra spectrum of region 2 (with the nucleus excised) is shown in Fig. 4b and contours in the $\dot M - N_{\rm H}$ plane are shown in Fig 4c.  

The Colour image shows where the multilayer absorption and emission is occurring as a 
darker colour compared with the lighter regions where it is minimal. It is not symmetrical and roughly matches the optical emission and absorption filamentation (see e.g. \cite{Fabian2017}). 
The Chandra HCF rates for regions 1, 2, 3, 4 and 5 are 0.24, $5.7\pm0.45$, 1.65, 0.85 and $<0.03\Msunpyr$, respectively.  The abundance rises from 0.67 in region 2 to 1.5 for region 5.  
The pattern shows that any simple deprojection is not appropriate and we defer any detailed multilayer analysis of the Chandra data for now.

The dark patch in Fig. 4a also  matches the low abundance region identified by \cite{Panagoulia2013}. Similar drops in central abundance were found and studied in other cool core clusters \citep{Panagoulia2015,Mernier17,Lakhchaura19a,Liu19a}. As discussed by \cite{Panagoulia2013}, stellar mass loss in the host galaxy should lead to central peaks in iron and other abundances. Drops in abundance were ascribed to outflows from ongoing feedback. Here we suggest that dust grains swept into the ongoing cooling flows may be another part of the explanation for the drops. They will of course contribute to the intrinsic absorption. Some of the "missing iron mass" of \cite{Panagoulia2013} may have been swept inward by the cooling flow.

\section{Emission from gas cooling below 2 Million K}
X-ray emission from OVII in cool cores was found in stacked RGS spectra by \cite{Sanders2011}. It was then studied in individual spectra of several elliptical galaxies by \cite{Pinto2016}. The ratio of FeXVII to OVII emission was larger then expected from simple cooling gas. A possible explanation for this result is that there is cold absorbing gas along our line of sight. The OVII doublet is at a wavelength just below the neutral O edge, so more sensitive to absorption. OVII emission from Centaurus was also reported by \citep{Fabian2016} at an implied level of about $1\Msunpyr$. This is likely due to the covering fraction not being  exactly unity (Fig. 2b), with small open patches.

FUV emission from OVI, expected from gas at $2.5 - 4\times 10^5K$ has been sought in cooling flows. As noted in HCF2, it was detected in the cluster A2597 by \cite{0egerle2001} and using FUSE in the elliptical galaxy NGC4636 by \cite{Bregman01}. Re-analysis by \cite{Bregman06} confirmed the A2597 result and revealed OVI emission in A1795 and the Perseus clusters.  \cite{Lecav04} found  limits on 2 more clusters (AWM7 and A3112). A tight limit on OVI emission corresponding to $<10\Msunpyr$ was also found using a small aperture COS observation with HST in RXJ1532 by \cite{Donahue17}. One the contrary, \cite{McDonald15} found OVI emission corresponding to $>1000\Msunpyr$ in the high luminosity Phoenix cluster.

OVI emission was detected by \cite{Bregman01} in the Virgo elliptical galaxy NGC4636 at a rate of $0.43\Msunpyr$. but gave only an upper limit of $0.3\Msunpyr$ on NGC1404. HCF in the ellipticals M49 and M84 reported in HCF2 at rates of 1 and $2\Msunpyr$, FIR detections \citep{Temi04, Temi2018}) and evidence of dust  (eg \cite{Goudfrooij95}) fit a picture of small absorbed cooling flows operating there.

The overall results on OVI may appear confusing with detections in some objects but not others. They are however consistent with the hidden cooling flow model in which some objects including A2597 (HCF3, also see \citep{Morris2005}) have unabsorbed components and A1795 (Section A3) having mass cooling rates consistent with the OVI fluxes. The dust extincton also appears patchy.   The detections of OVI emphasise that the gas can continue to cool below $10^6\K$ at a rate similar to that at higher X-ray emitting temperatures. As we shall see in the next Section, MIR lines should be  a better test of low temperature cooling.

\section{NIR and MIR from JWST: a decisive test of the hidden cooling flows framework}

The cooling flow mystery is the question why the cooled
gas is not detected in the soft X-ray band.
The “hidden cooling flows” model discussed
here and in \citep{2022MNRAS.515.3336F, 2023MNRAS.521.1794F, 2023MNRAS.524..716F} proposes that gas does cool down, but that there are sources of soft X-ray opacity that absorb soft X-ray emission. The JWST infrared detectors can see through any such opacity, so they will detect the cooled gas if it exists. We are developing decisive spectral tests
that will allow JWST to settle the matter.

\begin{figure}
    \centering    
\includegraphics[width=0.48\textwidth]{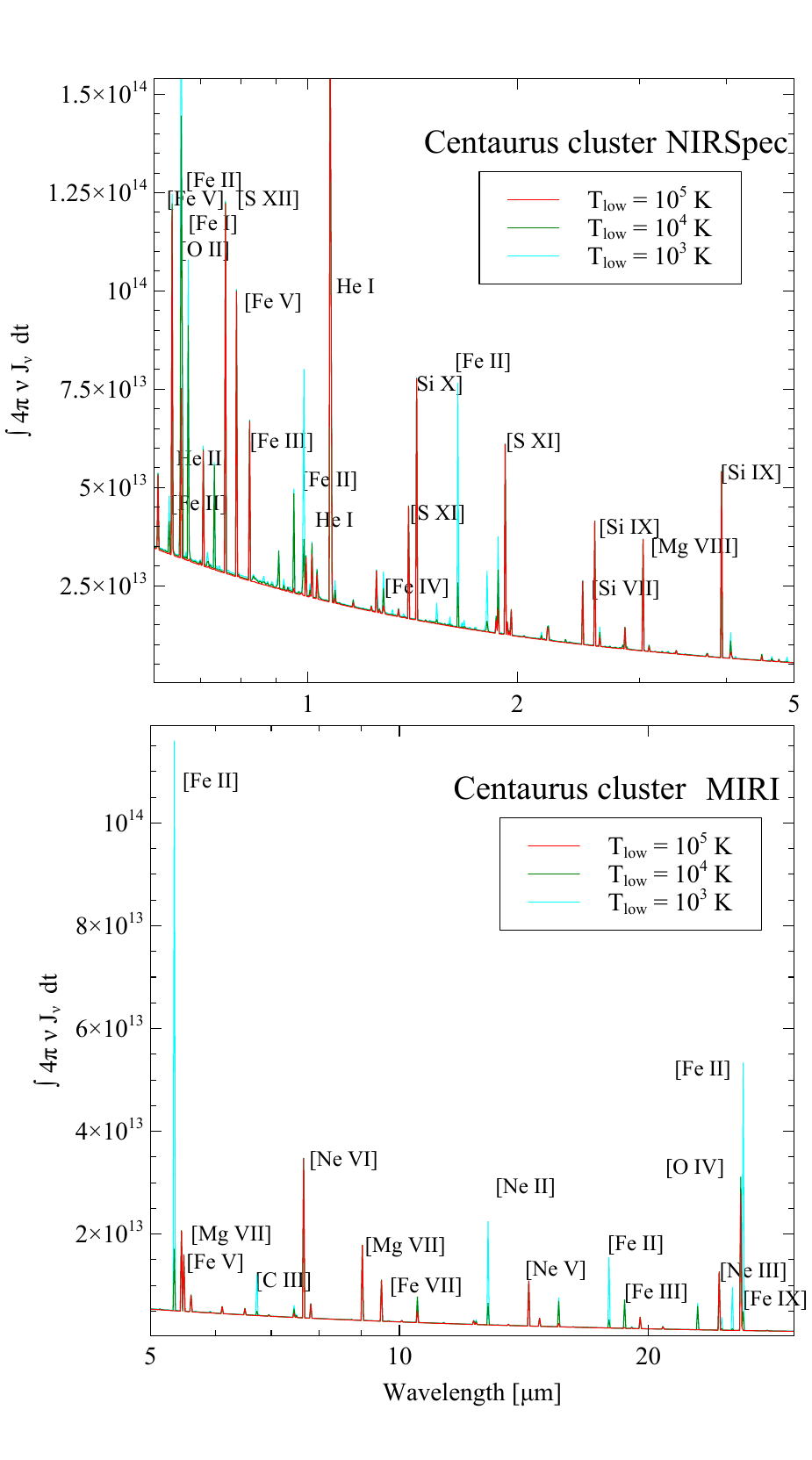}
    \caption{The predicted JWST spectrum of a
    low redshift cooling flow and a variety of lowest flow temperatures.
    The upper panel shows the wavelength range detected
    by NIRSpec ($0.6 - 5\mu$m) while the lower panel 
    shows MIRI ($5 - 28\mu$m).
    Very low ionization [Fe~II] and [N~I] lines are strong in flows that
    cool down to $\sim 10^3$~K but are absent in flows that are reheated
    at higher temperatures. The spectrum is normalised to a cooling flow of  1 $\Msunpyr$ at the distance of the Centaurus cluster $z=0.0104$. The results in Fig. 1 suggest that the flow is about $14 \Msunpyr$} 
    \label{fig:CfLinear}
\end{figure}

\begin{figure}
    \centering    
\includegraphics[width=0.48\textwidth]{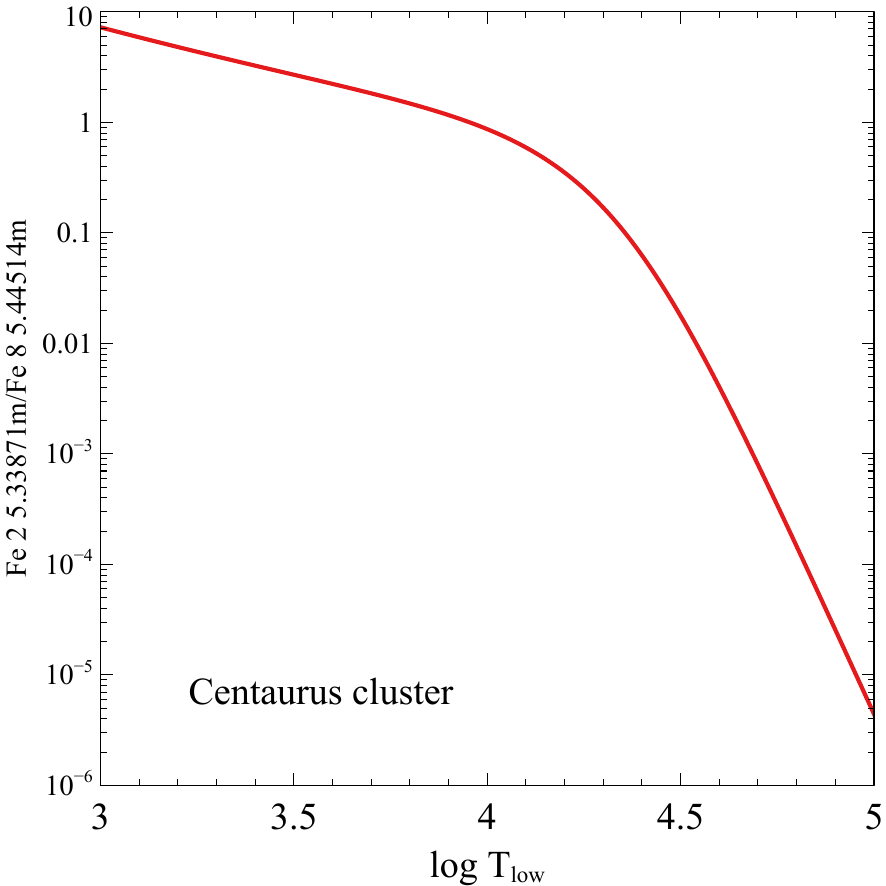} 
    \caption{Many lines of a wide range of ionization are present in
    Figure \ref{fig:CfLinear}.
    The vertical axis shows the predicted intensity ratio of a coronal line of [Fe~VIII] to
    a low-ionization line of [Fe~II].
    The temperature where the cooling flow stops is shown as the independent axis.
    Flows that are not allowed to cool do not produce low ionization
    emission. The line ratio is one of many that could have been shown and
    changes by $\sim 7$ dex as the lowest temperature varies.
    \label{fig:lrat}}
\end{figure}

\begin{figure}
\centering    
\includegraphics[width=0.48\textwidth]{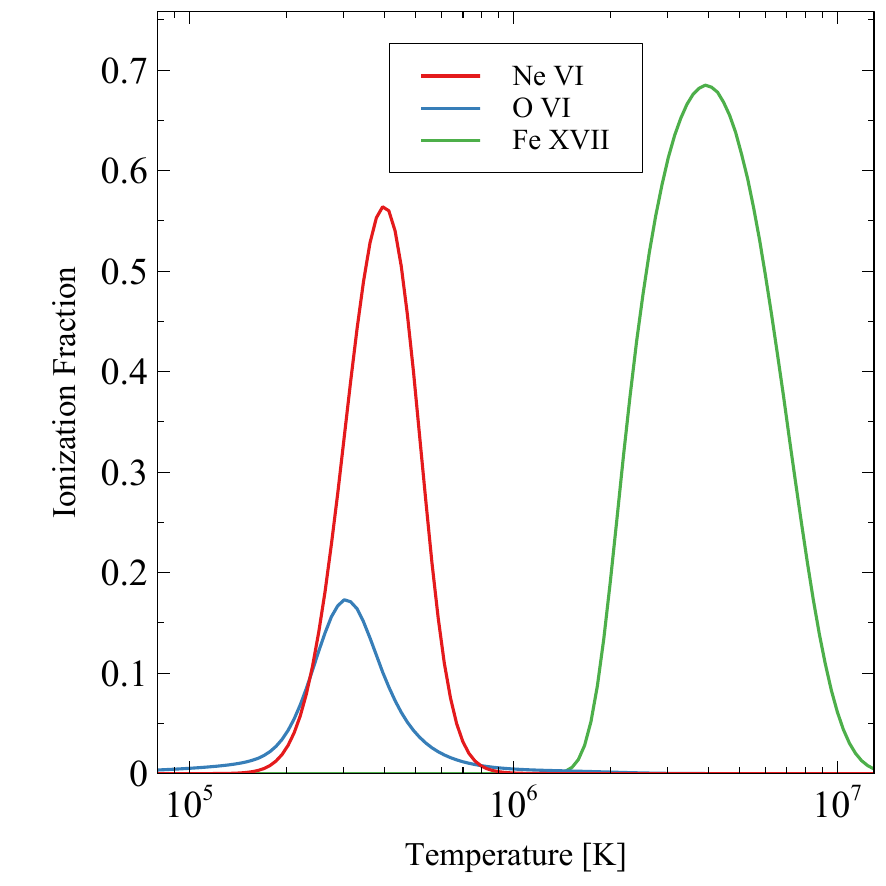} 
    \caption{The ionization fractions of cooling flow ions in the IR,
    Ne VI, UV (O VI), and X-ray (Fe XVII). This is a constant
    pressure cooling flow.}
\end{figure}

Figure \ref{fig:CfLinear} shows CLOUDY simulations of the 
spectrum of the Centaurus cluster cooling flow. 
Version 23.01 of CLoUDY is used \citep{2023RMxAA..59..327C, 2023RNAAS...7..246G}.

JWST's IR wavelength range of $\sim 1\mu$m to 30$\mu$m is shown.
This model is speculative and is based on X-ray observations
since the IR part of the spectrum has not yet been observed. 
The simulation follows a time-dependent non-equilibrium
cooling flow starting with hot gas, $T_{\rm kinetic}=3.4 \times 10^7 \K$,
which is mainly emitting in the X-ray.
It cools while maintaining constant gas pressure.
\citet{2015MNRAS.446.1234C} give computational details of
CLOUDY's treatment of a non-equilibrium cooling gas.

The microphysics in CLOUDY is solved for a cooling
unit cell, a cube 1 cm on a side and
the predictions are in units of emission
per unit area of cooling flow.
The results must be renormalized to account for the strength and 
distance of a real  flow. 
Initially we 
renormalize the spectrum to agree with the
observed XSPEC MKCFLOW prediction of the Fe XVII 15A lines from a $1\Msunpyr$ cooling flow at the distance of the Centaurus cluster of
$3.2\times 10^{-14}$ erg cm$^{-2}$ s$^{-1}$.


A key issue is how cool the gas becomes. Is it reheated, preventing any cold gas? Does it cool to temperatures where star formation is possible? Several spectra, with different lowest cooling flow temperatures $T_{\rm{low}}$, are shown in Figure \ref{fig:CfLinear}. The infrared spectra are quite different;  the simulations allowed to fully cool produce noticeably stronger low-ionization emission.

Figure \ref{fig:lrat} shows the intensity ratio of a pair of low and high-ionization iron lines. There are many such pairs in the infrared; this particular one was chosen as an example. It involves two lines of the same element, so the chemical composition need not be known. The high ionization coronal line [Fe VIII] forms in hot parts of the flow, whereas the [Fe II] line forms in cooled gas. The panel shows the predicted line ratio as a function of the lowest temperature in the cooling flow. The intensity ratio changes by nearly seven dex, showing that JWST can accurately determine the lowest temperature in the flow.

The simulations in the figure assumed that the gas maintained constant gas pressure as it cooled. Other Equations Of State (EOS) could be tested since JWST spectra reveal a wide range of ionization. The distribution of ionization states, for example, Fe~X, Fe~IX, Fe~VIII, etc.,
are sensitive to the EOS because it affects the density of the flow, its cooling rate, and hence the relative portion of each ion. 

As an example, if magnetic pressure contributes to the EOS and we assume flux freezing with a chaotic field, the field will increase linearly with the density, and the magnetic pressure will vary as the square of the density. The field will increase as the flow cools until the magnetic pressure dominates and the flow becomes constant density rather than constant gas pressure. The spectrum of a constant gas pressure flow can be distinguished from the constant-density case if a range of ionization is observed. JWST can solve this problem.

Strong [Fe~II] lines are ubiquitous in Figure \ref{fig:CfLinear} due
to the relatively high abundance of Fe and the dense
spacing of low-lying Fe$^+$ energy levels.
Cloudy's upgraded treatment of Fe~II 
emission is described in
\citet{2021ApJ...907...12S}.
[FeII] emission is a hallmark of a cooling flow in the infrared.

The calculations assume that grains do not
form in the cooling gas, so that iron remains in the gas phase throughout.
\citet{2009ASPC..414..453D} argues that
a significant fraction of 
grains in our galaxy form within interstellar
clouds rather than in stellar winds. 
Iron is heavily depleted when grains form
\citep{2009ApJ...700.1299J}.
Iron emission would
become significantly fainter in regions where iron-bearing grains exist.
This is a complication, but it is possible
that JWST cooling-flow observations could
shed insight into the circumstances where
grain formation in diffuse clouds becomes
important.

\subsection{NeVI}

NeVI emits a forbidden line at $7.65\mu$min the MIRI band that is produced by gas cooling over the temperature range of $6-1.5\times 10^{5}\K$ (Fig. 7).   Scaling to the flux predicted in Table 2, which is for a Solar abundance cooling flow of $1\Msunpyr$ at the distance of the Centaurus cluster ($z=0.0104$), should enable MIRI observations of cooling flows to be tested. For a flow of $1\Msunpyr$, the luminosity of [NeVI] $\lambda 7.65\mu$m is $5\times 10^{36}\ergps$.  For a $14\Msunpyr$ steady cooling flow in Centaurus we predict a [NeVI] line flux of $4\times 10^{-16}\ergpcmsqps$. 
As a noble gas, Ne is not involved in grain formation, of course.

It will be important to know how spread out the coolest emission from the flow is. In HCF1, To make an approximate estimate of the volume of such gas, $\Delta V$, we use the cooling equation:
\begin{equation}
   n^2\Delta V\Lambda= \frac{3}{2} \frac{\dot M k\Delta T}{\mu m}.
\end{equation}
Taking $\Lambda=8\times 10^{-23}\ergcmcups$ 
\citep{Lykins13} for the total cooling function at the appropriate metallicity (about half Solar) and  temperature range ($\sim5\times 10^5K$), and  pressure $nT=10^{6.5}\pcmK$, we find that $\Delta V=2\times 10^{63}\dot M\cmcu$. (The factor of 5/2 will be 3/2 at constant density.) Assuming a sphere, the equivalent radius is then
$8{\dot M}^{1/3} \pc$. This will have reduced from $200{\dot M}^{1/3} pc$ when the gas is at about $5\times 10^6\K$, due to the rapidly rising density and increase in emissivity. Assuming that the $14\Msunpyr$ cooling flow in Centaurus is steady and continuous means that [NeVI] emission originates from a volume equivalent to a sphere of radius 20 pc. 
We do not know the geometry of the inner flow, which may be in many small clouds, sheets or filaments (as in the HST and Chandra images, Fig. 4a).  

Detection and mapping of the NeVI MIR line can  validate the Hidden Cooling Flow model presented here. It can demonstrate whether or not the gas cools over 3 orders of magnitude in temperature from, say, 5 keV to 5eV, in a continuous steady fashion or not.

As well as direct line emission from the cooling gas, there will be line emission from the surrounding gas irradiated by the cooling gas. The results of CLOUDY simulations of this process are published in the work of \citep{Polles2021}.  Near and mid-infrared Spitzer spectra of filaments in the Centaurus and Perseus clusters have been reported by \citep{Johnstone07}.

\begin{table*}
\centering

{
\caption{Predicted fluxes for the ten brightest lines in the
MIRI and NIRSpec ranges for a $1\Msunpyr$ cooling flow in the Centaurus cluster, from the
simulations shown in Figure \ref{fig:CfLinear}.
The first three columns give the results for the flow stopping at 
10$^5$~K, and the remaining groups stop at 10$^4$~K and 10$^3$~K.
The first column indicates the species (Ne~6 is Ne~VI).
The second column gives the wavelength in microns, except for
the first two in Angstroms. 
The third column  is the line flux
[erg cm$^{-2}$ s$^{-1}$].
Low-ionization lines are absent from the flow that stops at 10$^5$~K, but are among the strongest lines in the IR spectrum for the 
flow that extends down to 10$^3$~K. For Centaurus, all fluxes should be multiplied by 14 if the flow is steady and persistent, based on the RGS fits discussed in Section 2. 
}
}

\begin{tabular}{ccccccccccc}
\hline
&T= 10$^5$~K&&&&T=10$^4$~K&&&&T=10$^3$~K&\\ \\ \hline
Fe17&15.0130A&2.52E-14&&Fe17&15.0130A&2.52E-14&&Fe17&15.0130A&2.52E-14\\ 
Fe17&15.2620A&6.81E-15&&Fe17&15.2620A&6.81E-15&&Fe17&15.2620A&6.81E-15\\ 
&&&&&&&&&&\\ 
&&&&&NIRSpec&&&&& \\
S  8&9.91E-01&9.62E-18&&Fe13&1.07461&1.49E-16&&C  1&9.85E-01&5.70E-17\\ 
He 2&1.01235&1.10E-17&&Fe13&1.0798&7.85E-17&&Fe13&1.07462&1.52E-16\\ 
Fe13&1.07462&1.52E-16&&S  9&1.25196&1.60E-17&&Fe13&1.07978&8.01E-17\\ 
Fe13&1.07978&8.01E-17&&S 11&1.39232&2.84E-17&&S 11&1.39232&2.84E-17\\ 
S 11&1.39232&2.84E-17&&Si10&1.43008&6.10E-17&&Si10&1.43008&6.10E-17\\ 
Si10&1.43008&6.10E-17&&S 11&1.91958&4.80E-17&&Fe 2&1.64355&6.10E-17\\ 
S 11&1.91958&4.80E-17&&Si 7&2.48071&1.61E-17&&S 11&1.91958&4.80E-17\\ 
Si 9&2.58394&3.15E-17&&Si 9&2.58394&3.15E-17&&Si 9&2.58394&3.15E-17\\ 
Mg 8&3.02764&2.82E-17&&Mg 8&3.02764&2.83E-17&&Mg 8&3.02764&2.82E-17\\ 
Si 9&3.9282&4.71E-17&&Si 9&3.9282&4.70E-17&&Si 9&3.9282&4.71E-17\\ 
&&&&&&&&&&\\ 
&&&&&MIRI&&&&& \\
Fe 8&5.44514&1.56E-17&&Fe 2&5.33871&1.20E-17&&Fe 2&5.33871&1.10E-16\\ 
Mg 7&5.4927&1.10E-17&&Fe 8&5.44514&1.56E-17&&Fe 8&5.44514&1.56E-17\\ 
Mg 5&5.607&3.28E-18&&Mg 7&5.4927&1.10E-17&&Ne 6&7.65019&3.09E-17\\ 
Ne 6&7.65019&3.09E-17&&Ne 6&7.65019&3.09E-17&&Mg 7&9.03094&1.46E-17\\ 
Mg 7&9.03094&1.46E-17&&Mg 7&9.03094&1.46E-17&&Ne 2&12.8101&1.99E-17\\ 
Fe 7&9.50763&7.98E-18&&Ne 5&24.3109&1.13E-17&&Fe 2&17.9311&1.36E-17\\ 
Ne 5&14.3178&8.72E-18&&O  4&25.8863&2.96E-17&&O  4&25.8863&2.96E-17\\ 
Ne 5&24.3109&1.13E-17&&Si 2&34.8046&1.38E-17&&Fe 2&25.9813&5.17E-17\\ 
O  4&25.8863&2.65E-17&&O  3&51.8004&9.13E-18&&Si 2&34.8046&1.43E-16\\ 
O  3&88.3323&2.93E-18&&O  3&88.3323&1.16E-17&&O  1&63.1679&2.63E-17\\ 
\hline
\end{tabular}
\end{table*}

\begin{table*}
	\centering
	\caption{Spectral Fitting Results. The units of column density $N_{\rm H}$ are $10^{22}\pcmsq$, the temperature $kT$ of the \textsc{apec} and maximum of \textsc{mkcflow} component $kT$ (which are the same) is in $\keV$, $Z$ is  abundance  relative to Solar and $\dot M$ is in $\Msunpyr$. $\dot M_{\rm u}$ is the uncovered rate (i.e. with no absorption).  (f) means that a parameter is fixed. All uncertainties correspond to the 90\% confidence level.}
	\begin{tabular}{lcccccccccr} 
		\hline
		Cluster & $N_{\rm H}$ &  $kT$ & $Z$ &$z$ & $Norm$ &  $CFrac$ & ${N_{\rm H}}^{'}$ &  $\dot M$ & $\chi^2/{\rm dof}$ &$\dot M_{\rm u}$\\
		\hline
		 & $10^{22}\cmsq$ & $\keV$ & $Z_{\odot}$ & & &  & $10^{22}\cmsq$ & $\Msunpyr$ & & $\Msunpyr$ \\
		 \hline
     HCG62 & 3.8e-2& 0.98 & 0.26 &1.45e-2 &1.42e-3  & >0.98& $1.09^{+0.16}_{-0.5}$& $15\pm3$ & 747/649  & --\\
   A1068 & 1.62e-2 & 2.39 & 0.22 & 0.138 & 8.6e-3 & 1 & 4.76 &$630\pm{550}$ &528/508  & --\\
   A1664 &0.128 & 2.52& 0.25 &0.118 &6.32e-3 &>0.85 &$0.48^{+0.17}_{-0.23}$ &$145^{+65}_{-35}$ &1017/997 & --\\
   A1795 & 0.01 & 4.18 & 0.36 & 6.38e-2&2.5e-2 & -- & -- & unconstrained &1017/997 & $15.7^{+11}_{-8}$ \\
   A3112 &1.1e-2& 3.51 & 0.42 &7.61e-2 & 1.55e-2 &>0.9 2 & >0.7 & $99^{+128}_{-84}$ & 1544/1510 & -- \\
   A3581 &4.4e-2&1.51&0.29 & 2.09e-2 &7.3e-2& >0.88 & $1.33^{+0.3}_{-0.7}$ & $10.8^{+1.2}_{-1.8}$ & 1143/1175 & -- \\
   RXJ1504 &8.3e-2 & 7.44 & 0.504 & 0.216 & 2.28e-2 & >0.62 & 0.25&$757^{+78}_{-287}$ & 1583 1509 &-- \\
   Cyg A & 0.12 & 6.0 &0.4 & 5.36e-2 & 3.2e-3 &>0.94 & >1.7 &$350^{+200}_{-70}$&567/452 & -- \\
   Hydra A &4.04e-2 & 2.95 & 0.25 & 5.4e-2 & 2.23e-2 & >0.9 & $0.47^{+0.4}_{-0.1}$ & $17^{+13}_{-6}$ & 1824/1671 & --\\
   A1835 & 2e-2 & 5.9 & 0.34 & 0.252 &1.33e-2  & --& 5 & >70 & 1483/1512 & 116\\
   NGC533 &0.33& 1.08& 0.23&0.019 &1.35e-3 & >0.95 & $0.8\pm0.3$ & $11.5^{+5.5}_{-3.5}$ & 209/190&  -\\
		\hline
	\end{tabular}
\end{table*}

\section{The emerging overall picture of Hidden Cooling Flows}

Including the objects in Appendix A, We now have mass cooling values for 38 clusters, groups and elliptical galaxies. In all cases the RGS spectra are consistent with and show better spectral fits with the simple hidden cooling flow model. This has just 3 new parameters: total cooling rate $N_{\rm H}$, Mdot and covering fraction $f$ (which just measures the partition between absorbed and unabsorbed cooling).

\begin{figure}
    \centering    
\includegraphics[width=0.48\textwidth]{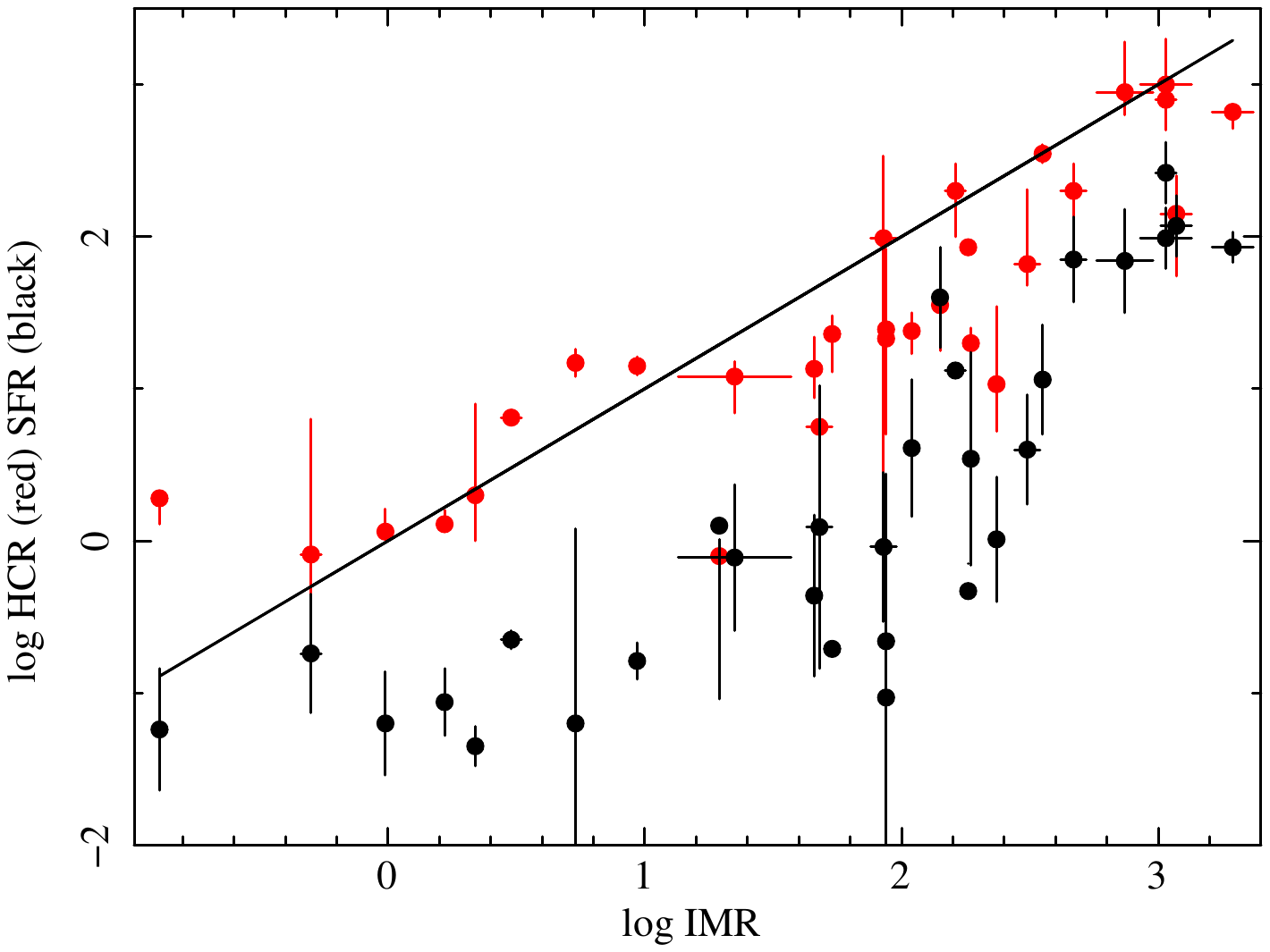} 
\includegraphics[width=0.48\textwidth]{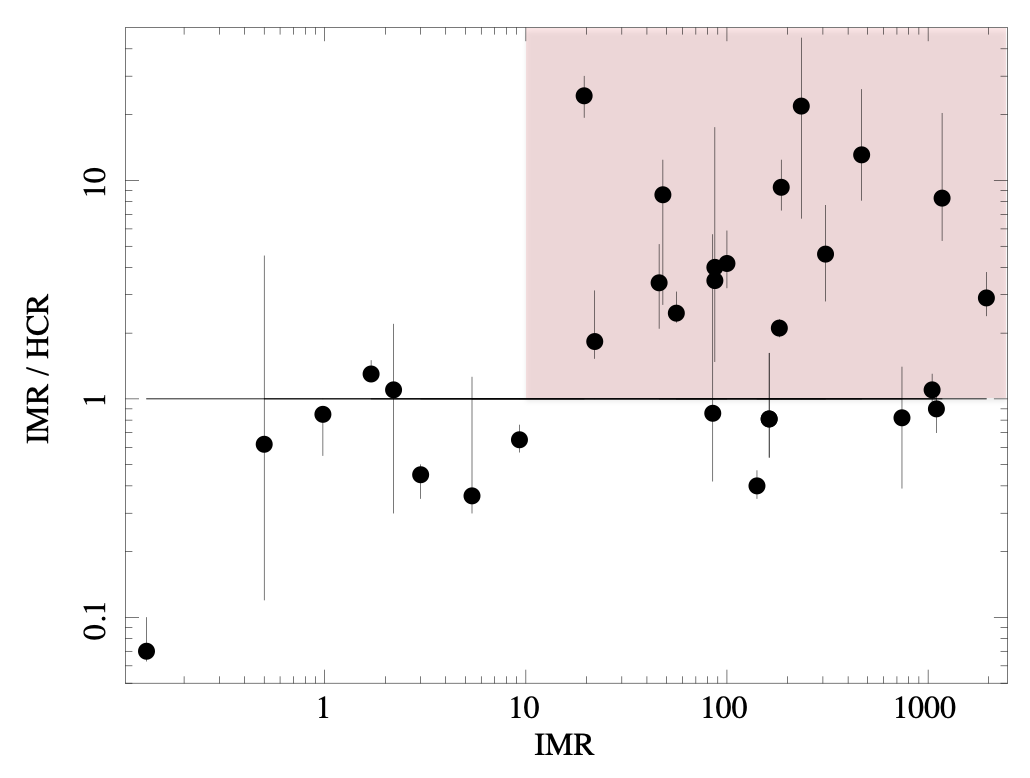} 
    \caption{ Top:  The imaging cooling flow rates (IMR) and star formation rates (SFR) from \citep{McDonald2018} plotted against the hidden cooling flow rates (HCR). Note that for many objects the HCR and IMR are about a factor of 20 above the SFR. Bottom: The ratio of imaging to Hidden mass cooling rates (IMR/HCR) plotted against imaging rates. When ${\rm IMR}<10\Msunpyr$  the two rates are mostly comparable, but above that value the ${\rm IMR} > {\rm HCR}$ by up to a factor of 10 or more. This zone in parameter space is coloured pink. The difference between IMR and HCR rates, which cover different size regions (the imaging cooling region  being much larger than the Hidden cooling one) is presumably made up by AGN Feedback heating and offsetting cooling of outer gas. Objects from left to right and where relevant top to bottom are M84, NGC1600, NGC4472, NGC5846, NGC5813, HGC62, A262, Cen, M87, NGC3581, A2052, A2199, A496, NGC5044, A85, A3112, Hydra A, Cyg A, A1068, 2A0335, S159, A2597, Per (upper limit), ZW3146, R1931, R1532, A1835 (upper limit)  and RXJ1504.}
\end{figure}

\begin{figure}
    \centering    
\includegraphics[width=0.48\textwidth]{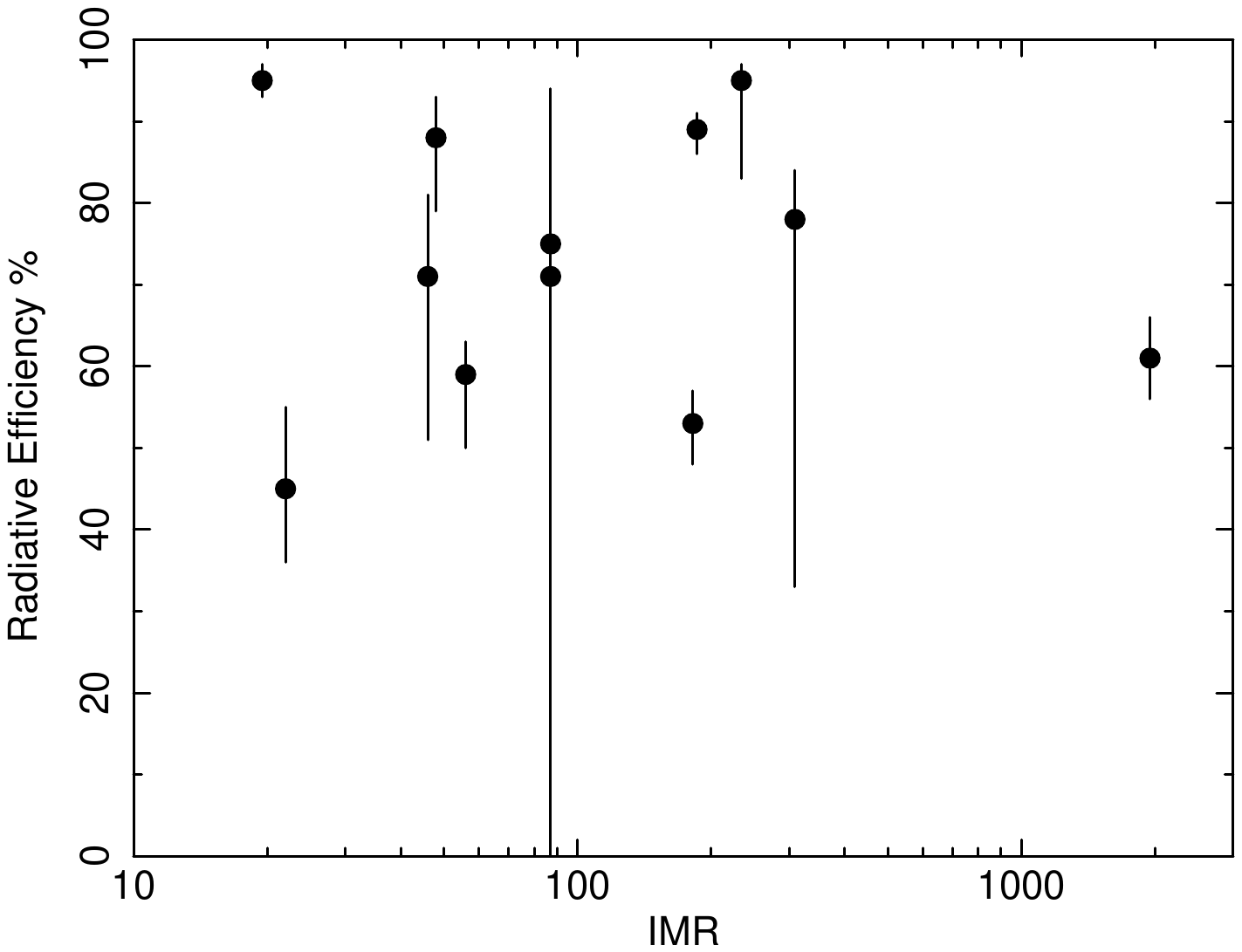}
\caption{The percentage feedback efficiency, defined in the text, plotted against imaging mass cooling rate. The clusters shown from left to right (and where relevant top to bottom) are M87, NGC3581, A2052, A2199, A496, NGC5044, A85, 2A0335, S159, A2597, RXJ1504.}
\end{figure}

We now compare results of the 29 objects which have unabsorbed cooling flow estimates from Chandra imaging data  published by \citep{McDonald2018}, here labelled IMR. (The excluded objects are nearby elliptical galaxies.)  Such estimates require a fiducial cooling time which is less  than 7 or 8 Gyr, on the assumption that the age of cluster cores, which are still forming, is at least the look back time to redshift $z=1$. \cite{McDonald2018} argue for a shorter time of 3Gyr based on the size of the associated optical nebulae. Their results include most of our galaxies and also give estimates of the Star Formation Rate (SFR) due to normal star formation. References to the original  measurements used are given in the Tables of \citep{McDonald2018}.  

Before proceeding, we note that the IMR values are now at least 4 times less than the pre-2000 estimates due to the shorter cooling time (factor of $\sim 2$ and the change in Hubble constant $H_0$ from 50 to $70 \kmpspmpc$).  There is however still an issue with the cooled gas mass which we addressed in depth in HCFI and here is Section 5.1.

Fig. 8 shows the HCR and SFR as a function of imaging rate IMR. 
The HCR are mostly consistent with the IMR when they are less than about $15\Msunpyr$, which is where the elliptical galaxies and many groups lie. From IMR rates of 15 to $300\Msunpyr$ the HCR lie about a factor of several below. This is the range for regular rich clusters where AGN Feedback may be reducing the cooling in the outer parts of the cool core, mainly beyond the bubbles. 
The HCR then increase back close to the line at higher cooling rates. The SFR all lie below the HCR.

This is emphasised in the lower panel of Fig.8 where the ratio of IMR/HCR is  plotted against IMR. Ignoring the point on the lower left, which is the Virgo elliptical galaxy M84, the objects with IMR less than 10 roughly have equal values of IMR and HCR (IMR/HCR = 1). Above IMR of 10 the mean ratio is 5.8, but there is a wide spread of values. As suggested above, this could be the action of AGN feedback. We  note that the mass deposition rate of cooling flows determined from early imaging analyses generally  showed $\dot M \propto r$, where $r$ is the radius (see e.g. \cite{Fabian1994}).  The  aperture used for the RGS spectra are typically much smaller than that used for imaging measurements of cooling rate (it also depends on source distance). 

We summarise the results for 12 clusters with Imaging rates greater than $15\Msunpyr$ and where ${\rm IMR}>{\rm HCR}$ in Fig.  8. Here we define the percentage radiative efficiency as $100 \times (1-{\rm HCR}/{\rm IMR}).$ This is then plotted against the imaging rate ${\rm IMR}.$
It shows that the efficiency of AGN Feedback exceeds 50 per cent for many clusters. There is however a significant fraction of massive clusters that appear immune to feedback heating. This could be intrinsic or a time-dependent behaviour. We find no simple correlation between the above feedback efficiency and the position of a cluster in the Cavity Power -- Cooling Luminosity plane (e.g.\cite{Rafferty06, Hlavacek12}).

In practice, assuming a universal cooling age of 3 Gyr may be inappropriate. If too short then the values of IMR are underestimates. If so, then the feedback efficiencies will increase and vice versa if the age is too long.  The values, especially for the extreme clusters with very high cooling rates may also change over time, sometimes abruptly if the central AGN has large increases in its power (e.g. H1821+643 \citep{Russell24}).

The viability of jetted feedback to completely cancel soft X-ray cooling is highly unlikely. Heating by jets and bubbles principally occurs at and beyond the ends of jets.  

\subsection{Low-mass star formation}

We attribute the lack of any normal star formation in most cooling flows, despite substantial mass cooling rates,  to low-mass star formation (e.g HCF1, \cite{Fabian24}). High-pressure, typical of cooling flows ($nT>10^{6.5}\pcmcuK$), leads to a low Jeans mass ($\sim 0.01\Msun$ and therefore low-mass protostars (\cite{Jura1977,Fabian1982,Sarazin1983,Ferland1994}). Elliptical galaxies, such as the hosts of cooling flows, are observed to have a bottom-heavy Initial Mass Function (IMF, \cite{vDC2010, Mehrgan2019, Gu22, VanDokkum24}).  These issues are probably coupled. 

The development of an IMF is due to accretion of surrounding gas onto protostars (\cite{Krumholz16}). As explained in that paper, in order to remain low mass and give a bottom-heavy IMF, further accretion needs to be stopped or disrupted. This may naturally  happen at the centres of elliptical galaxies where the stellar number density and velocity dispersion are both very high. Fast interactions with the numerous low-mass stars and protostars could be responsible for stopping further accretion. We suggest that, in this way, the IMF from the cooled material may accumulate into a peaked distribution of objects centred on the Jeans mass. 

\section{Conclusions}
The further analyses presented here show that the general picture for Centaurus is robust. Soft X-ray cooling is spread over the inner 30 arcsec, approximately where absorption is seen in optical images and coincident with the central drop in iron abundance. 

JWST observations of the centres of cool cores will allow decisive checks of further cooling of gas to $10^5\K$ and below. Of particular note is the detection and mapping of MIR emission [NeVI] at $7.65 \mu$m. 

The current set of 38 objects shows that absorbed cooling flows are common. The operation of AGN Feedback reduces cooling of outer gas near and beyond the bubbles  in more than 50 per cent of cooling core clusters. At the same time, soft X-ray cooling  persists in the innermost regions of all cool cores in clusters, groups and in elliptical galaxies.

The overall lack of normal star formation in these innermost regions can be explained by a mode of low-mass star formation. This contributes to the bottom-heavy initial mass function now commonly inferred for the central parts of massive elliptical galaxies. 

\begin{table*}
\centering
\caption{Details of Objects and RGS exposures}
\begin{tabular}{lcccp{5cm}}
\hline
Cluster & RA (deg) & Dec (deg) & Exposure (ks) & OBSIDs \\ \hline
Centaurus & $192.2052$ & $-41.3108$ & $326.6$ & 0046340101  0406200101 \\
Perseus & $49.49507$ & $41.5117$ & $363.1$ & 0085110101  0305780101 \\
HCG 62 & $193.2736$ & $-9.2043$ & $328.6$ & 0112270701 0504780501 0504780601 \\
A1068 & $160.1855$ & $39.9532$ & $119.6$ & 0147630101 0827330401 \\
A1664 & $195.9279$ & $-24.2447$ & $501.5$ & 0840580401 0840580301 \\
A1795 & $207.2208$ & $26.5956$ & $85.7$ & 0097820101 \\
A3112 & $49.4903$ & $-44.2381$ & $407.5$ & 0105660101 0603050101 0603050201 \\
A3581 & $211.8741$ & $-27.0179$ & $314.9$ & 0205990101 0504780301 0504780401 \\
RXJ1504 & $226.03114$ & $-2.8046$ & $576.1$ & 0401040101 0840580101 0840580201 \\
Cyg A & $299.8682$ & $40.7339$ & $82.2$ & 0302800101 0302800201 \\
Hydra A & $139.5236$ & $-12.0955$ & $940.7$ & 0109980301 0109980501 0504260101 0843890101 0843890201 0843890301 \\
A1835 & $210.2581$ & $2.8788$ & $574.2$ & 0098010101 0147330201 0551830101 0551830201 \\ 
NGC 533  & $21.3807$ & $1.7591$ & $109.5$ & 0109860101  0109860201 0109860401 \\ 
\hline
\end{tabular} 
\end{table*}

\begin{figure}
    \centering    
\includegraphics[width=0.48\textwidth]{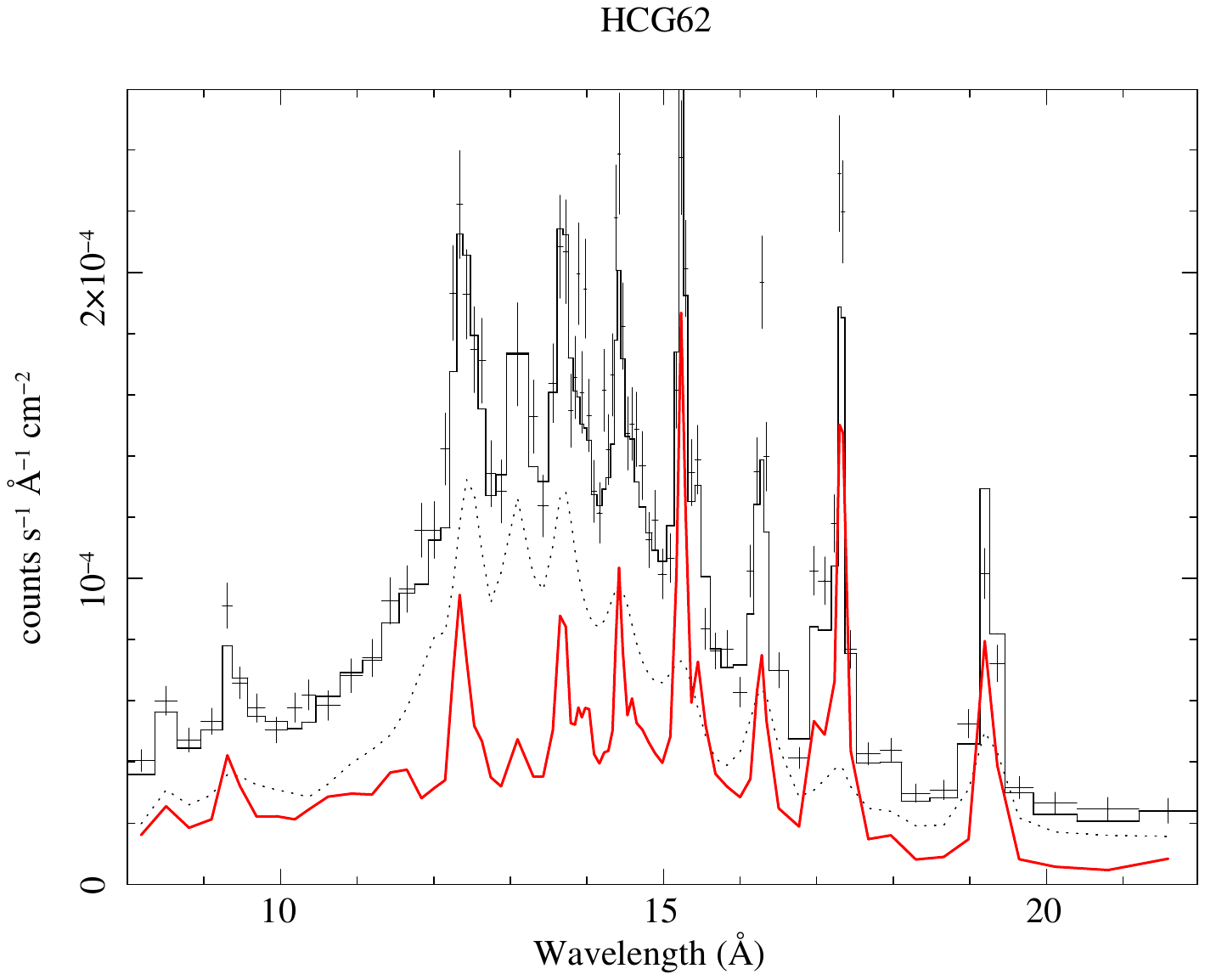} 
\includegraphics[width=0.48\textwidth]{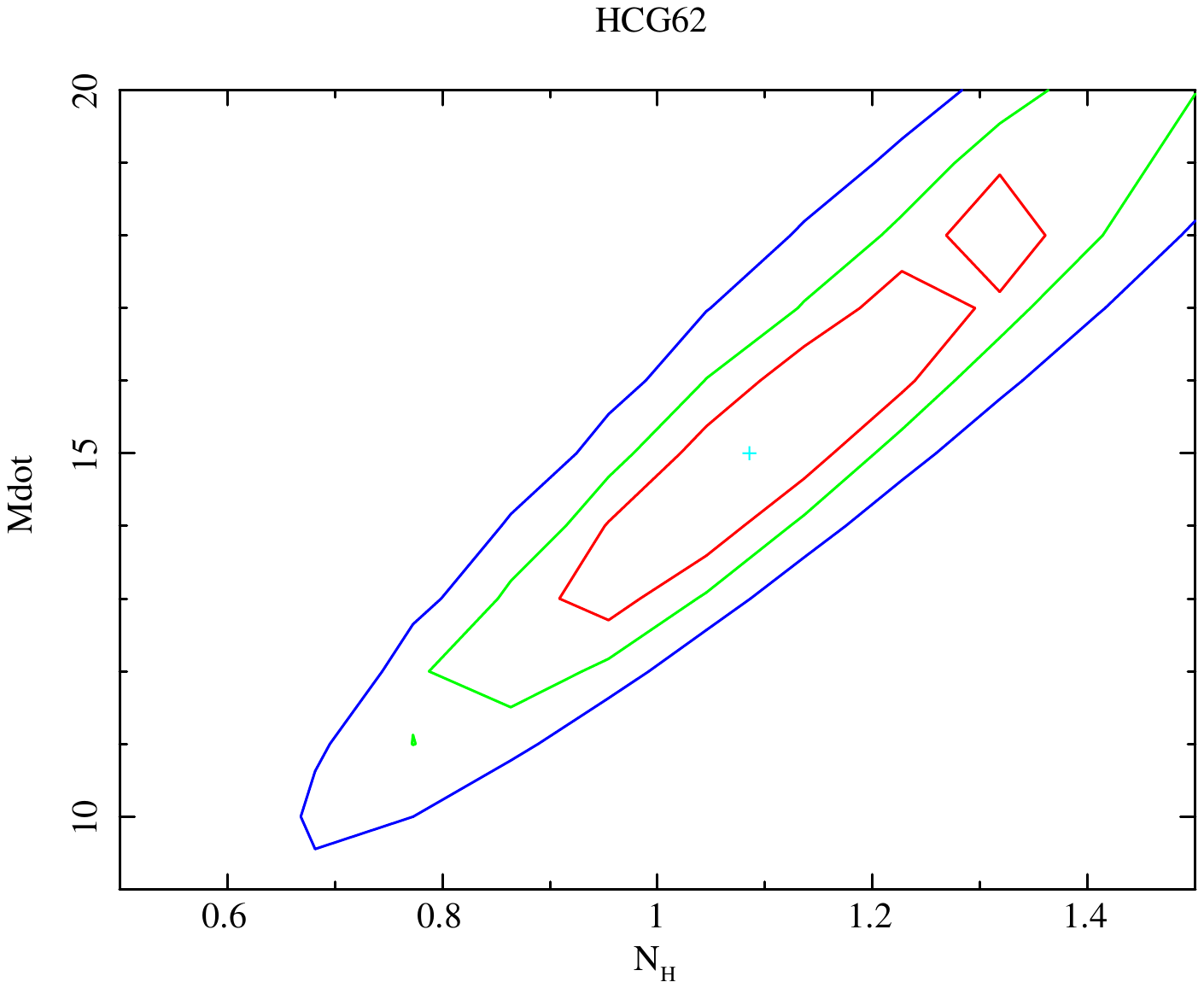}
\includegraphics[width=0.48\textwidth]{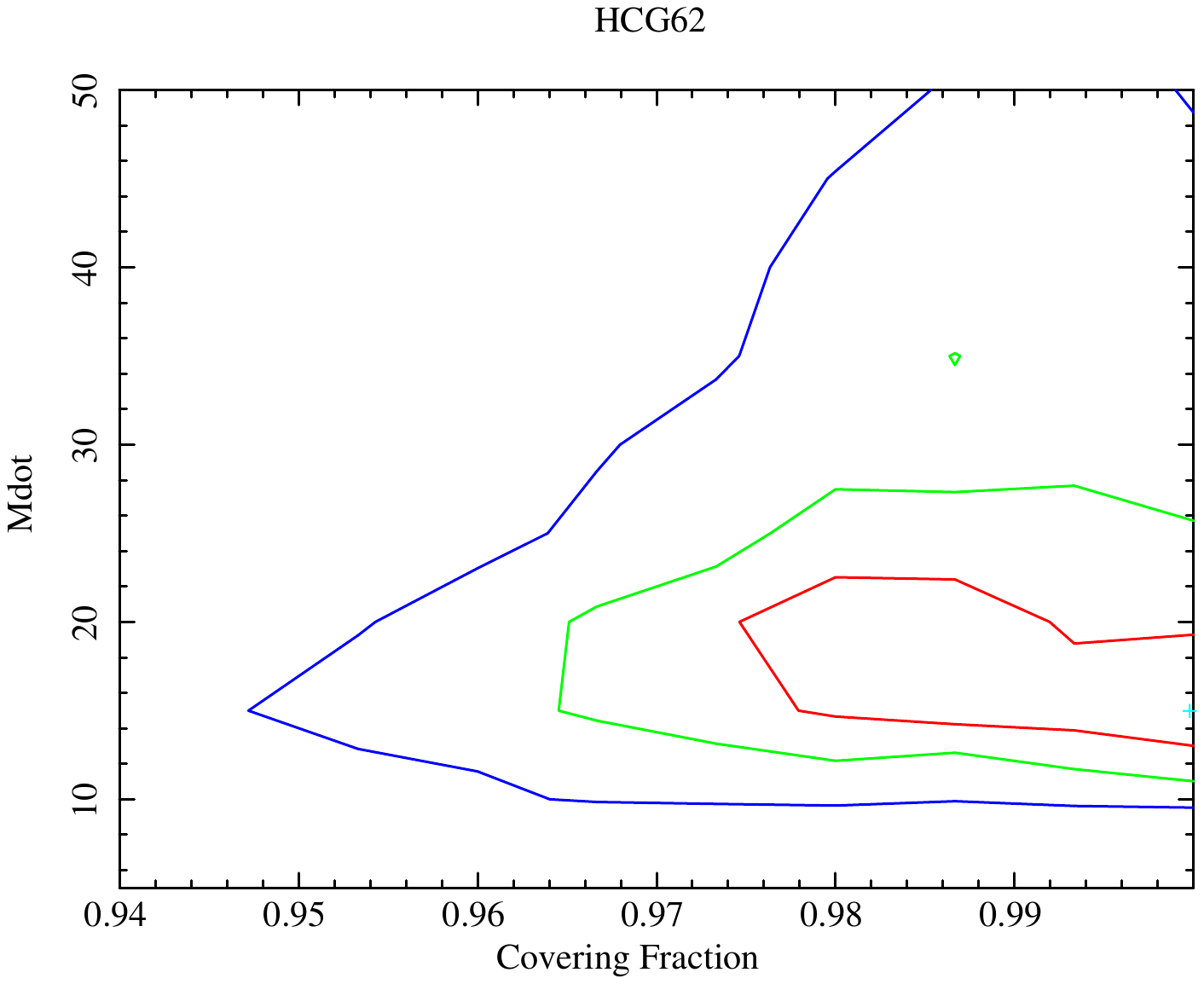}
    \caption{ a) RGS spectrum, b) hidden mass cooling rate versus total interleaved column density and c) versus Covering fraction. In a), note the strong FeXVII lines at 15 and 17A charateristic of gas at temperatures of 0.35--0.8 keV and OVIII at 19A (all rest wavelength, the spectra here are plotted at the observed wavelength). Identifications of all lines shown and emissivities of the brighter ones can be found in \citep{Sanders2011}} 
\end{figure}

\begin{figure}
    \centering    
\includegraphics[width=0.48\textwidth]{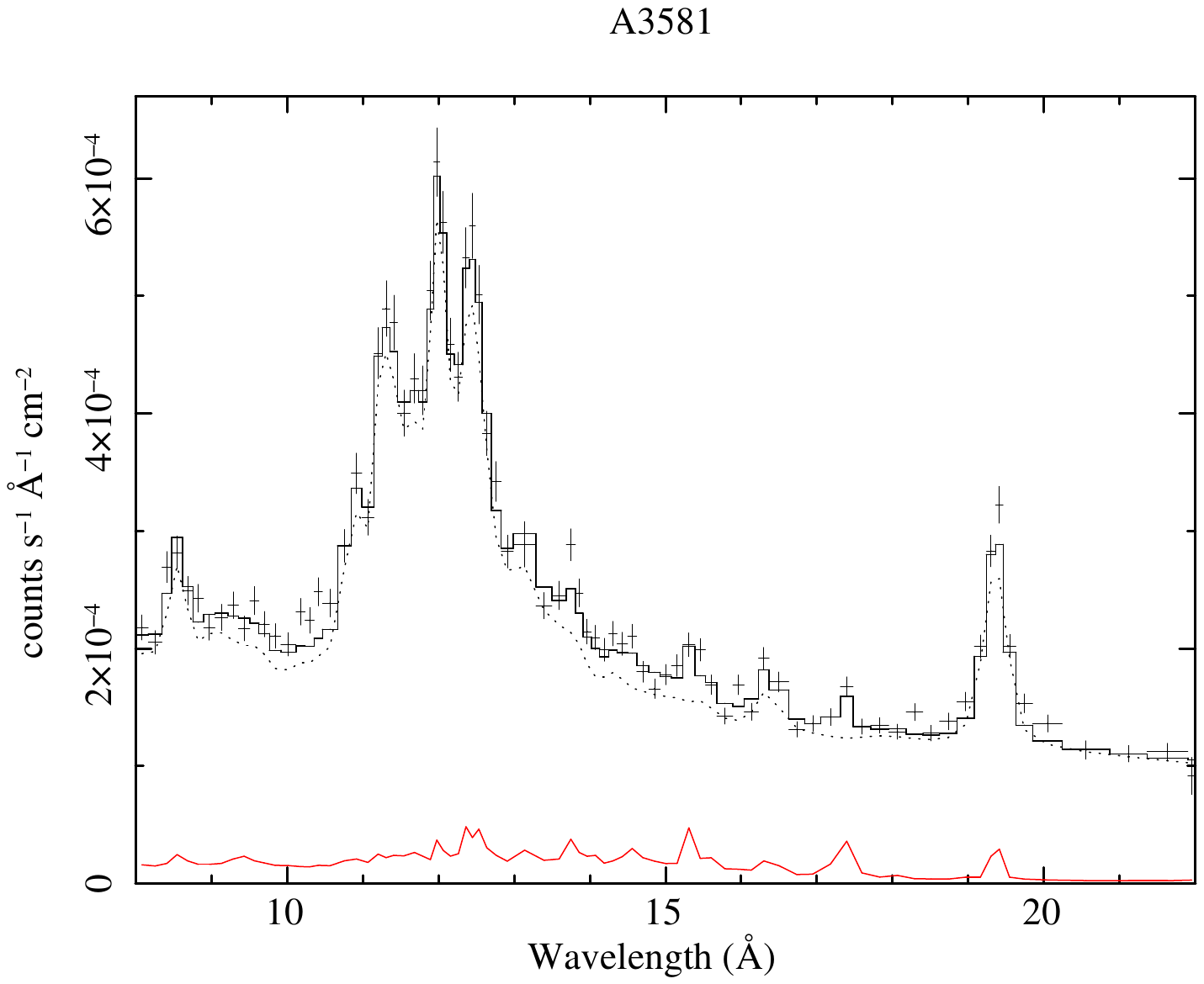} 
\includegraphics[width=0.48\textwidth]{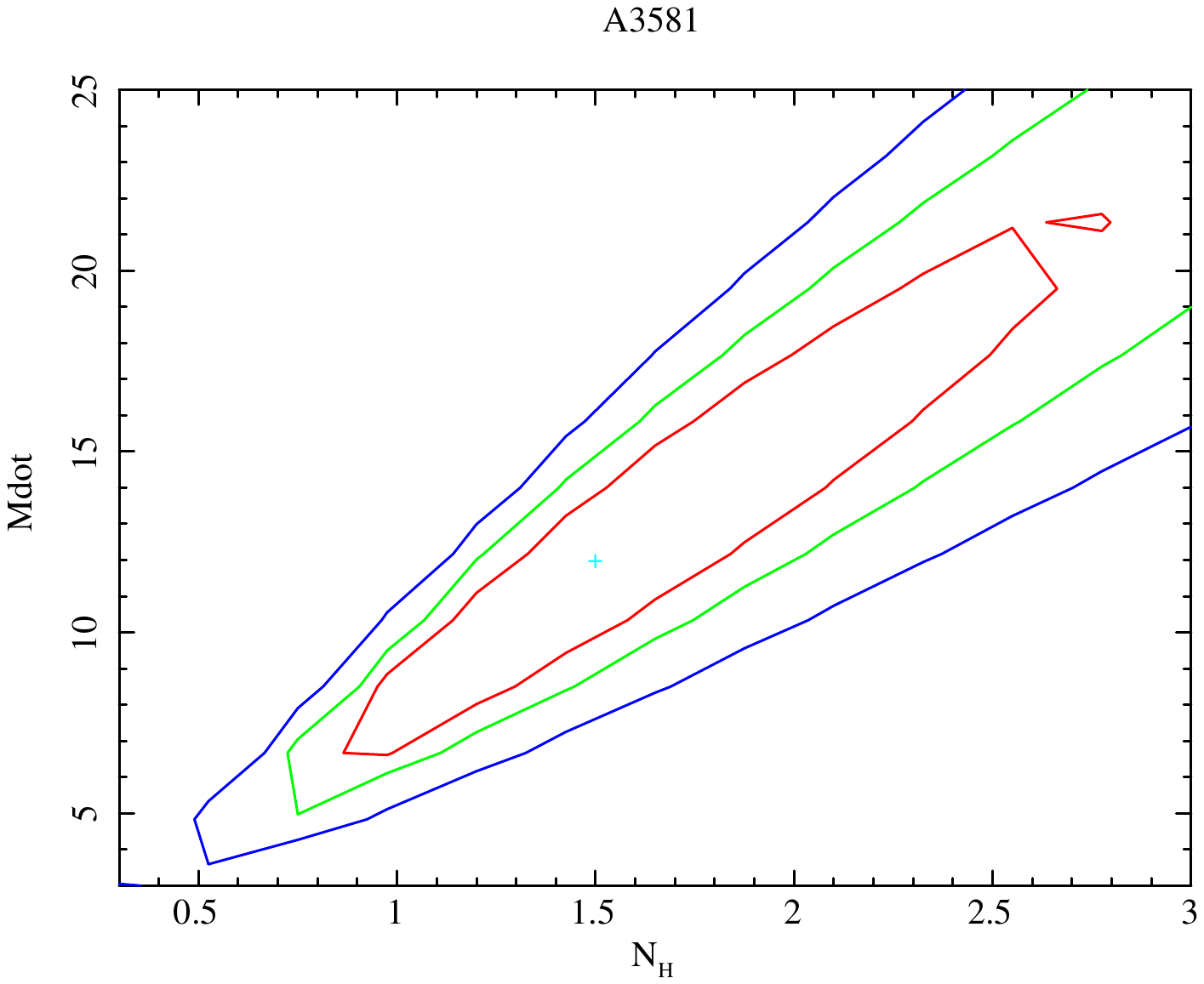}
\includegraphics[width=0.48\textwidth]{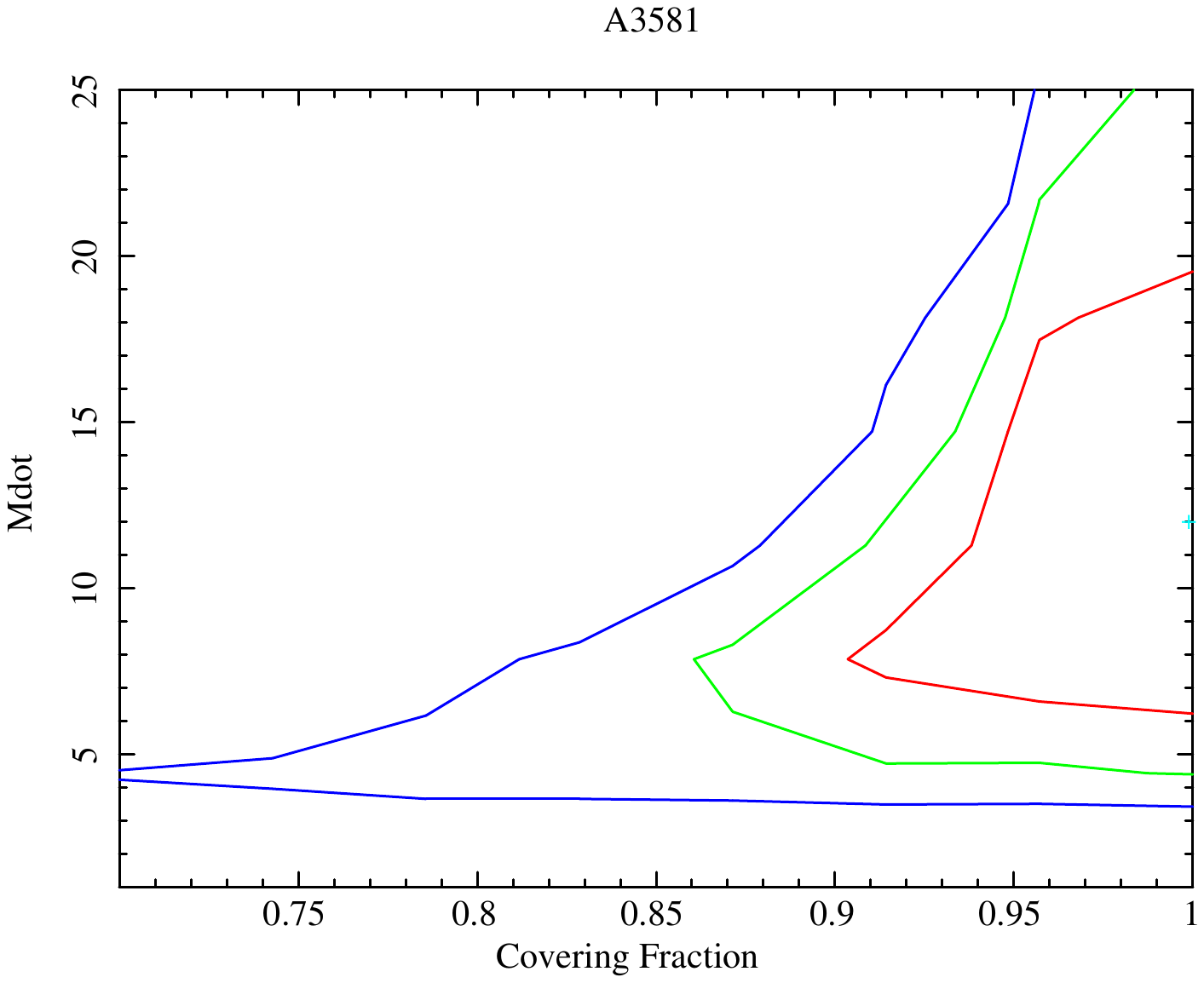}
    \caption{ a) RGS spectrum, b) hidden mass cooling rate versus total interleaved column density and c) versus Covering fraction. }
\end{figure}

\begin{figure}
    \centering    
\includegraphics[width=0.48\textwidth]{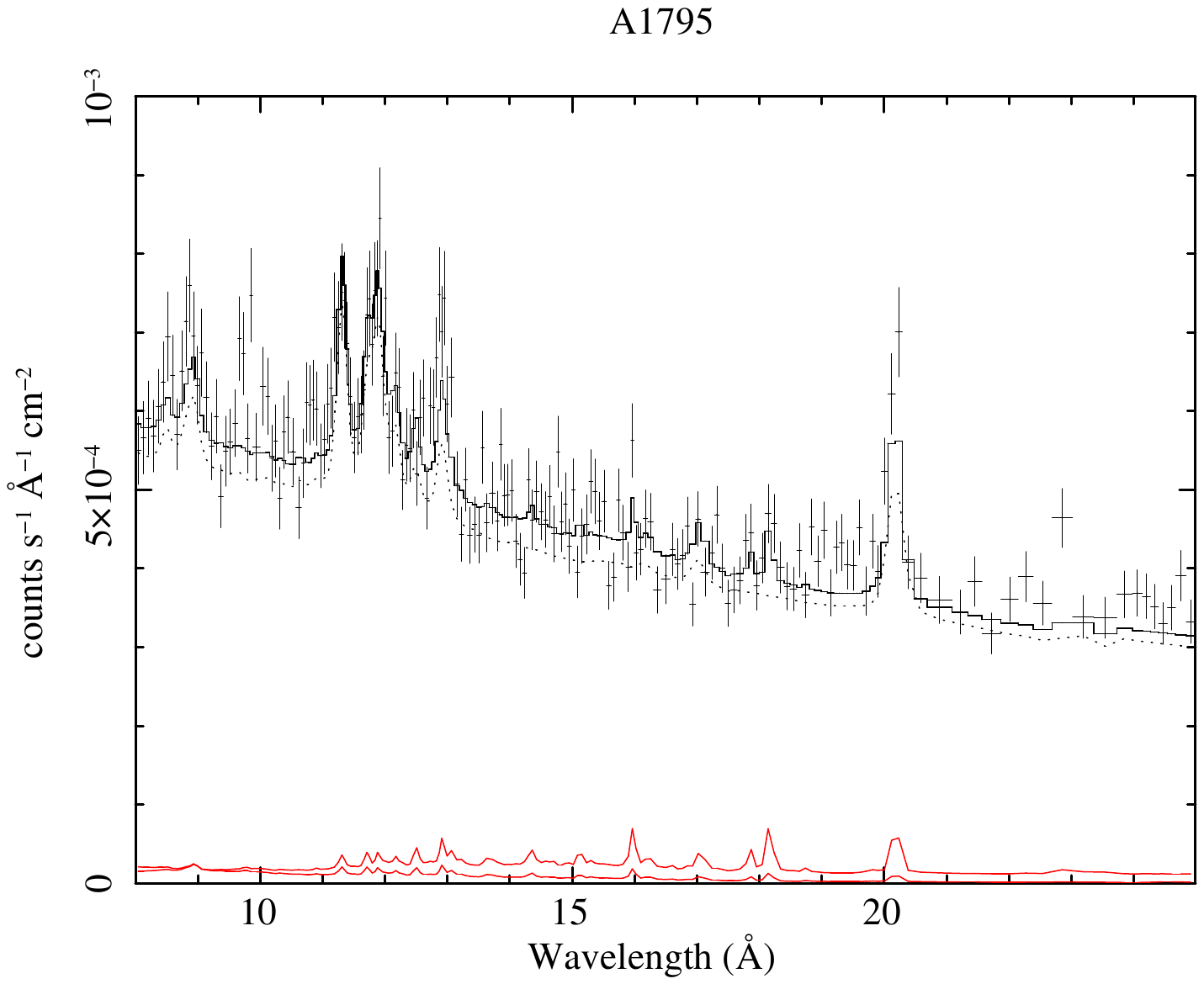} 
\includegraphics[width=0.48\textwidth]{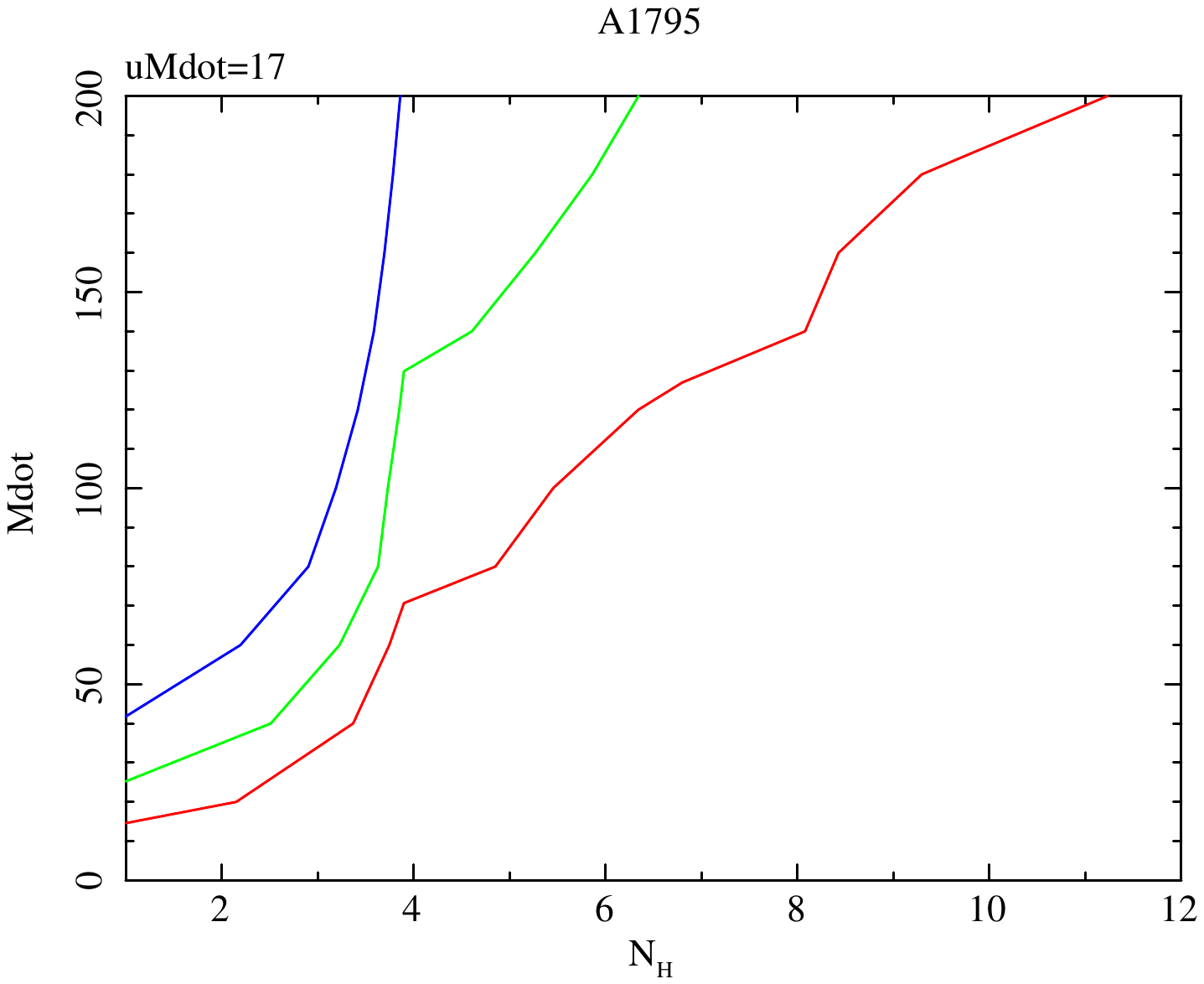}
\includegraphics[width=0.48\textwidth]{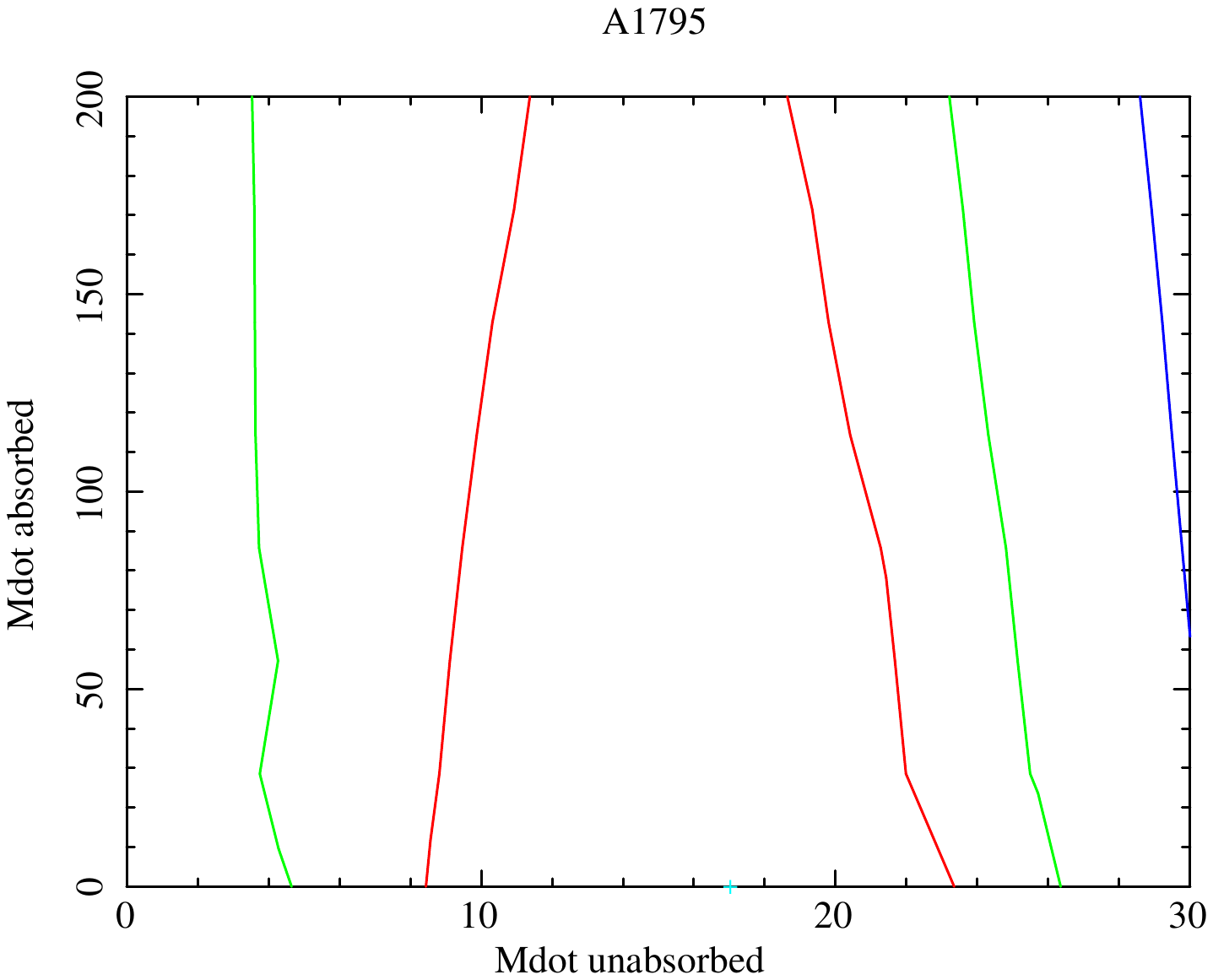}
    \caption{ a) RGS spectrum, b) hidden mass cooling rate versus total interleaved column density and c) versus Covering fraction. }
\end{figure}

\begin{figure}
    \centering    
\includegraphics[width=0.48\textwidth]{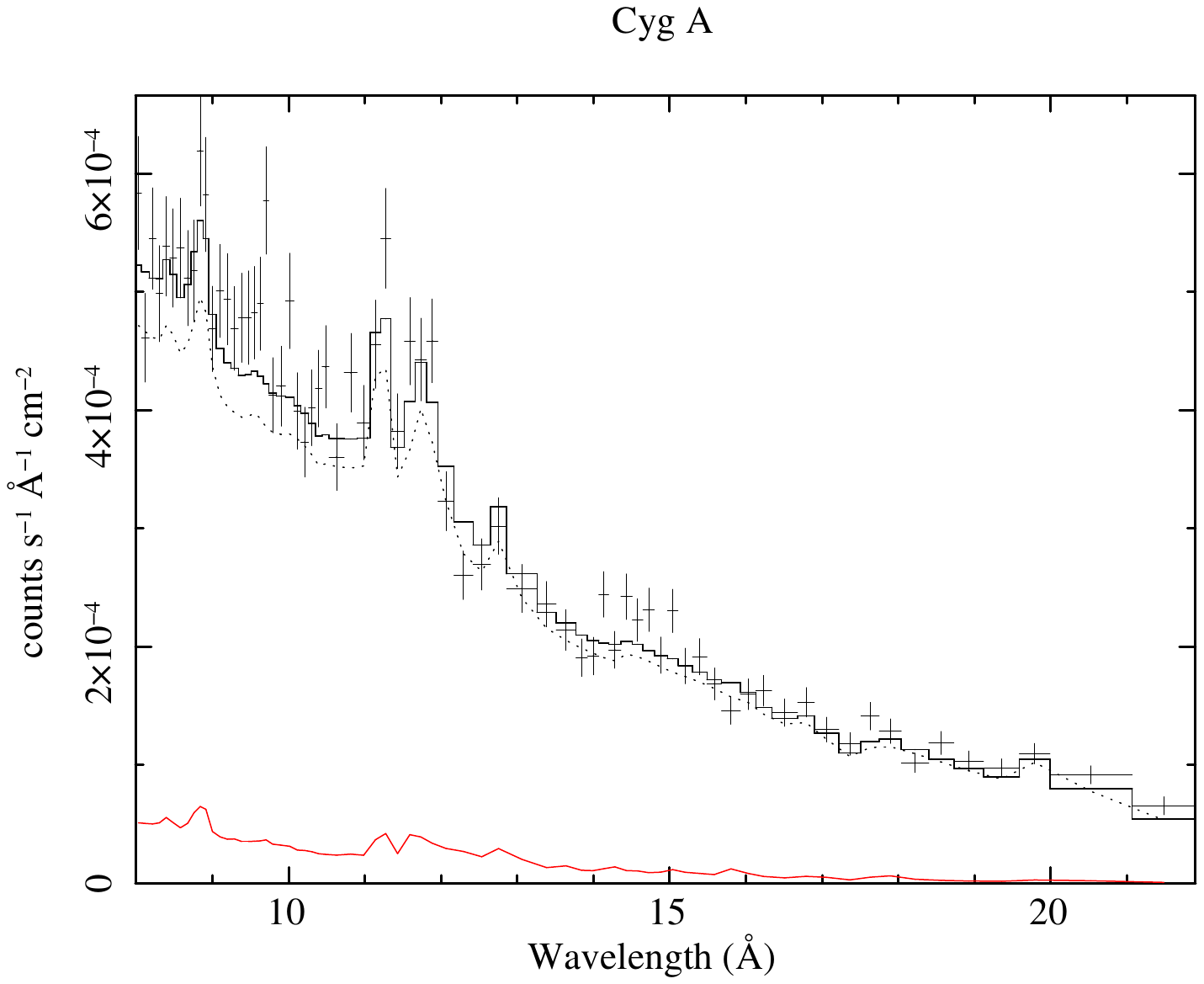} 
\includegraphics[width=0.47\textwidth]{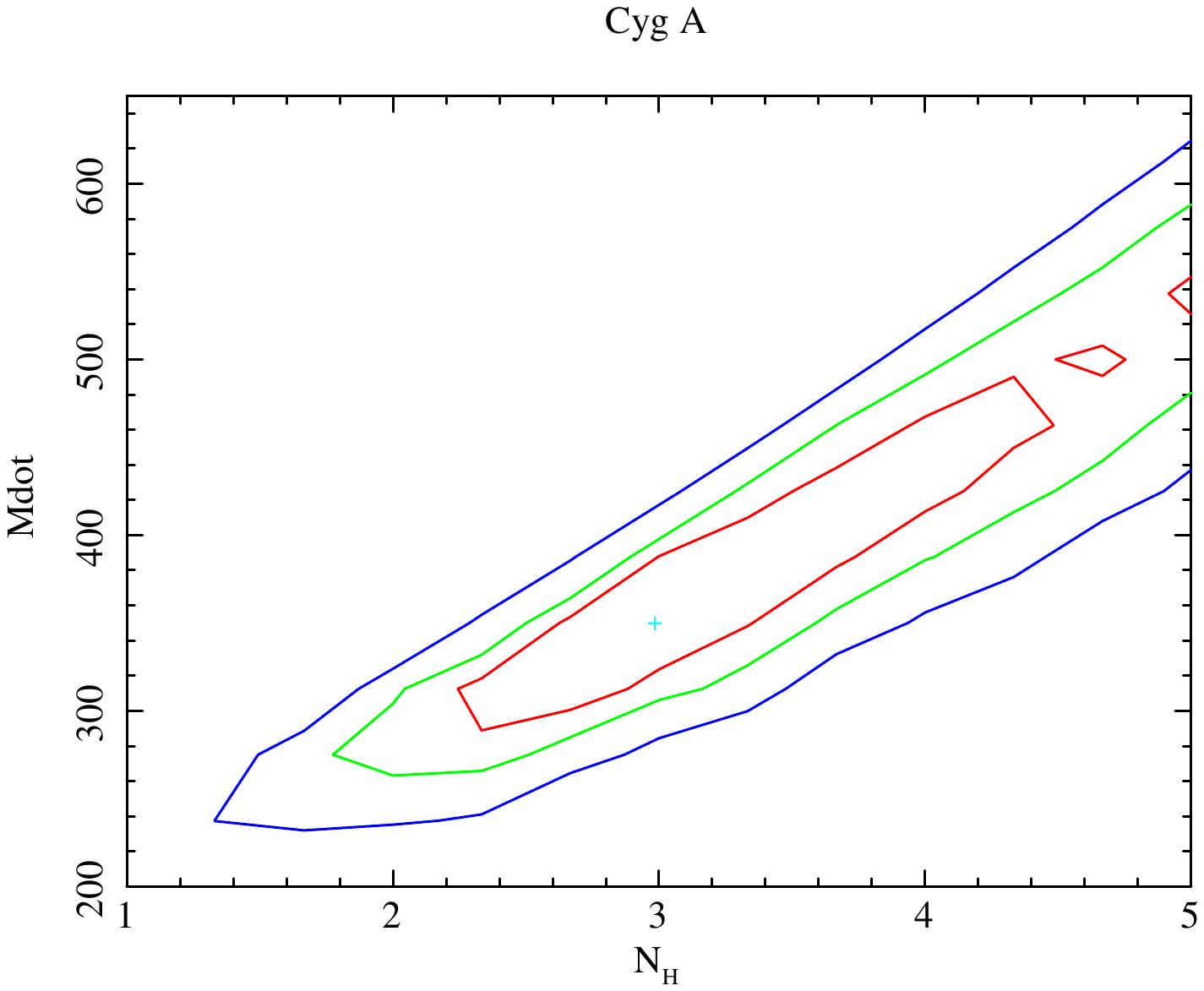}
\includegraphics[width=0.47\textwidth]{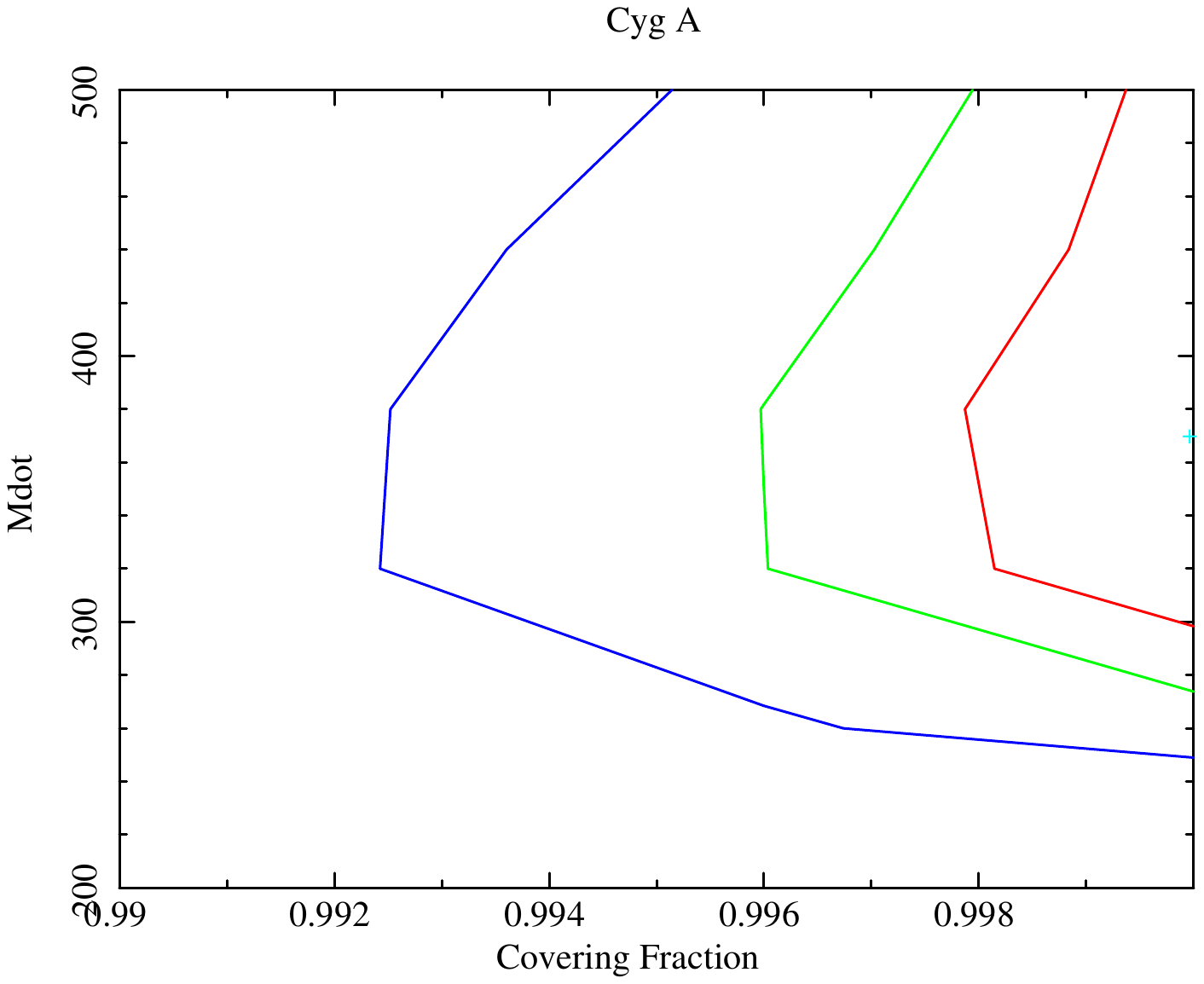}
    \caption{ a) RGS spectrum, b) hidden mass cooling rate versus total interleaved column density and c) versus Covering fraction.}
\end{figure}

\begin{figure}
\centering  
\includegraphics[width=0.48\textwidth]{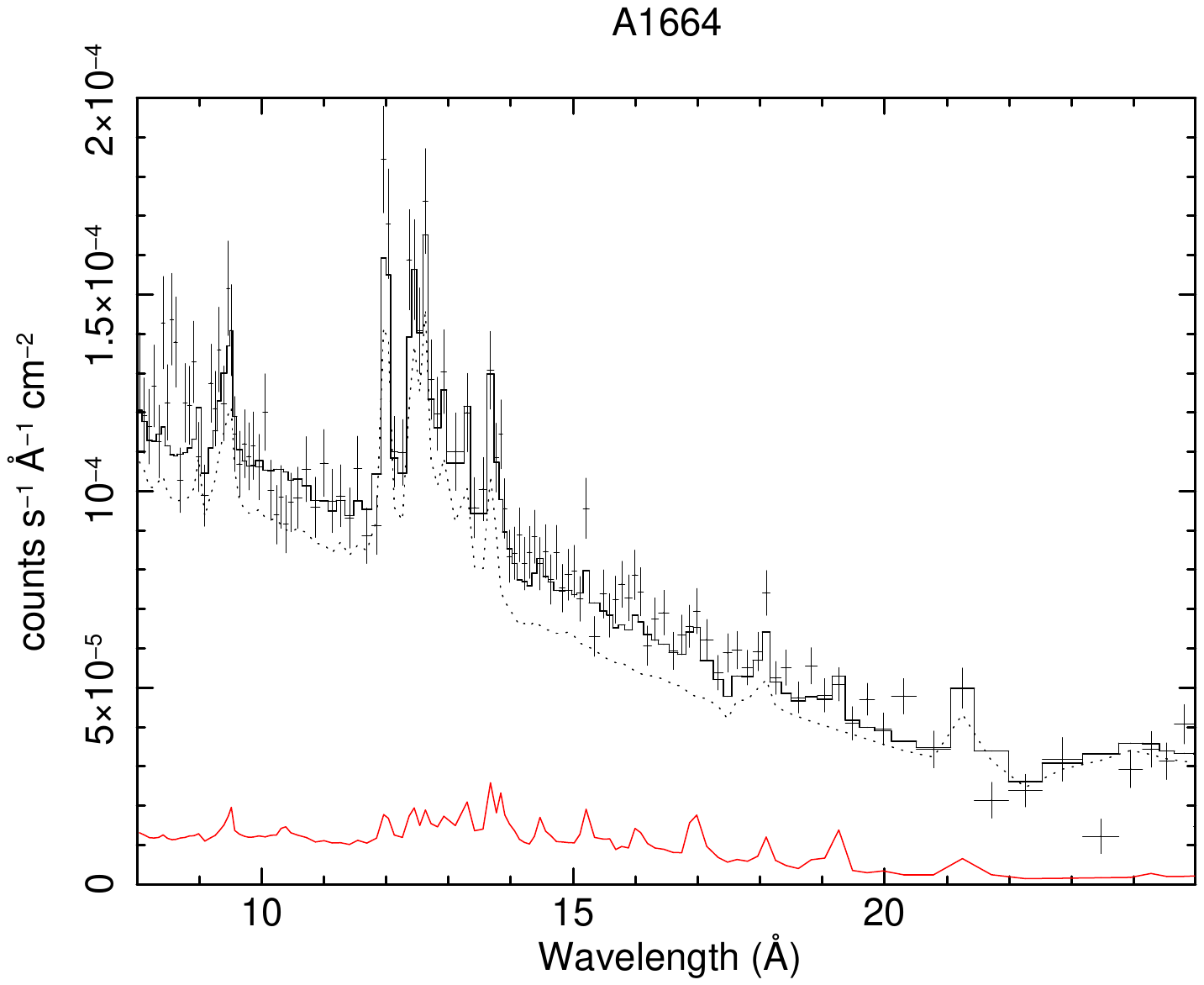}   
\includegraphics[width=0.48\textwidth]{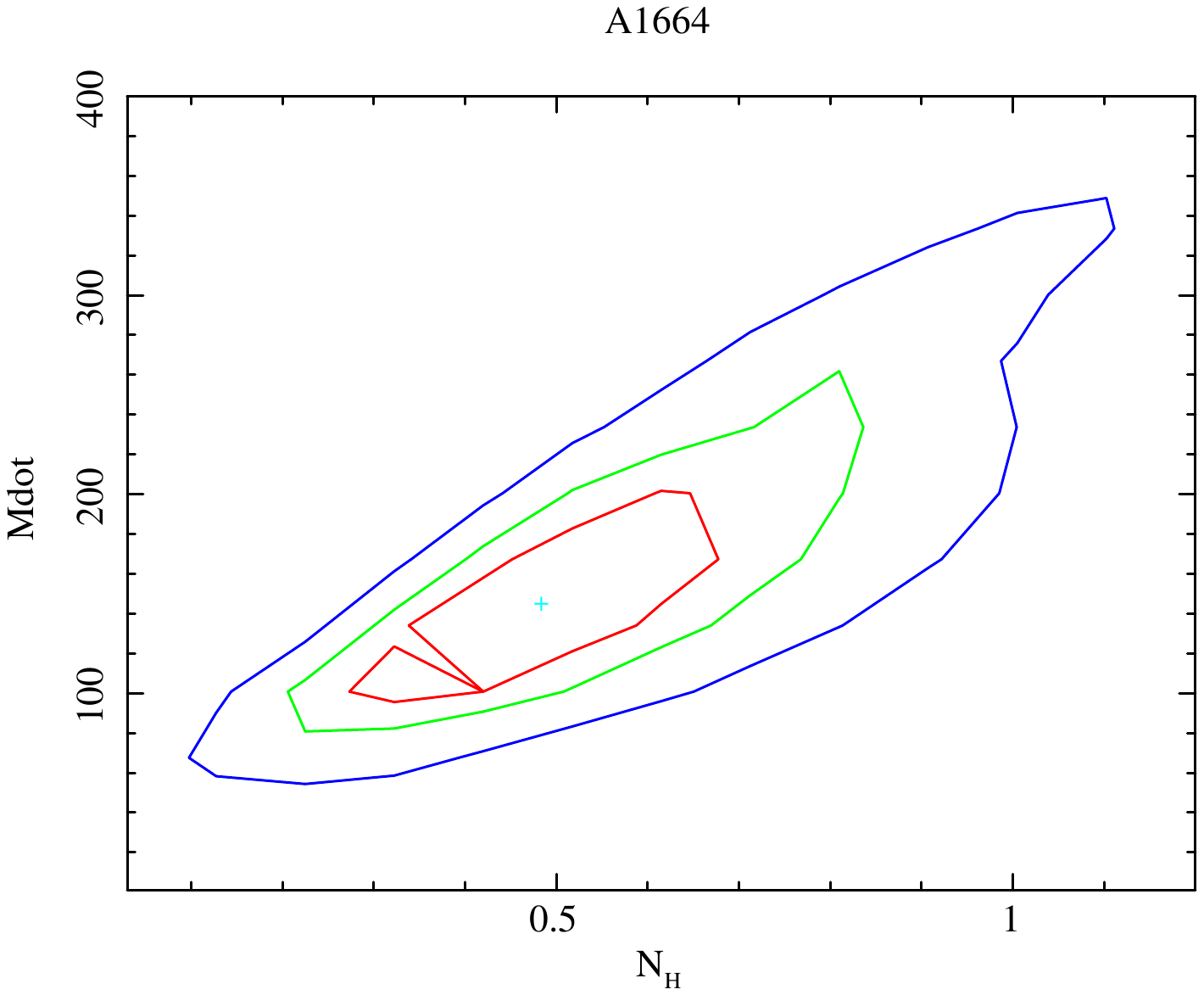} 
\includegraphics[width=0.48\textwidth]{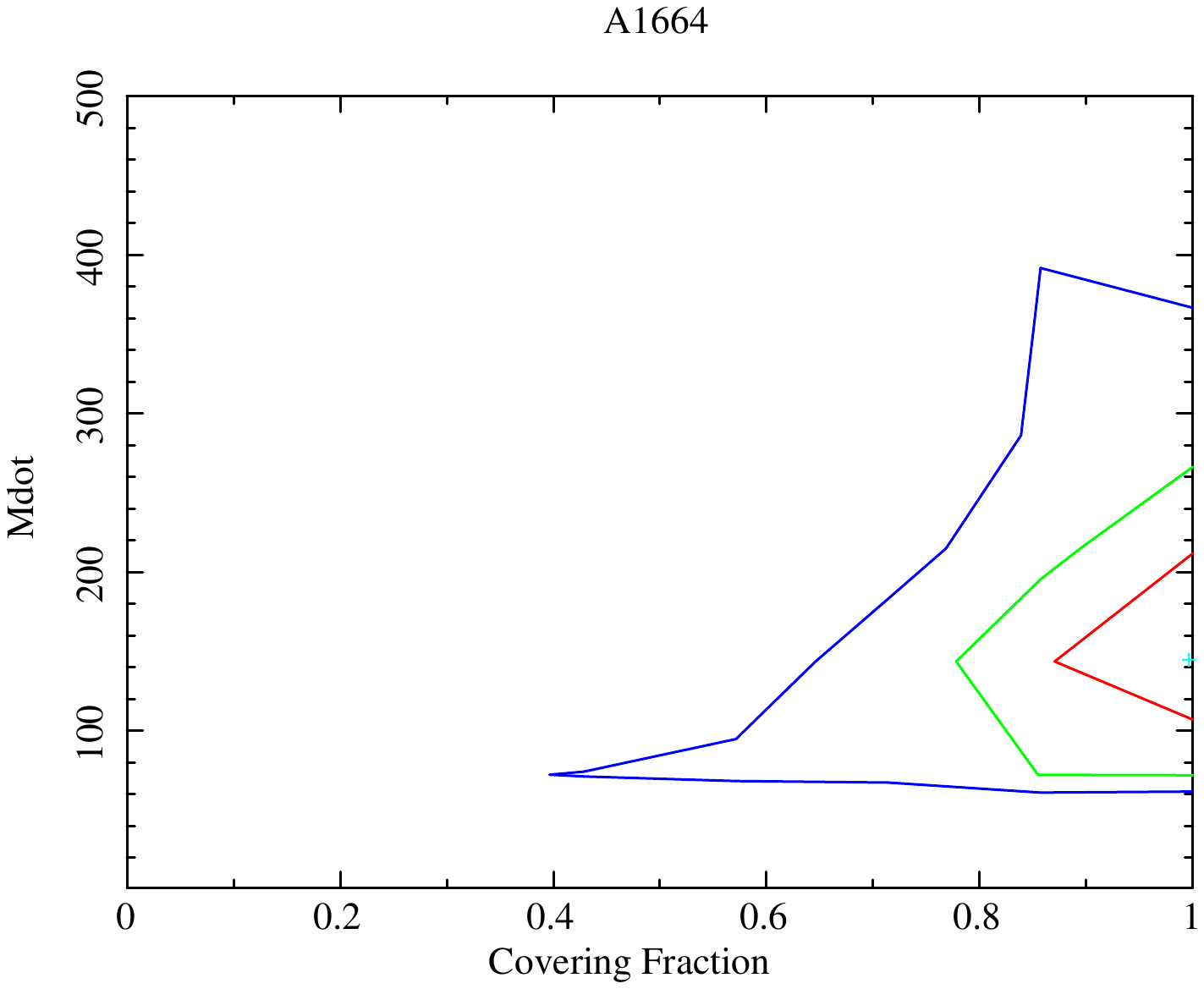} 
    \caption{  a) RGS spectrum, b) hidden mass cooling rate versus total interleaved column density and c) versus Covering fraction.}
\end{figure}

\begin{figure}
\centering  
\includegraphics[width=0.48\textwidth]{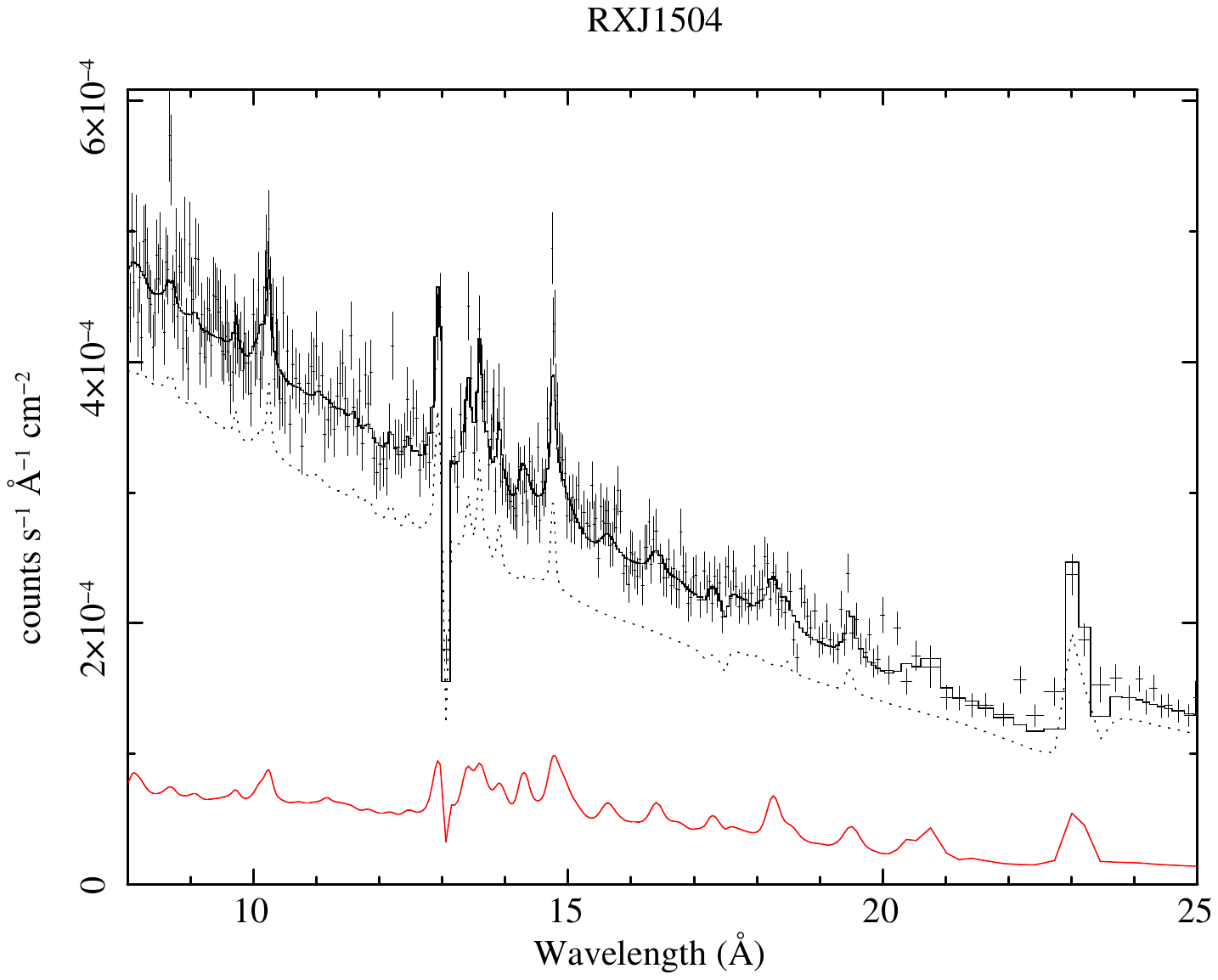}   
\includegraphics[width=0.48\textwidth]{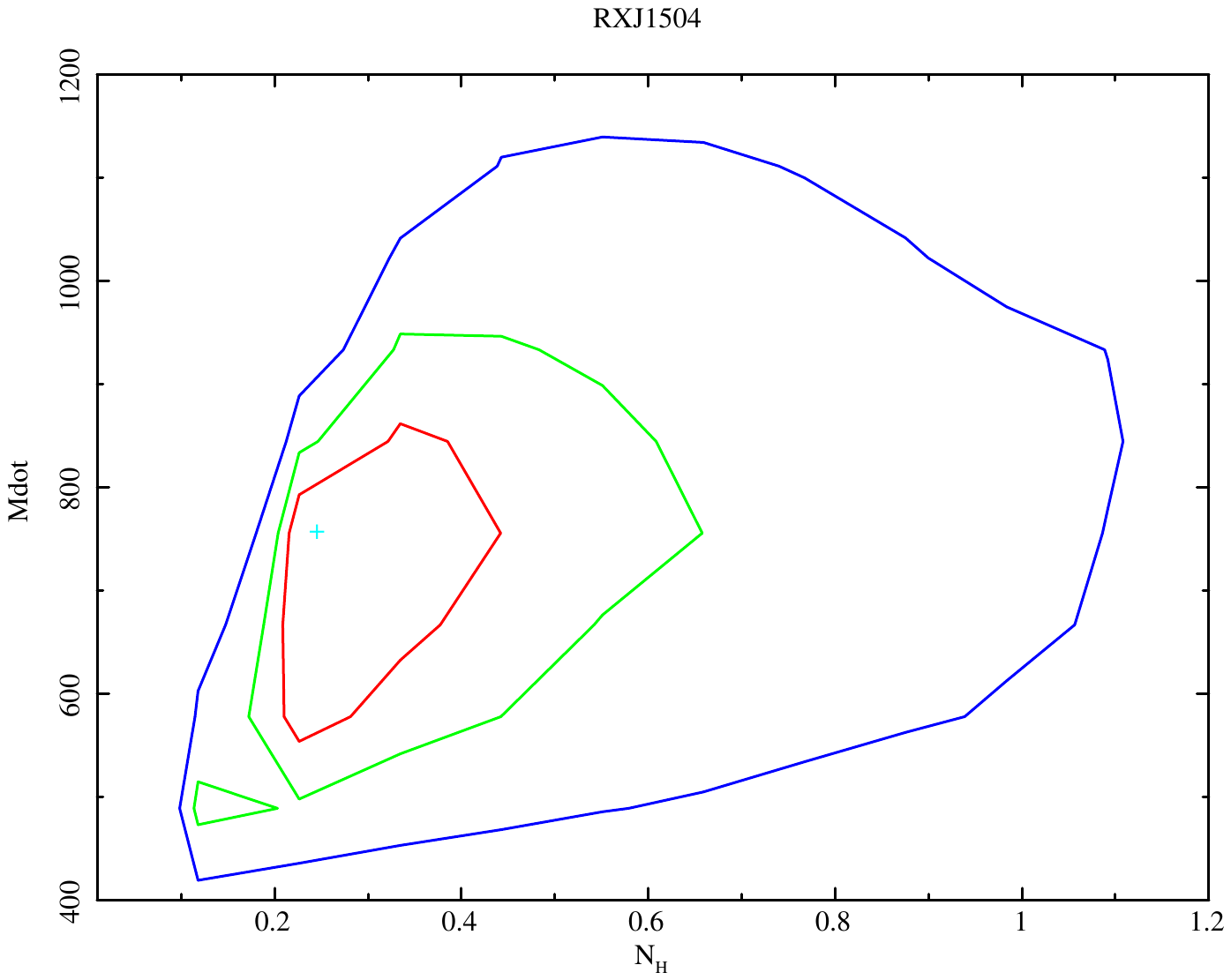} 
\includegraphics[width=0.48\textwidth]{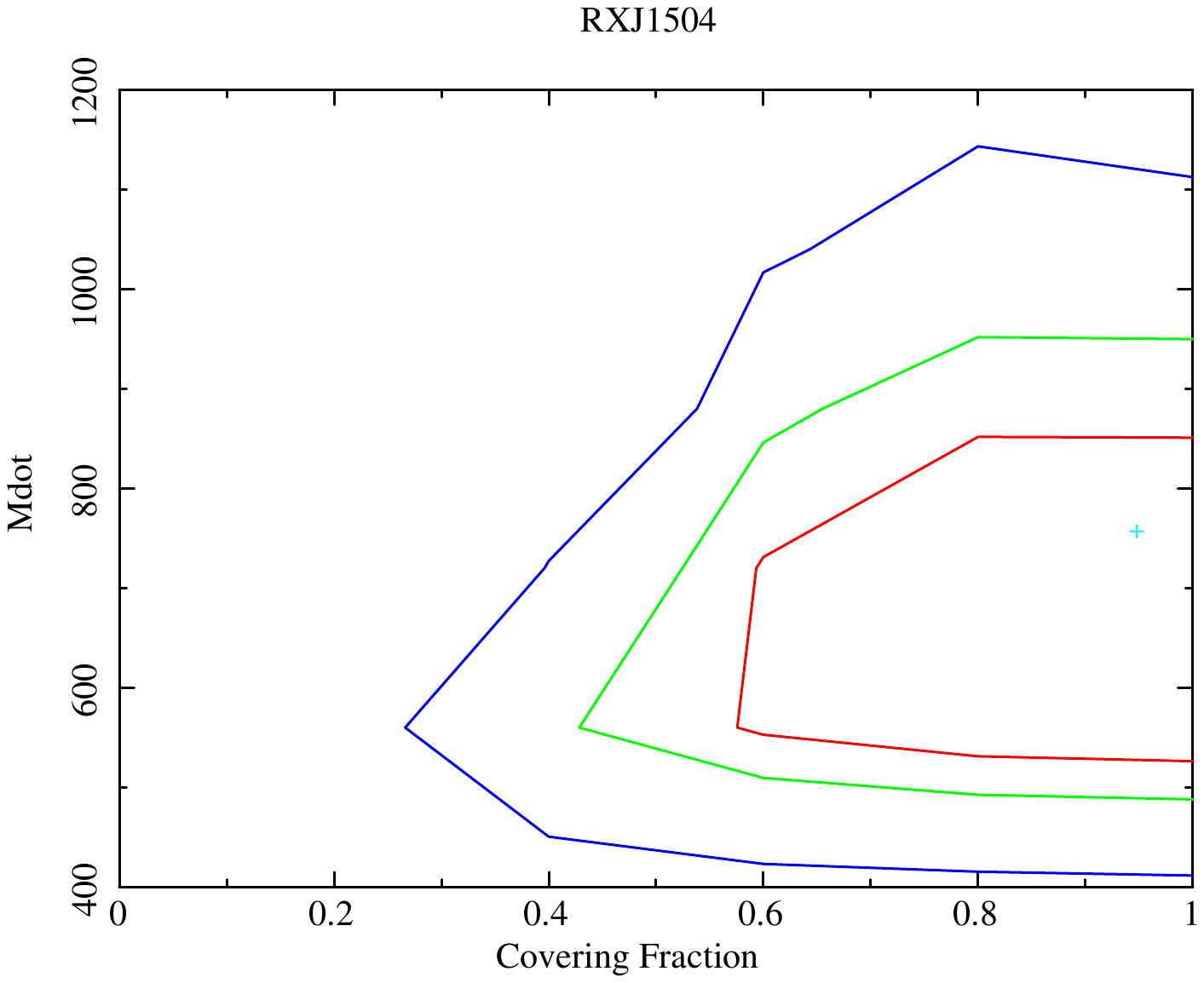} 
    \caption{  a) RGS spectrum, b) hidden mass cooling rate versus total interleaved column density and c) versus Covering fraction.}
\end{figure}

\begin{figure}
    \centering  
\includegraphics[width=0.48\textwidth]{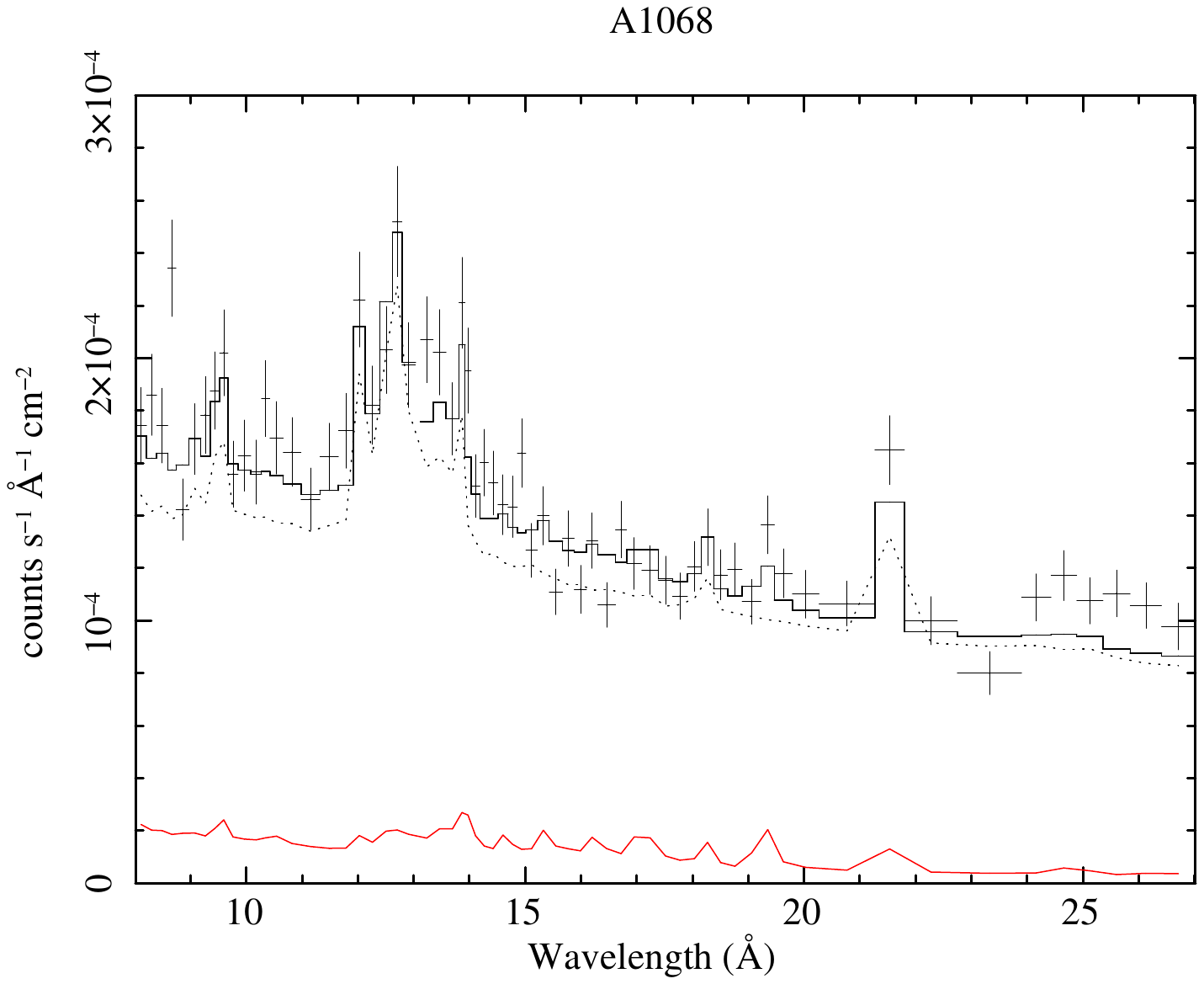}   
\includegraphics[width=0.48\textwidth]{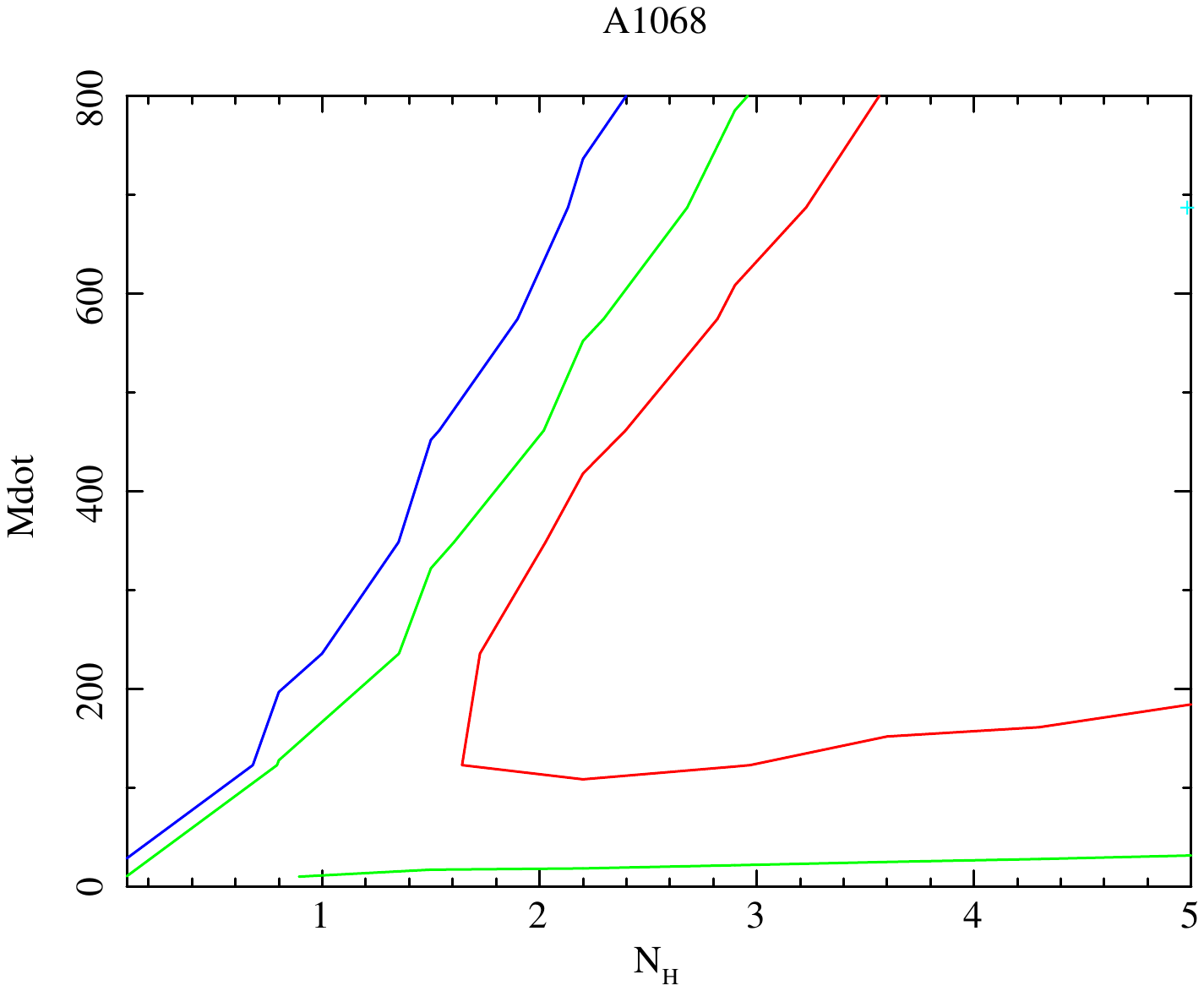} 
\includegraphics[width=0.48\textwidth]{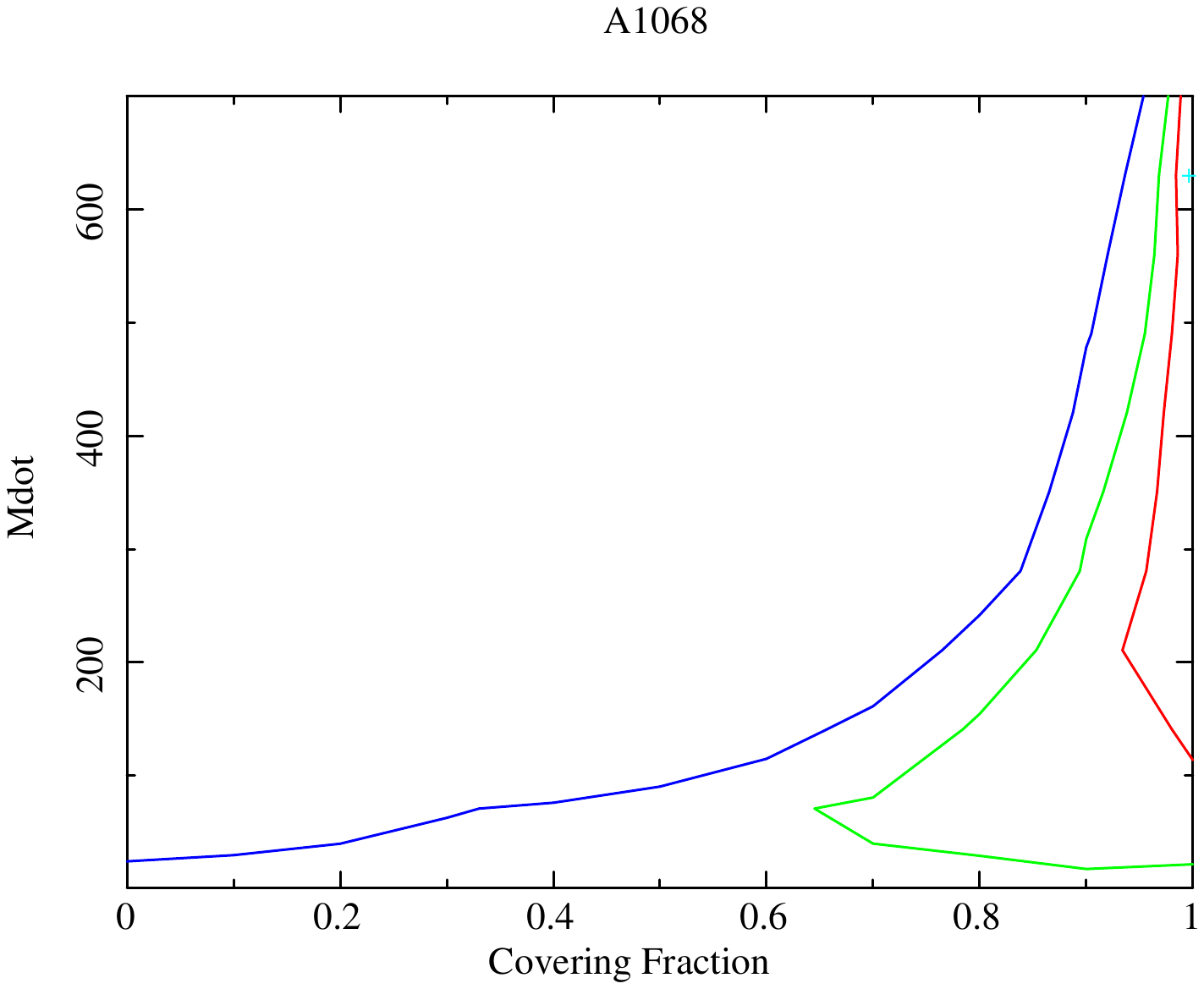} 
    \caption{  a) RGS spectrum, b) hidden mass cooling rate versus total interleaved column density and c) versus Covering fraction.}
\end{figure}

\begin{figure}
    \centering  
\includegraphics[width=0.48\textwidth]{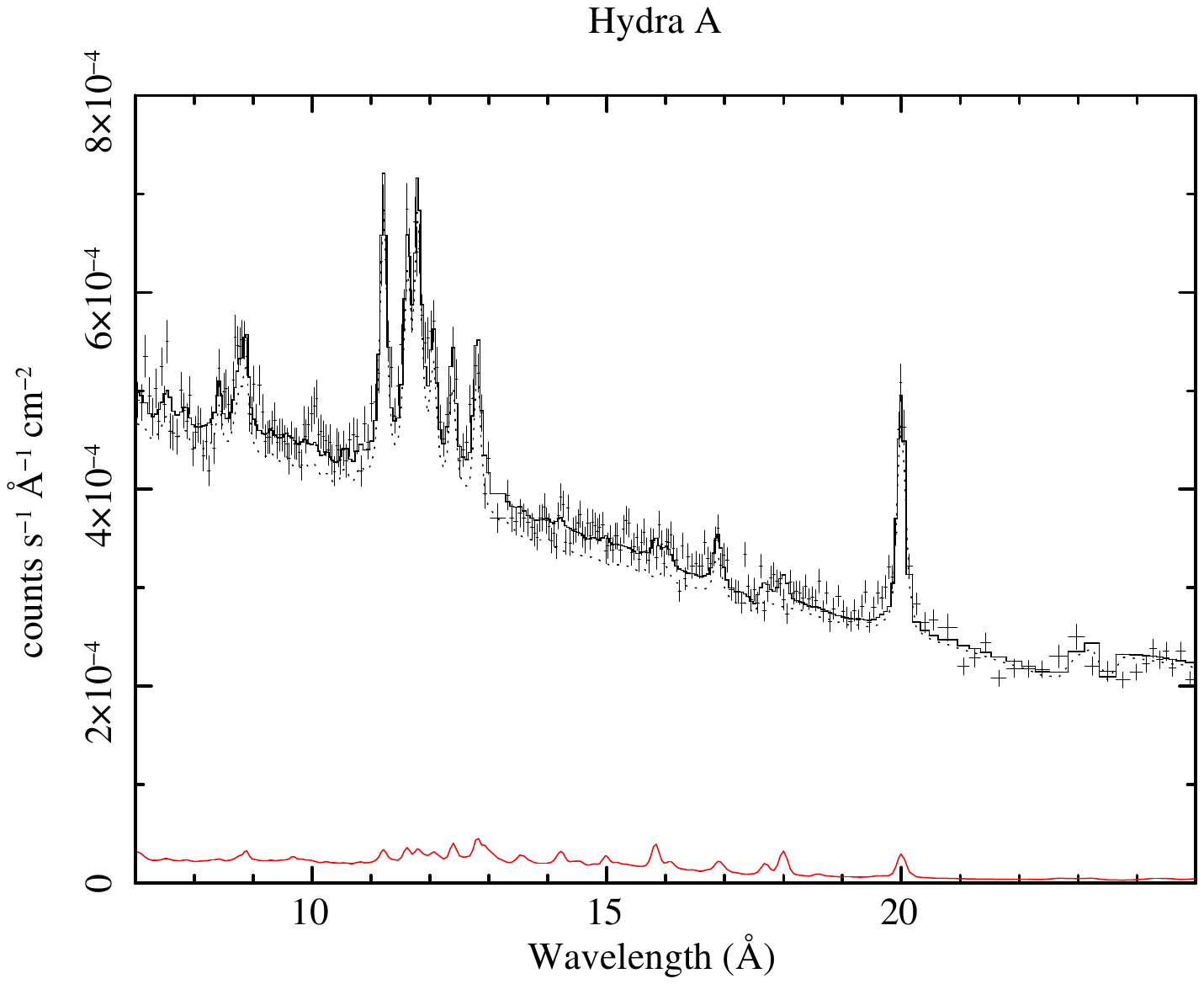}   
\includegraphics[width=0.48\textwidth]{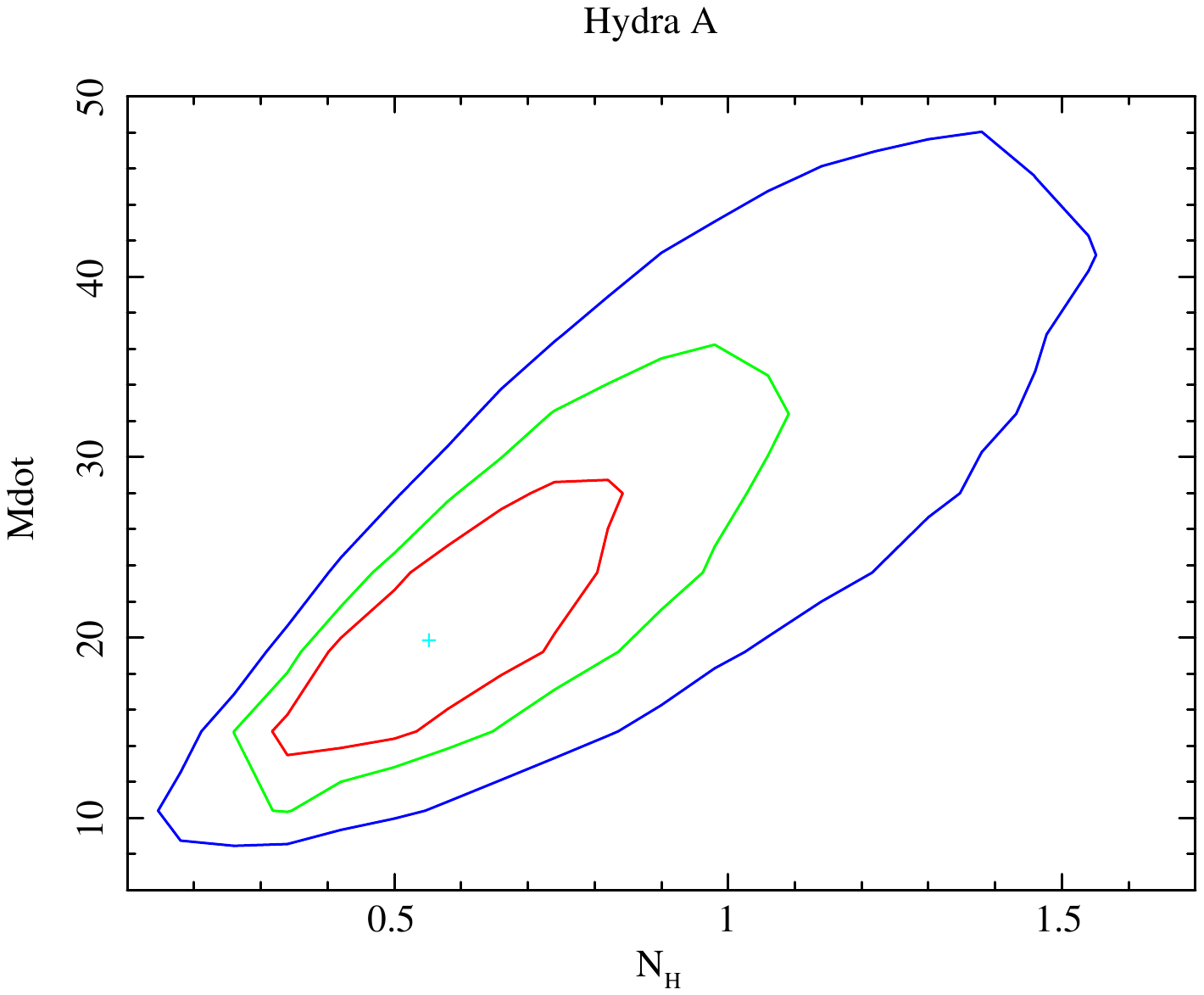} 
\includegraphics[width=0.48\textwidth]{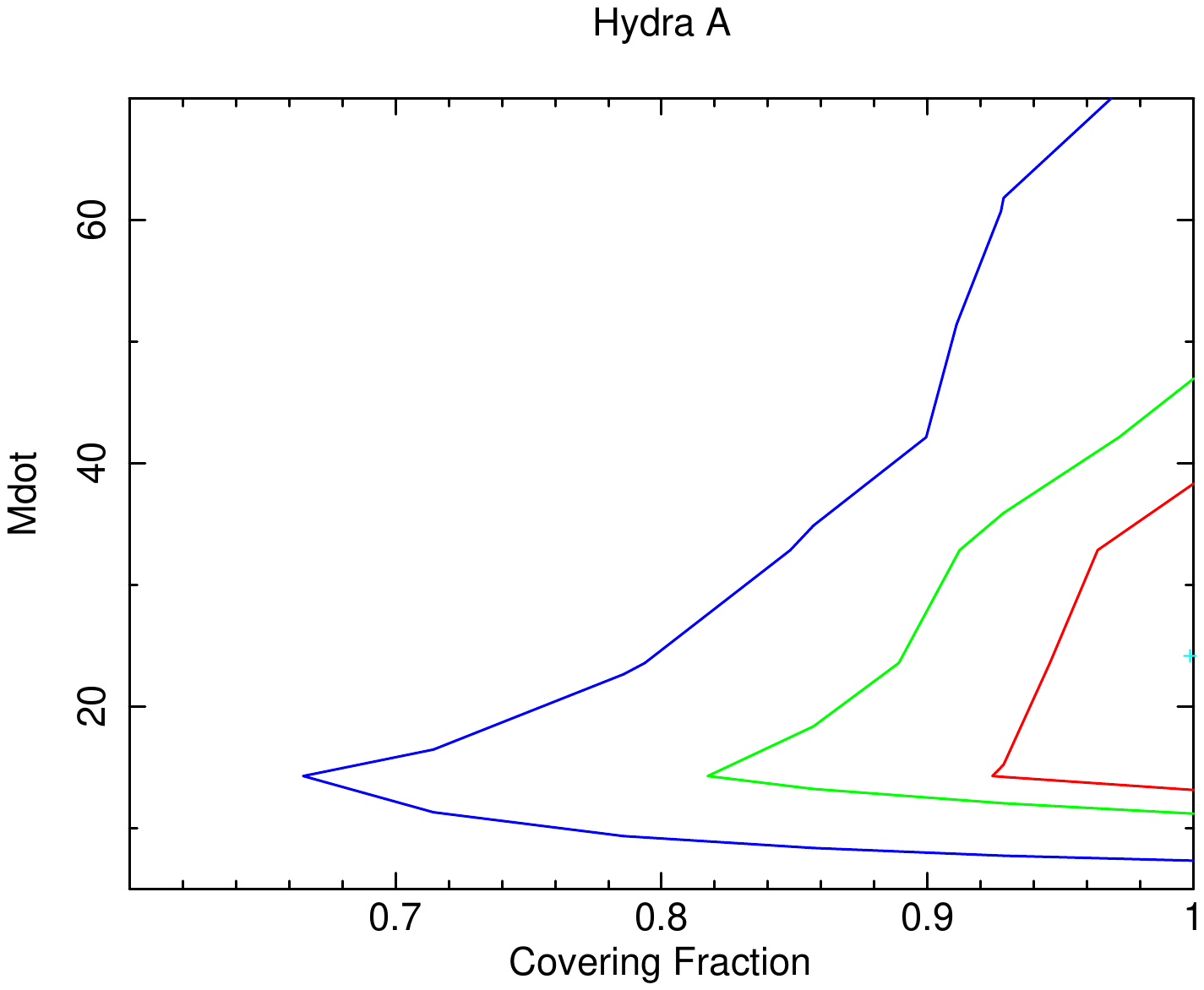} 
    \caption{  a) RGS spectrum, b) hidden mass cooling rate versus total interleaved column density and c) versus Covering fraction.}
\end{figure}

\begin{figure}
    \centering    
\includegraphics[width=0.48\textwidth]{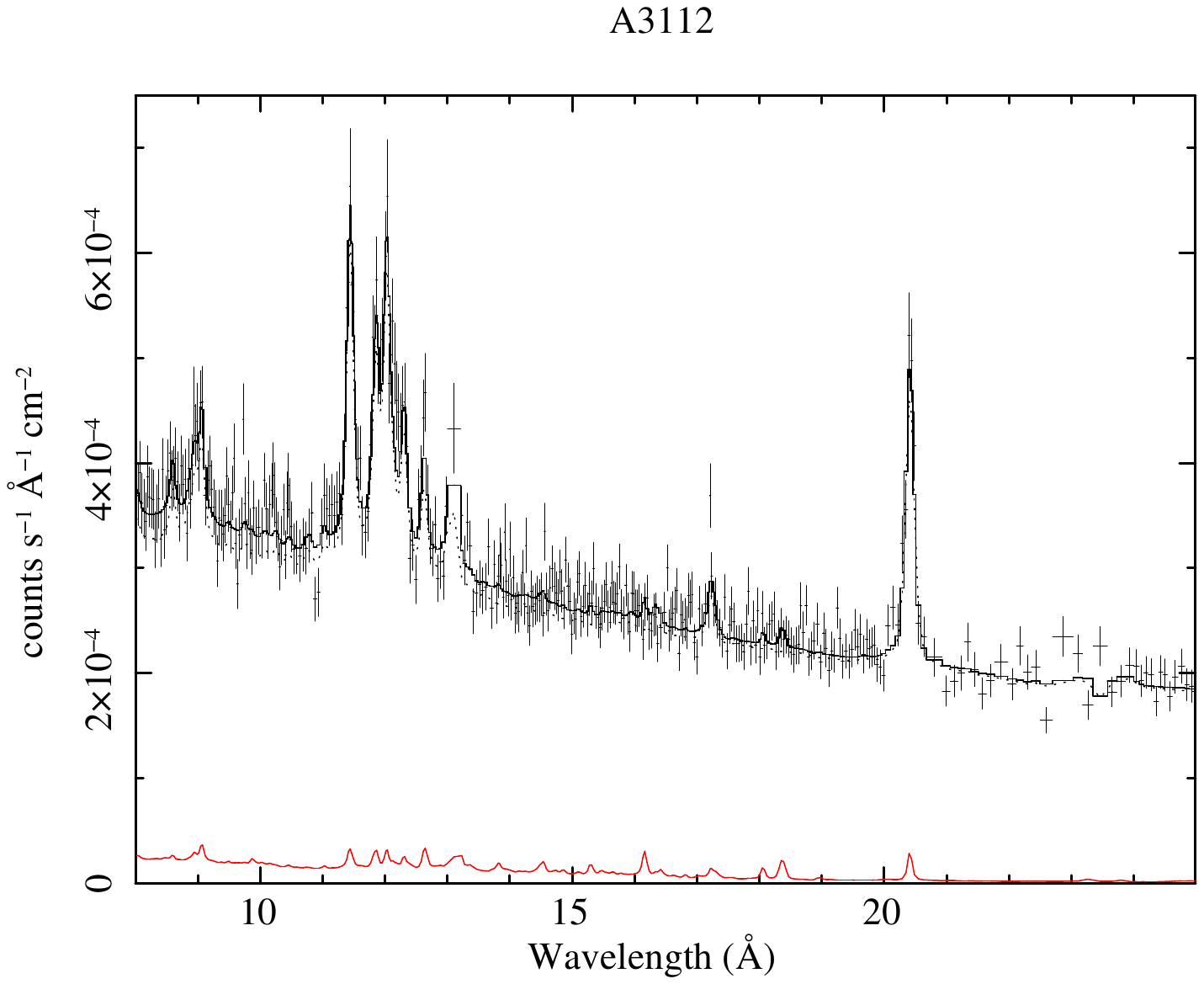}
\includegraphics[width=0.48\textwidth]{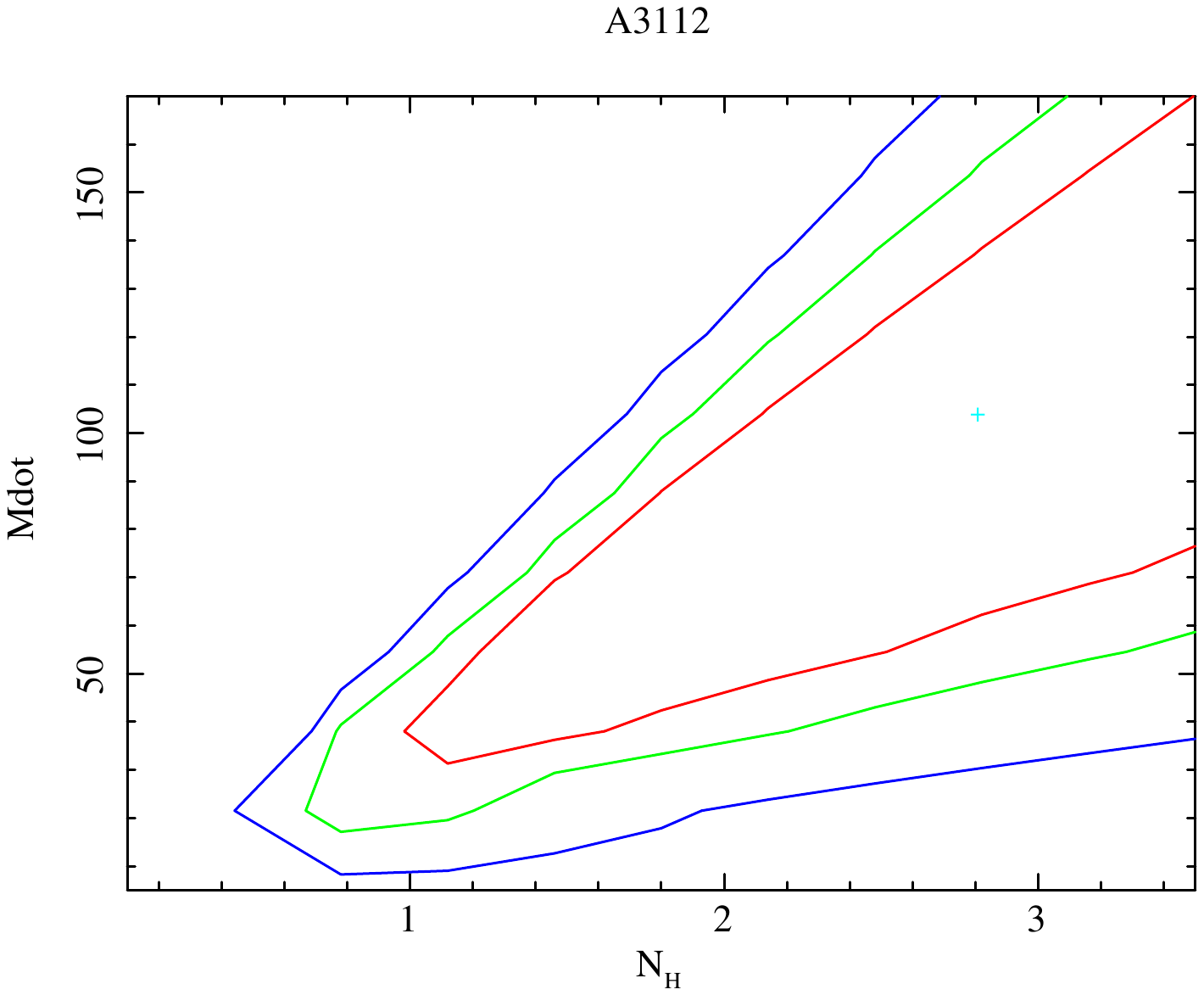} 
\includegraphics[width=0.48\textwidth]{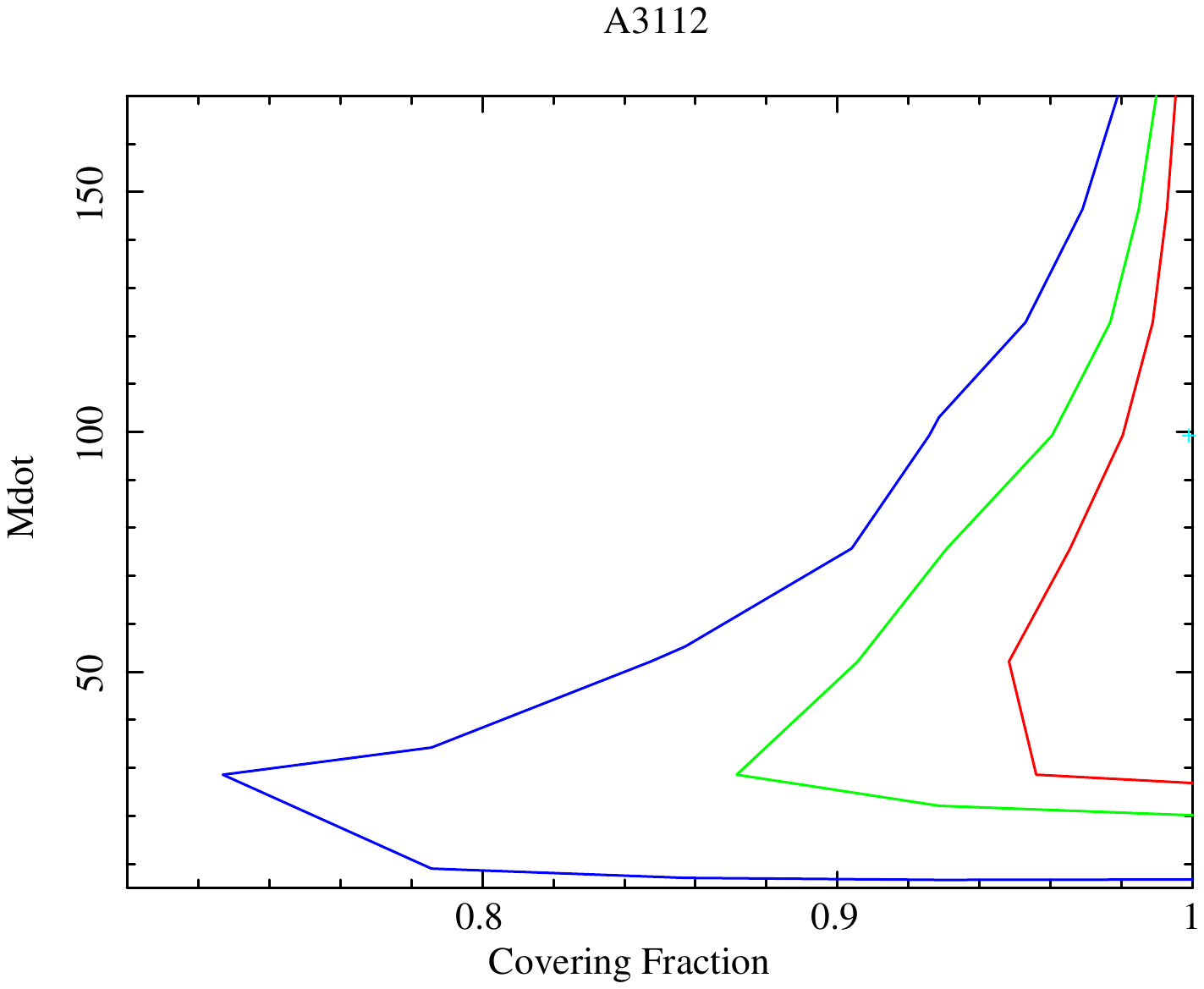} 
    \caption{  a) RGS spectrum, b) hidden mass cooling rate versus total interleaved column density and c) versus Covering fraction.}
\end{figure}

\begin{figure}
    \centering    
\includegraphics[width=0.48\textwidth]{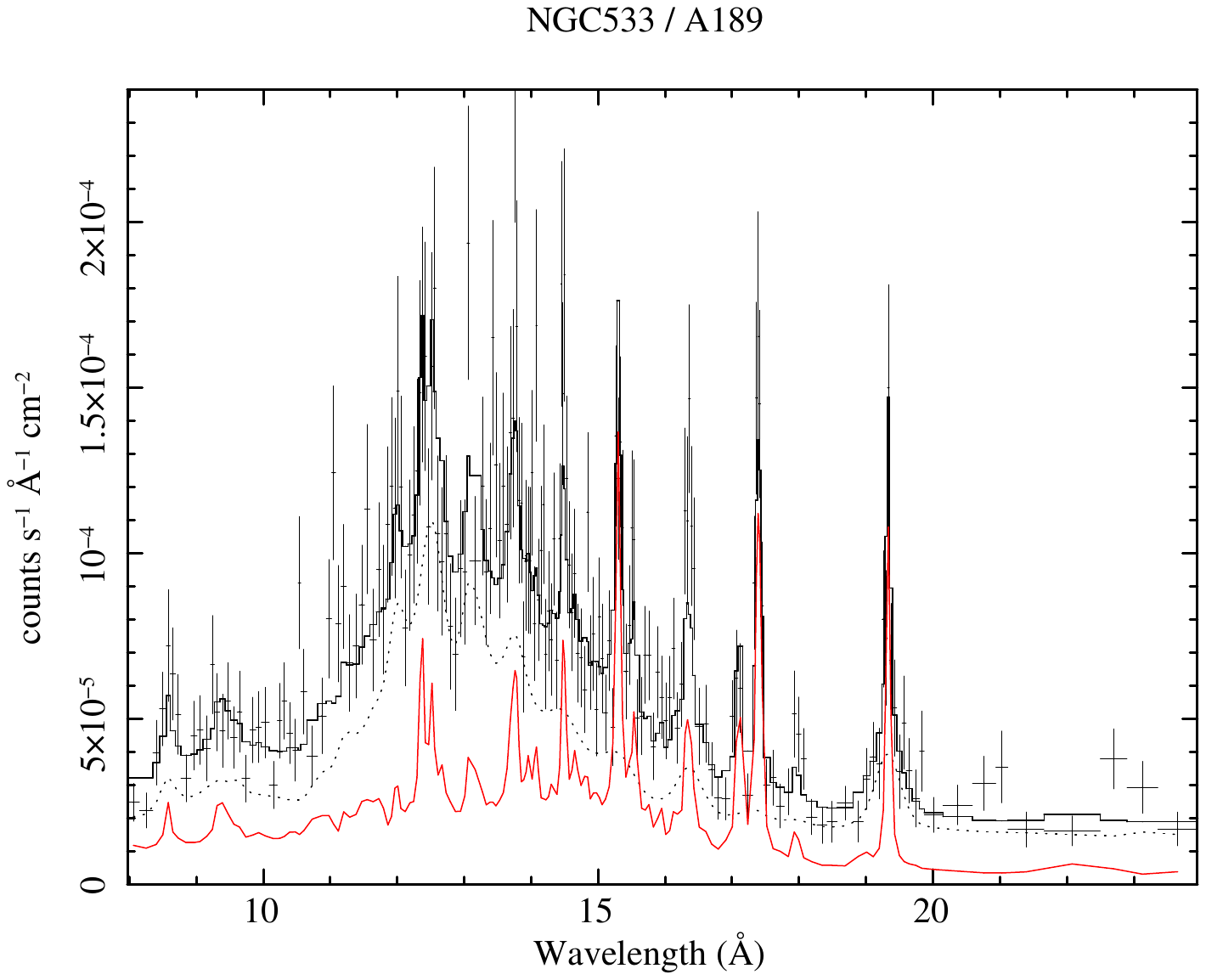}
\includegraphics[width=0.48\textwidth]{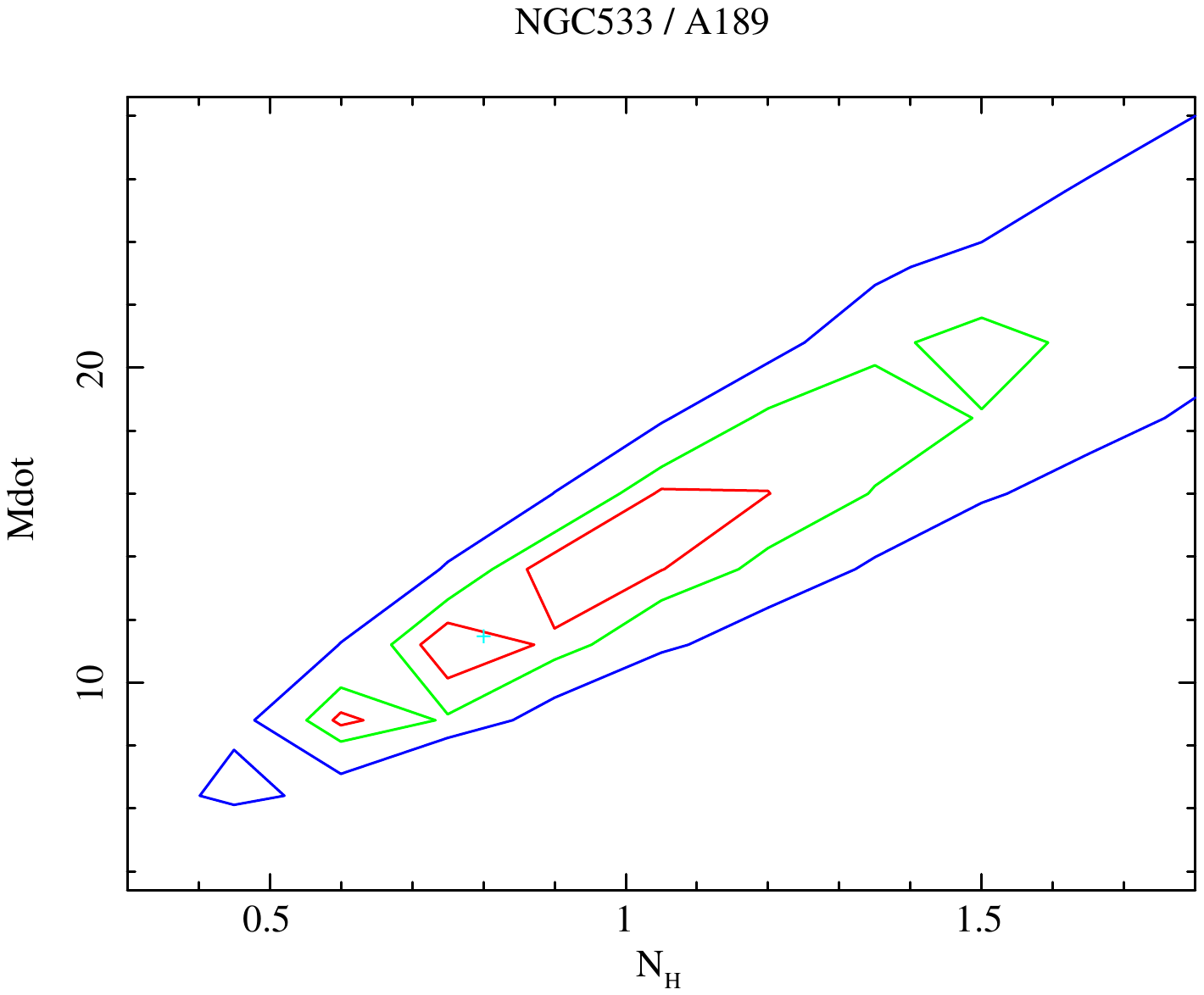} 
\includegraphics[width=0.48\textwidth]{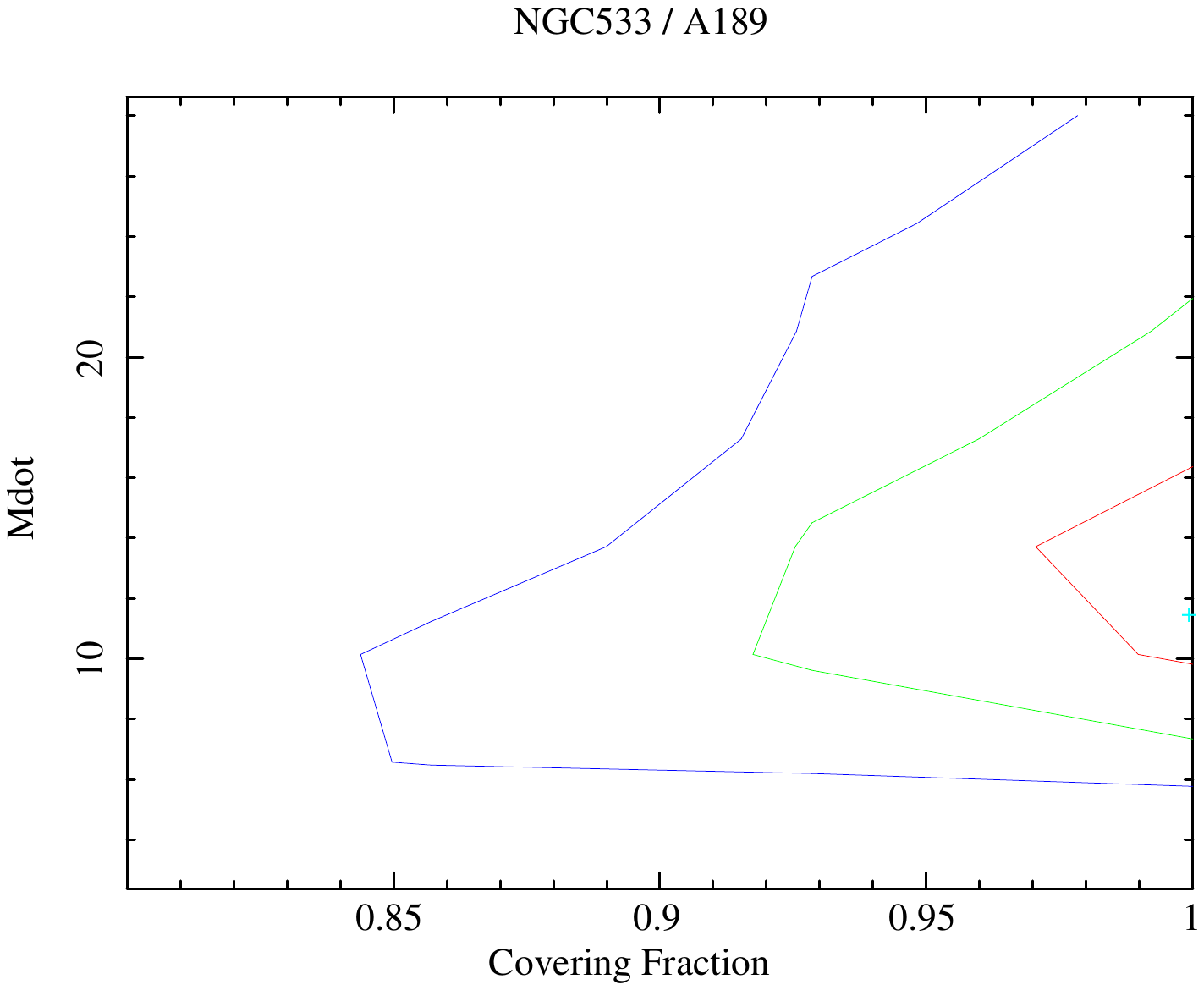} 
    \caption{  a) RGS spectrum, b) hidden mass cooling rate versus total interleaved column density and c) versus Covering fraction.}
\end{figure}

\begin{figure}
    \centering    
\includegraphics[width=0.48\textwidth]{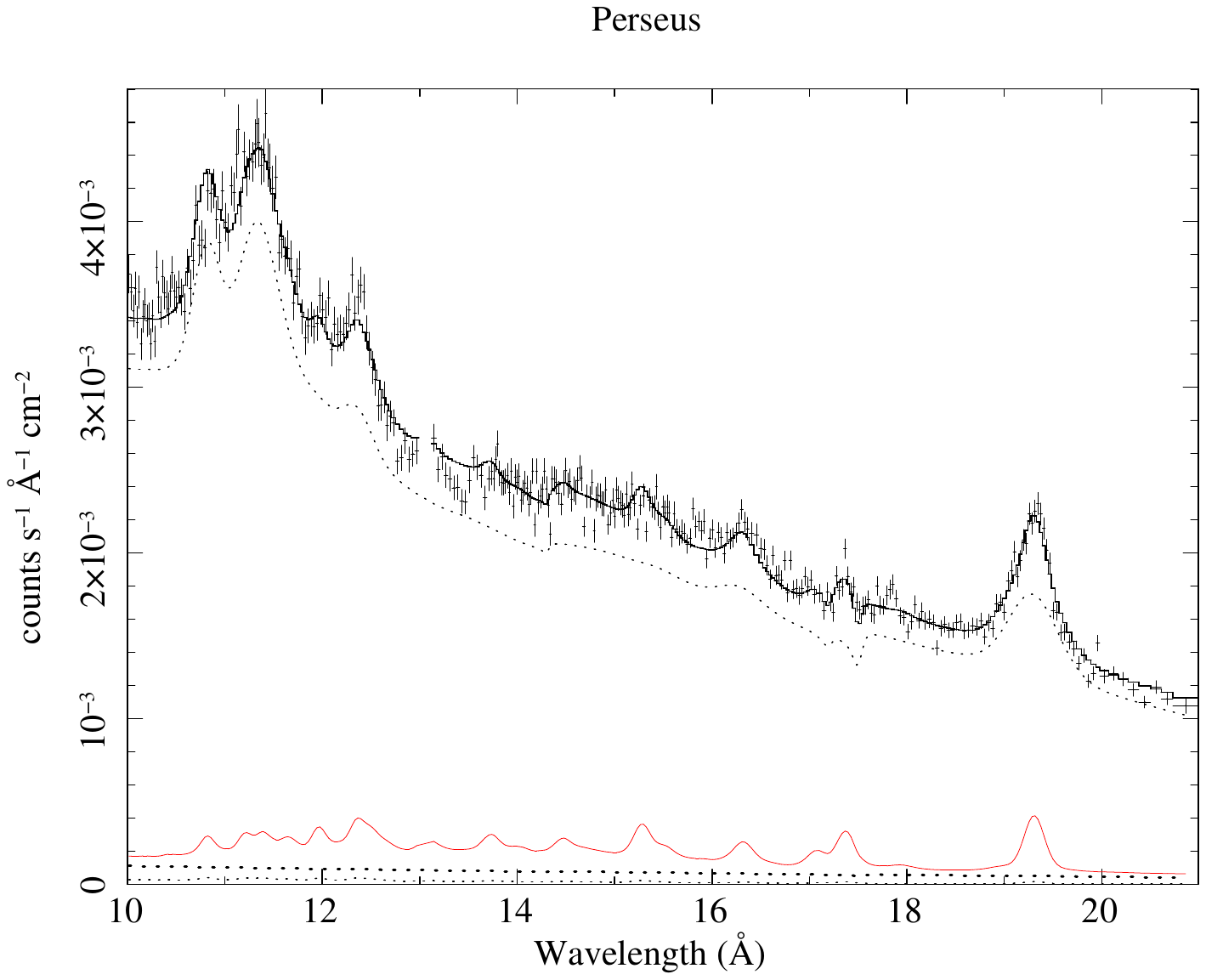} 
\includegraphics[width=0.48\textwidth]{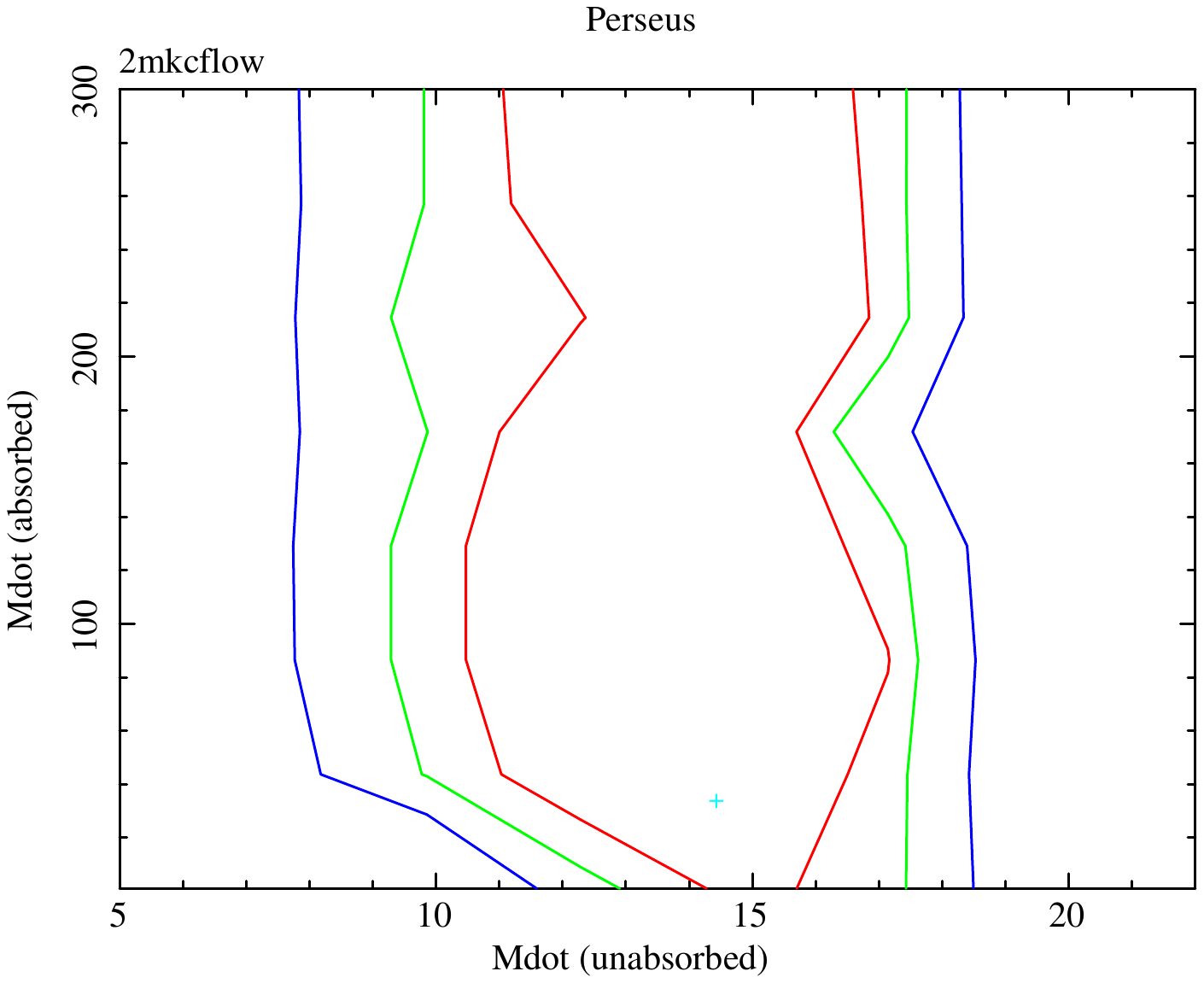} 
    \caption{ a) RGS sepctrum and b) Unabsorbed cooling rate plotted against Hidden rate (HCR) for the Perseus cluster after the maximum value of absorbing column density is raised from $5\times 10^{22}$ to $10^{23}\pcmsq$.   }
\end{figure}

\begin{figure}
    \centering    
\includegraphics[width=0.48\textwidth]{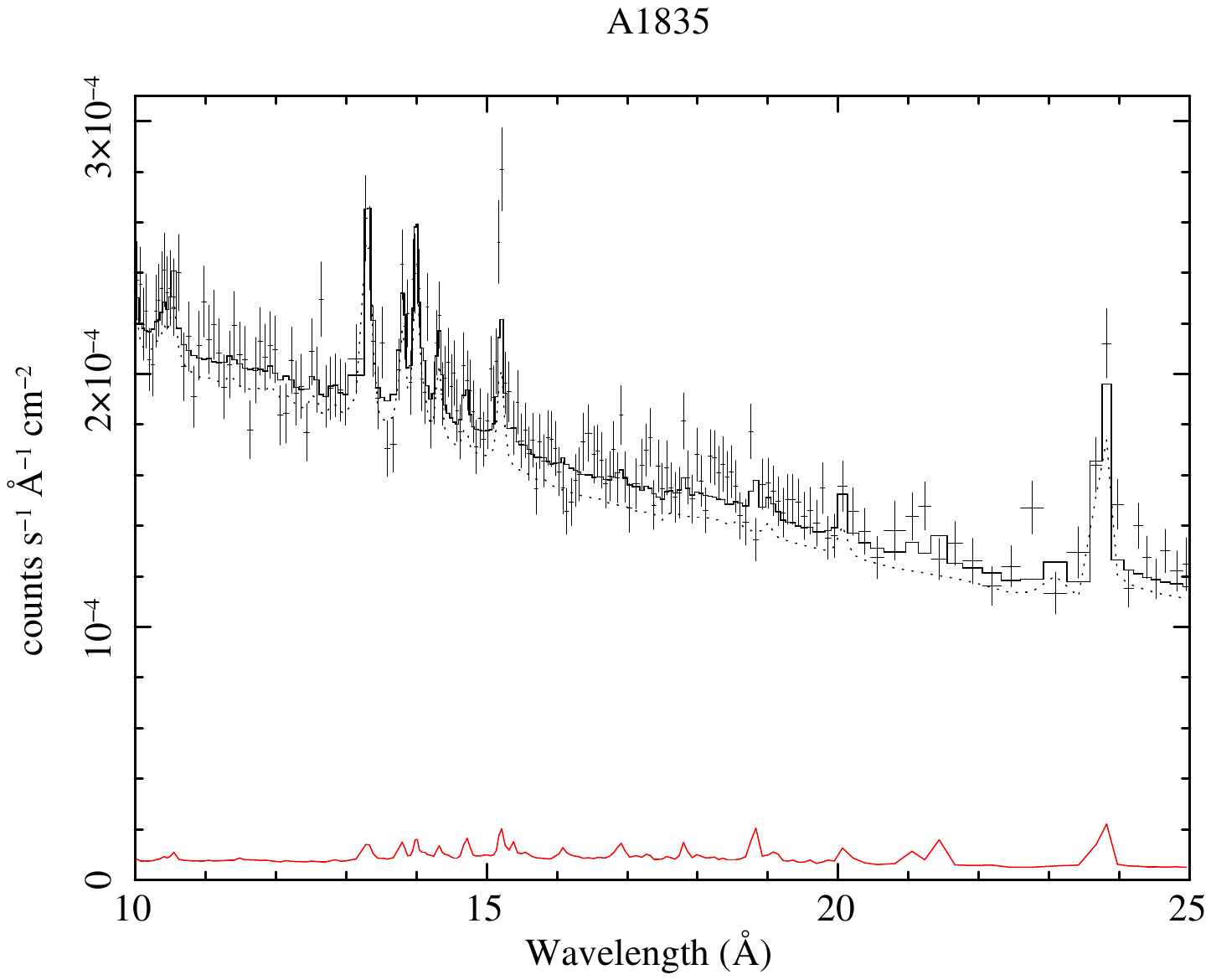}
\includegraphics[width=0.48\textwidth]{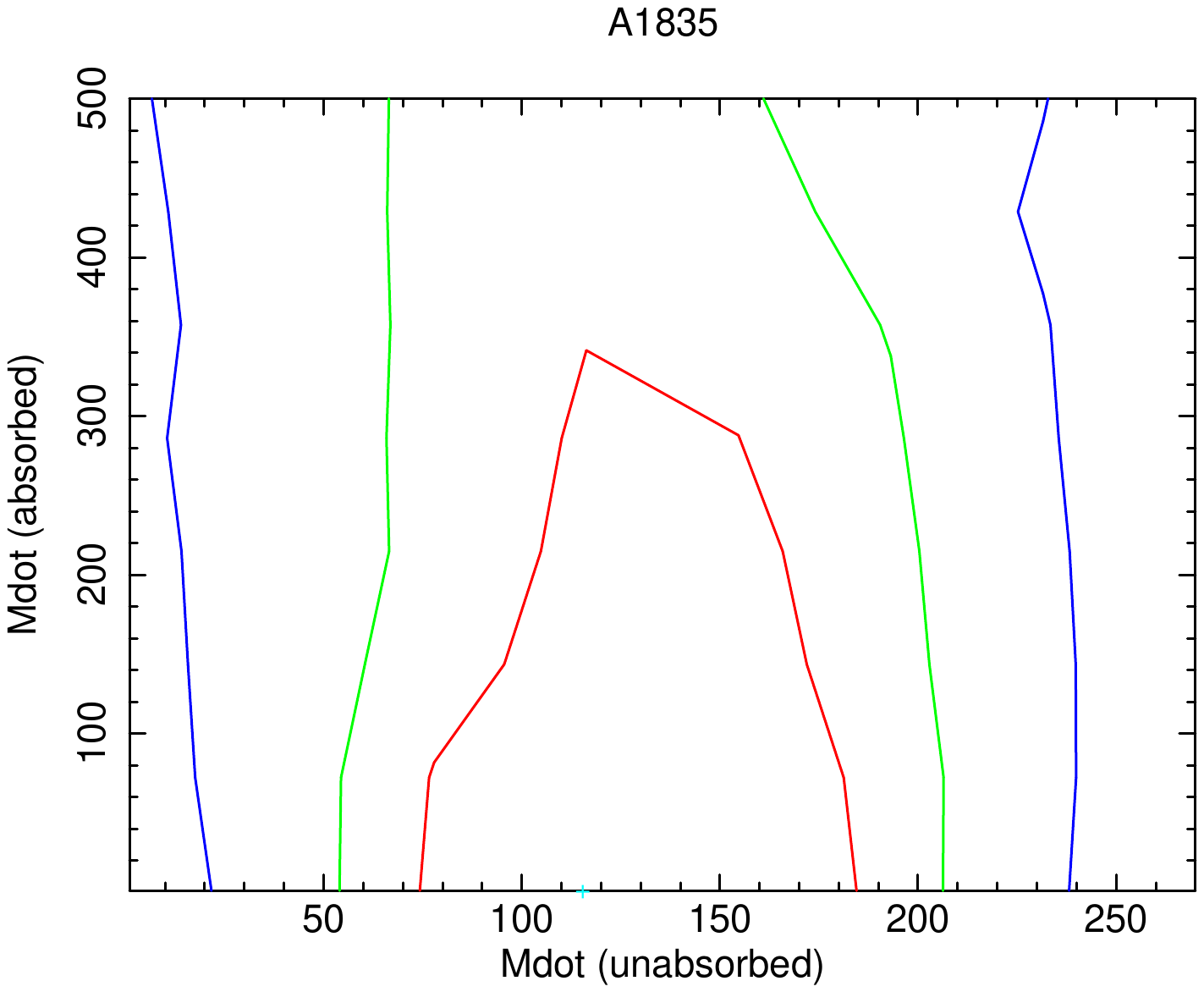} 
    \caption{ a) RGS spectrum and b) Unabsorbed cooling rate plotted against Hidden rate (HCR) for A1835. }
\end{figure}



\section*{Acknowledgements}
ACF acknowledges a discussion with  Mark Krumholz.

\section*{Data Availability}

All data presented here is available on the XMM and Chandra archives.



\bibliographystyle{mnras}
\bibliography{cool_core_Mdot_24} 




\appendix

\section{Further cool core clusters and groups}

\subsection{HCG62}
Hickson Compact Group 62 is X-ray luminous. It has been well observed with Chandra \citep{Morita06, Rafferty13} and shows a cool core disturbed by shocks or more likely a cold front due to a merger. Cavities in the X-ray emission caused by a central jetted radio source are evident. No recent star  formation has been reported.

\subsection{A3581} 
A3581  hosts the most luminous cool core of any nearby group and shows active radio mode feedback in the from of bubbles from the supermassive black hole in its brightest group galaxy, IC 4374. Chandra observations are discussed in detail by \cite{Canning13}. Multiple outbursts are inferred to have occurred.  No young  stars are seen except in a star cluster. 

\subsection{A1795} 
The X-ray cool core cluster A1795 has a 46 kpc long cool X-ray filaments extending 46kpc to the South of the BCG \citep{Crawford05b} which represents the peak of soft X-ray emission. Further Chandra data (it is a Chandra calibration source) are discussed by \citep{Ehlert15}. Star formation is occurring in the long filament.  RGS spectra are presented by \citep{Tamura2001}. Our RGS results show that there is both a modest unabsorbed cooling flow of about $17\Msunpyr$ together with a possibly much larger hidden flow.

\subsection{Cygnus A}
The famous radio galaxy Cygnus A lies at the centre of a cool core cluster of galaxies. Chandra data of the nuclear region were presented by \cite{Young02}, revealing an AGN absorbed by a column density of $N=10^{23}\pcmsq$. This means negligible X-ray emission from the nucleus in the RGS band used here. A dust lane is seen across the centre \citep{Riffel21, Carilli22}). The plot of column density versus mass cooling rate was made with unit covering fraction. 

\subsection{A1664}
Chandra data on A1664 are presented by \citep{Calzadilla19}. 
\citep{Liu2021} analyse the RGS data of A1664. The ALMA molecular mass is $10^{10}\Msun.$

\subsection{RXJ1504}
RXC J1504.1-0248 is a high luminosity massive cluster with a bright cool core accounting for 70 per cent of its X-ray luminosity \citep{Bohringer05}. Chandra observations are discussed by \cite{Vantyghem2018} together with ALMA and HST data revealing $2\times 10^{10}\Msunpyr$ of molecular gas and a system of dusty filaments. \cite{Ogrean2010} finds a star formation rate of $136\Msunpyr$ from GALLEX UV data. \cite{Liu2021} analysed the RGS X-ray spectrum and deduced a hidden cooling flow rate of $520\pm 30\Msunpyr$. 

\subsection{A1068}
Chandra data is presented in \citep{Wise04}, with further analysis by \cite{Rafferty08}. \citep{Edge10} present Herschel FIR data and give a CO gas mass of $4.1\times 10^{10}\Msun$. The plot of column density versus mass cooling rate was made with unit covering fraction. 

\subsection{Hydra A}
Observed by Chandra \citep{McNamara00} and \citep{Gitti11}. \cite{Olivares19} give a molecular mass of $5.5\times 10^9\Msun.$ \citep{Rose19} finds molecular absorption in the form of a dense  disc.

\subsection{A3112}
The cool core cluster A3112 has been discussed by\citep{Calzadilla22}, \cite{Quillen08} and \citep{Rose19}.

\subsection{NGC533}
NGC533 is a massive elliptical galaxy at a distance of about 70 Mpc, at the centre of a group or poor cluster, sometimes confused with the more distant cluster A189. 

It was studied by \citep{Peterson2003} in a cooling flow study  of 14 objects observed with the XMM RGS. A cooling flow model was fitted to successive temperature intervals showing that Mdot dropped from $5\pm 0.4\Msunpyr$ to values consistent with zero below 0.4 keV. No intrinsic absorption was considered. \citep{Lakhchaura2018} have deprojected the Chandra data to find that the radiative cooling time within the central 600 pc is 30 Myr.  The luminosity of h$\alpha$ emission is $3.24 \times 10^{40}\ergps$ \citep{Macchetto96}. \citep{Grossova2022} show radio emission contours revealing a low power FRII double source with lobes E-W at radii of  1 kpc. 

Our RGS analysis (Fig. 9) yields an excellent fit to a mass cooling rate of about $12\Msunpyr$ with intrinsic absorption of $1.2\times 10^{22}\psqcm$.

\citep{Gu22} have studied the stellar IMF in the centre of 41 massive early-type galaxies, including NGC533. All have an IMF steeper than a Milky Way (MW, Kroupa) IMF, i.e. a bottom heavy IMF. NGC533 requires, within its innermost 4 kpc radius, a mass to light ratio $74\pm0.15$ per cent larger than would be expected from a MW IMF.  This could be the repository for much of the cooled gas. 
 
\subsection{A1835 and the Perseus Cluster}
This massive cool core cluster was discussed in HCF1. It is shown again here with a clearer spectrum. The analysis of the Perseus cluster shown in HCFI (Fig. 5a) allowed a maximum  value of $N_{\rm H}$ of $5\times 10^{22} \pcmsq$. Extending that limit to $10^{23}\psqcm$ allows higher values of Mdot (Fig. 5b). The high FIR luminosity from the cluster  means that we have no strong soft X-ray constraint on the total level of cooling. The values found from Chandra of $\sim 400\Msunpyr$ are possible.


\bsp	
\label{lastpage}
\end{document}